\PassOptionsToPackage{safe}{pbalance}
\documentclass[sigconf,nonacm,pbalance]{acmart}

\usepackage{pvldb}

\ExplSyntaxOn
\cs_if_exist:NT \ignoreprimitiveerror
  {
    \cs_set_protected:Npn \__mark_vbox_set_split_to_maxdimen:NN #1#2
      {
        \tl_set:Ne \l__mark_saved_parameters_tl
          { \ignoreprimitiveerror \int_use:N \ignoreprimitiveerror \scan_stop: }
        \ignoreprimitiveerror 1 \scan_stop:
        \vbox_set_split_to_ht:NNn #1 #2 { \c_max_dim }
        \l__mark_saved_parameters_tl
      }
  }
\ExplSyntaxOff

\renewcommand\vldbdoi{XX.XX/XXX.XX}
\renewcommand\vldbpages{XXX-XXX}
\renewcommand\vldbavailabilityurl{https://github.com/Texera/bobflow}

\usepackage{pifont}

\newcommand{\xmark}{\ding{55}}
\usepackage{xspace}
\usepackage[inline]{enumitem}
\usepackage{multirow}
\newcommand{\sys}{BobFlow\xspace}
\newcommand{\agent}{\textit{Bob}\xspace}
\newcommand{\boldstart}[1]{\par\noindent\textbf{#1}\ }

\usepackage[most]{tcolorbox}
\usepackage{array}
\usepackage{tabularx}
\definecolor{cardline}{RGB}{70,90,100}
\definecolor{cardback}{RGB}{246,247,248}
\newtcolorbox{experimentobservation}[1]{
  enhanced, colback=white, colframe=black!55, boxrule=0.45pt, arc=2pt,
  boxsep=0pt, left=4pt, right=4pt, top=2pt, bottom=2pt, width=\linewidth,
  before skip=2pt, after skip=2pt, fontupper=\small,
  before upper={\textbf{Observation~#1.}\ }}

\usepackage{xcolor}
\definecolor{labelbg}{RGB}{235,240,245}

\usepackage{tikz}

\newcommand{\parlabel}[1]{%
  \tikz[baseline=(label.base)]{
    \node[
      fill=labelbg,
      rounded corners=2pt,
      inner xsep=2pt,
      inner ysep=0.5pt,
      outer sep=0pt
    ] (label) {\emph{#1}};
  }%
  \hspace{2pt}%
}

\usepackage{placeins}
\usepackage{needspace}

\usepackage{booktabs}

\usepackage{graphicx}

\usepackage{amsthm}
\theoremstyle{definition}
\newtheorem{definition}{Definition}[section]

\usetikzlibrary{arrows.meta, positioning, fit, calc, backgrounds, decorations.pathmorphing, shapes.geometric}
\usetikzlibrary{shapes.multipart}
\definecolor{opheader}{RGB}{234,240,244}
\definecolor{opparam}{RGB}{54,81,99}
\tikzset{
  bfop/.style={rectangle split,rectangle split parts=2,
    rectangle split part fill={opheader,white},
    rectangle split part align={center,center},
    draw=cardline,line width=.55pt,rounded corners=1pt,
    inner xsep=3pt,inner ysep=2pt,
    font=\fontsize{9}{10}\selectfont\sffamily\bfseries,text=black},
  bfid/.style={font=\fontsize{7.5}{8.5}\selectfont\ttfamily,
    text=cardline,inner sep=1pt},
  bfedge/.style={-{Stealth[length=1.4mm]},draw=cardline,line width=.6pt}
}

\newcommand{\bfports}[2]{%
  \coordinate (#1-out) at (#1.east);
  \draw[draw=black!65,fill=white,line width=.55pt] (#1-out) circle[radius=.055];
  \ifnum#2=1
    \coordinate (#1-in1) at (#1.west);
    \draw[draw=black!65,fill=white,line width=.55pt] (#1-in1) circle[radius=.055];
  \fi
  \ifnum#2=2
    \coordinate (#1-in1) at ($(#1.west)+(0,.16)$);
    \coordinate (#1-in2) at ($(#1.west)+(0,-.16)$);
    \draw[draw=black!65,fill=white,line width=.55pt] (#1-in1) circle[radius=.055];
    \draw[draw=black!65,fill=white,line width=.55pt] (#1-in2) circle[radius=.055];
  \fi}
\newcommand{\bfconnect}[3]{%
  \draw[bfedge,shorten <=1.6pt,shorten >=1.6pt] (#1-out) -- (#2-in#3);}
\definecolor{opadded}{RGB}{23,122,63}
\definecolor{opupdated}{RGB}{193,141,0}

\newcommand{\bfmarkadded}{%
  \fill[opadded] (0,0) circle[radius=.125];
  \draw[white,line width=.85pt,line cap=round,line join=round]
    (-.058,.004) -- (-.020,-.044) -- (.062,.046);}
\newcommand{\bfmarkupdated}{%
  \fill[opupdated] (0,0) circle[radius=.125];
  \begin{scope}[rotate=-40]
    \fill[white] (-.085,-.029) -- (.016,-.029) -- (.086,0) -- (.016,.029)
      -- (-.085,.029) -- cycle;
    \draw[opupdated,line width=.35pt] (.016,-.028) -- (.016,.028);
  \end{scope}}

\begin{document}

\title{\sys{}: Towards Human--Agent Collaborative Data Science Using Dataflows}

\author{Jiadong Bai}
\affiliation{%
  \institution{University of California, Irvine}
  \city{Irvine}
  \state{California}
  \country{USA}
}
\email{jiadongb@ics.uci.edu}

\author{Yicong Huang}
\affiliation{%
  \institution{University of Massachusetts Amherst}
  \city{Amherst}
  \state{Massachusetts}
  \country{USA}
}
\email{yiconghuang@cs.umass.edu}

\author{Chen Li}
\affiliation{%
  \institution{University of California, Irvine}
  \city{Irvine}
  \state{California}
  \country{USA}
}
\email{chenli@ics.uci.edu}

\renewcommand{\shortauthors}{Bai et al.}

\begin{abstract}
Recent advances in large language models (LLMs) have established human--agent collaboration as a promising paradigm for data science. Effective collaboration requires a shared computational abstraction that enables agents to reason about processing logic while allowing human analysts to understand, guide, and refine the analysis.
Although scripts written in Python or SQL provide precise control, their syntax and execution details can be difficult for analysts to understand and modify.
In this paper, we argue that dataflows are an effective abstraction for human--agent collaboration in data science.
We introduce \sys, a system that combines the dataflow abstraction with the ReAct paradigm.
It enables agents to iteratively construct and execute dataflows for data science tasks while allowing analysts to inspect the analysis and provide feedback.
Our user study shows that \sys helps analysts understand, inspect, and verify agent-generated analyses. A thorough experimental evaluation across proprietary and open-weight LLMs shows accuracy score gains of up to 18.6 points at just over 1/3 of a script-based ReAct agent's inference cost.
\end{abstract}

\maketitle

\vldbtopmatter

\section{Introduction}
\label{sec:introduction}
Recent advances in large language models (LLMs) are fundamentally transforming how humans and AI agents collaborate on data science tasks~\cite{luo2026dataagents,you2025datawiseagent}.
Most commonly, agent-assisted data science has followed a human-led paradigm, where human analysts drive the analytical process, and AI agents assist with individual tasks, such as code generation, data transformation, and result interpretation~\cite{dataformulator2,conf/uist/Kazemitabaar0DG24}.
Recently, a growing number of LLM-powered agents have demonstrated the ability to plan and execute complex analytical tasks autonomously.
Systems such as Claude Science~\cite{claude-science}, OpenAI's in-house data agent~\cite{openai-data-agent}, and Google's multi-agent Co-Scientist~\cite{gottweis2025towards} exemplify this trend.
As a result, data science is shifting toward an agent-led, human-guided paradigm, in which analysts increasingly supervise agent-driven analyses, evaluate analytical decisions, and provide high-level guidance.

\looseness=-1
This shift introduces new challenges for effective human--agent collaboration~\cite{conf/uist/Kazemitabaar0DG24,xie2024waitgpt}.
In traditional human-led analytics, analysts maintain a natural understanding of the analytical process because they directly perform and control its individual steps.
In contrast, when agents autonomously carry out complex, multi-step analyses, potentially involving multiple collaborating agents~\cite{you2025datawiseagent,gottweis2025towards,autogen2024}, human analysts may no longer be directly involved in the decisions and operations that produce the results.
To effectively supervise these processes, humans must be able to quickly understand what agents have done, evaluate their analytical decisions, and provide timely feedback to guide subsequent steps~\cite{zeng2025airepr,schmidgall2025agentlaboratory}.
Moreover, analysts may wish to intervene in ongoing analyses or revisit earlier analytical states to explore alternative directions.
This calls for interaction mechanisms that let humans efficiently understand, evaluate, and steer agent-driven analyses while preserving agents' autonomy.

\boldstart{The Need for a Shared Computational Abstraction.}
A central design choice for such human--agent collaboration is the \emph{computational abstraction}, an intermediate representation through which humans and agents conduct and interact with an analysis.
Beyond expressing computational intent, this abstraction determines how analytical steps and their dependencies are exposed, how humans can inspect and modify agent-generated analyses, and how agents can reason about intermediate results and plan subsequent actions.
It must support human comprehension and agent reasoning.

Existing computational abstractions offer different trade-offs.
Natural language is accessible but often lacks precise semantics~\cite{xu2026prepbench}.
General-purpose programming languages, such as Python, are expressive and precise, but require analysts to understand program syntax, variables, and execution state to inspect or modify an analysis~\cite{b2notebooks2020,conf/icde/TaylorHHLCWL24}.
SQL provides a higher-level declarative abstraction with well-defined relational semantics, but does not naturally expose intermediate steps as independently inspectable and editable components.
These abstractions express computations effectively, but none is designed specifically for the iterative inspection, modification, and feedback that human--agent collaboration requires.

\begin{figure*}[htbp]
  \centering
  \scalebox{1}[0.9]{\includegraphics[width=\textwidth,trim=0 29bp 0 14bp,clip]{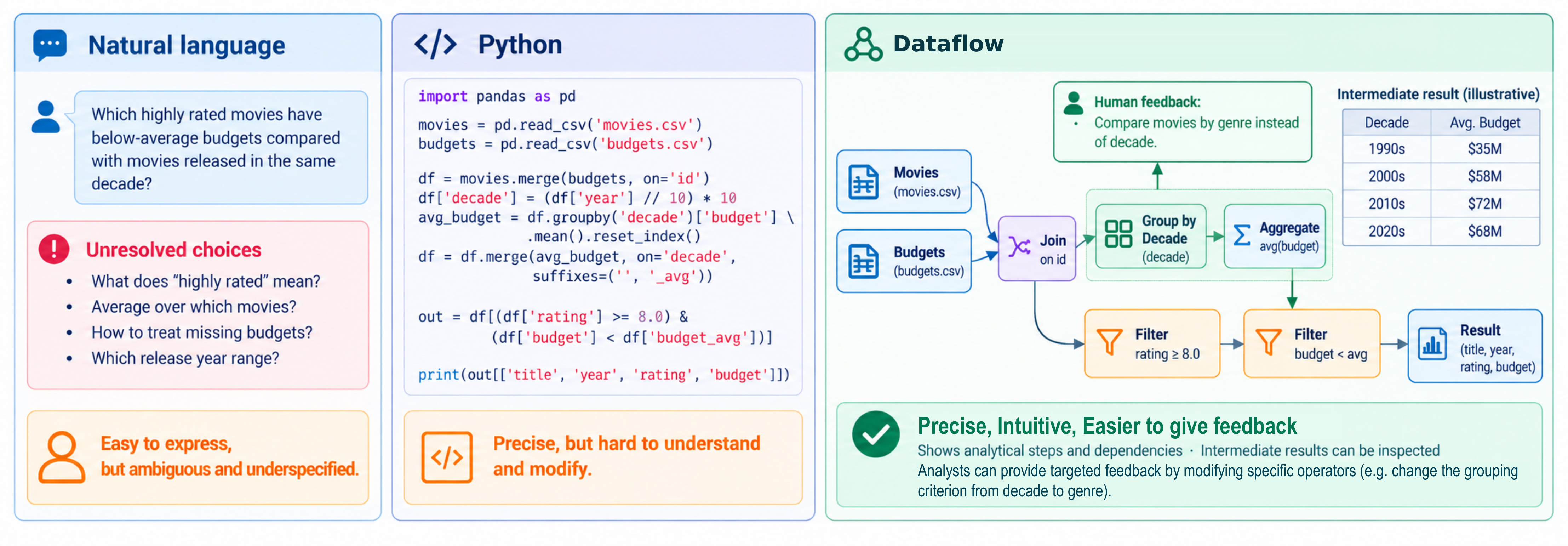}}
\par
  \caption{\protect Three abstractions of the same movie-analysis task. Natural language leaves decisions unspecified; Python is precise but harder to understand and modify; dataflows expose operators, dependencies, and intermediate results for targeted feedback.}
  \label{fig:movie-abstractions}
\end{figure*}

For example, consider a movie-analysis task involving two CSV files:
\texttt{movies.csv}, containing movie titles, genres, ratings, and release years,
and \texttt{budgets.csv}, containing production budgets.
An analyst asks, \emph{``Which highly rated movies have below-average budgets compared with movies released in the same decade?''}
As illustrated in \autoref{fig:movie-abstractions},
natural language expresses the analytical intent concisely but leaves important decisions unspecified, such as the rating threshold and how to handle missing budgets.
Python can precisely implement the analysis by joining the files, grouping movies by release decade, computing average budgets, and filtering the results.
However, understanding or modifying the agent-generated script requires the analyst to navigate the code and reason about intermediate variables and their dependencies.

\boldstart{Dataflows as a Shared Computational Abstraction.}
We argue that \emph{dataflows}, represented as directed acyclic graphs (DAGs) of declarative operators and data dependencies, provide an effective shared abstraction for human--agent collaboration.
Each operator represents an analytical step, such as scanning, filtering, joining, or aggregating data, while directed edges capture the dependencies between these steps.
By combining well-defined computational semantics with an intuitive visual representation, dataflows enable agents to construct and reason about analyses while allowing human analysts to understand, inspect, and modify them.

As illustrated in \autoref{fig:movie-abstractions}, dataflows offer three key benefits.
\emph{First, precise:} each operator explicitly specifies an analytical operation, such as filtering movies with ratings above a threshold or computing the average budget for each release decade.
\emph{Second, intuitive:} the visual representation exposes analytical steps and their dependencies, allowing analysts to follow the analysis without navigating program syntax or intermediate variables.
\emph{Third, easier to give feedback:} analysts can inspect intermediate results and modify individual operators to refine the analysis.
For example, an analyst can examine the average budget for each decade and ask the agent to compare movies by genre instead, allowing the agent to revise the relevant operators while preserving the remaining analysis.

Importantly, the same abstraction can benefit \emph{agents} as well.
\looseness=-1
Operators, schemas, intermediate results, and data dependencies provide structured context for reasoning about the analysis.
High-level operator declarations can avoid repeatedly presenting lengthy implementation scripts to the LLM, and dependencies can help select relevant observations during later interactions.
Our experiments show that, compared with the script abstraction, the dataflow abstraction can improve task accuracy scores and reduce inference cost simultaneously in most settings, with the largest observed gains reaching double-digit points in accuracy score and over 2.8$\times$ in inference-cost efficiency.
These benefits can improve performance in human--agent collaboration on data science tasks.

\looseness=-1
Building on this abstraction, we present \emph{\sys}, a data science system that enables human analysts and LLM agents to construct, execute, inspect, and refine dataflows collaboratively.
It grounds agent actions in declarative operators and integrates them with an execution engine, retained results, conversation history, and a visual interface.
Our contributions are:

\begin{itemize}[leftmargin=*, nosep]
    \item \textbf{Dataflow abstraction (\autoref{sec:abstraction}).}
    We identify key human--agent collaboration requirements and formalize dataflows with declarative operators and explicit data dependencies.

    \item {\emergencystretch=1em
    \textbf{Dataflow ReAct framework (\autoref{sec:components}).} It lets agents incrementally construct, modify, and execute dataflows through \mbox{iterative reasoning} and actions.\par}

    \item \textbf{Human--agent interaction (\autoref{sec:context-tracking}).}
    Our context tracker maintains conversation and dataflow snapshots for interruption and exploration from historical states.

    \item {\emergencystretch=1em
    \textbf{Dataflow-aware context optimizations (\autoref{sec:optimization}).}
    For agent reasoning, we summarize intermediate results with samples and statistics and prune irrelevant historical observations \mbox{using dataflow} dependencies.\par}

    \item \textbf{Experimental evaluation (\autoref{sec:experiments}).}
    We comprehensively evaluate \sys and the underlying dataflow abstraction with a user study and experiments comparing \sys with a baseline script-based agent across proprietary and open-weight LLMs.

\end{itemize}

\par
\section{Preliminaries}
\label{sec:preliminaries}

\par
\subsection{Data Science Tasks and Computational Abstractions}
\label{sec:pre-task}

A \emph{data science task} is a natural-language instruction that asks a question or requests an analysis over given data.
For example, a human analyst submits this task over two CSV files:
\begin{quote}
    ``Using \texttt{movies.csv} and \texttt{rating.csv}, which Comedy movies receive a 5-star rating?''
\end{quote}
This task asks for Comedy movies with at least one 5-star rating.

To solve a data science task, an analyst uses a {\em computational abstraction} to declare data-processing steps, execute them on an engine, and store and retrieve their results.
Example abstractions include Python scripts and SQL queries.
With Python, the analyst writes code specifying the processing steps, and a Python interpreter executes it.
Results are retained in variables or saved to files and loaded later.
With SQL, the analyst writes a query specifying the desired result, and a query engine executes a physical plan to compute it.
Results are returned or stored in a table for later queries.

\par
\subsection{LLM-Based Agents for Data Science}
\label{sec:pre-react}

\begin{figure}[!htbp]
\centering
\begingroup%
\definecolor{psuser}{RGB}{146,84,119}%
\definecolor{psblue}{RGB}{47,104,161}%
\definecolor{pscontrol}{RGB}{153,101,28}%
\begin{tikzpicture}[
  x=1cm,y=0.88cm,text=black!85,line cap=round,line join=round,
  font=\normalfont\sffamily\fontsize{8}{9.5}\selectfont,
  every node/.style={inner sep=0pt},
  bold/.style={font=\normalfont\sffamily\bfseries\fontsize{8.5}{10}\selectfont},
  layer/.style={rounded corners=2.5pt,line width=.6pt},
  fixed/.style={text height=1.6ex,text depth=.4ex},
  component/.style={fixed,draw=black!35,fill=white,rounded corners=2pt,line width=.5pt,
    minimum width=2.35cm,minimum height=.46cm,text=black!80},
  action/.style={fixed,draw=psblue!45,fill=white,rounded corners=2pt,line width=.5pt,
    minimum width=1.16cm,minimum height=.30cm,text=psblue!75!black},
  channel/.style={fixed,rounded corners=2pt,line width=.5pt,minimum width=1.98cm,
    minimum height=.30cm,font=\normalfont\sffamily\bfseries\fontsize{8}{9.5}\selectfont},
  twoway/.style={{Stealth[length=1.8mm,width=1.45mm]}-{Stealth[length=1.8mm,width=1.45mm]},
    line width=.7pt},
  oneway/.style={-{Stealth[length=1.8mm,width=1.45mm]},line width=.7pt},
  pics/user/.style={code={
    \draw[psuser,fill=white,line width=.65pt] (0,.12) circle[radius=.095];
    \draw[psuser,fill=white,rounded corners=1pt,line width=.65pt]
      (-.20,-.18)--(-.20,-.09)
      .. controls (-.20,.04) and (.20,.04) .. (.20,-.09)
      --(.20,-.18)--cycle;
  }},
  pics/robot/.style={code={
    \draw[black!70,line width=.6pt] (0,.13)--(0,.22);
    \fill[black!70] (0,.24) circle[radius=.027];
    \draw[black!70,fill=white,rounded corners=1.5pt,line width=.65pt]
      (-.20,-.14) rectangle (.20,.13);
    \fill[black!70] (-.08,.02) circle[radius=.028];
    \fill[black!70] (.08,.02) circle[radius=.028];
    \draw[black!70,line width=.6pt] (-.07,-.07)--(.07,-.07);
    \draw[black!70,line width=.8pt]
      (-.24,-.055)--(-.24,.045) (.24,-.055)--(.24,.045);
  }}
]
\pic[scale=1.6] at (2.235,2.87) {user};
\pic[scale=1.6] at (6.215,2.87) {robot};
\node[bold,text=psuser,anchor=east,align=right] at (1.785,2.87) {Human\\analyst};
\node[bold,anchor=west,align=left] at (6.735,2.87) {LLM\\agent};

\draw[twoway,draw=black!60] (2.695,3.07)--(5.685,3.07);
\node[channel,draw=black!40,fill=white,text=black!75] at (4.190,3.07) {Conversation};
\draw[oneway,draw=pscontrol] (2.695,2.67)--(5.685,2.67);
\node[channel,draw=pscontrol!45,fill=pscontrol!6,text=pscontrol] at (4.190,2.67) {Control};

\draw[twoway,draw=psblue] (2.235,2.51)--(2.235,1.47);
\draw[twoway,draw=psblue] (6.215,2.55)--(6.215,1.47);
\draw[fill=psblue!7,draw=psblue!30,line width=.45pt,rounded corners=2.5pt]
  (.815,1.79) rectangle (7.635,2.19);
\node[bold,text=psblue,anchor=west] at (1.015,1.99) {Actions};
\node[action] at (3.015,1.99) {Declare};
\node[action] at (4.295,1.99) {Inspect};
\node[action] at (5.575,1.99) {Execute};
\node[action] at (6.855,1.99) {Retrieve};

\draw[layer,draw=psblue!45,fill=psblue!6] (0,.94) rectangle (8.45,1.44);
\node[bold] at (4.225,1.19) {Computational Abstraction};

\draw[layer,draw=black!35,fill=black!4] (0,0) rectangle (8.45,.80);
\node[bold,anchor=west,align=left] at (.20,.40) {Computational\\Environment};
\node[component] at (4.15,.40) {Compute Engine};
\node[component] at (6.82,.40) {Data Storage};
\end{tikzpicture}%
\endgroup%
\par
\caption{Human--agent collaboration on data science tasks.}
\label{fig:problem-setting}
\end{figure}
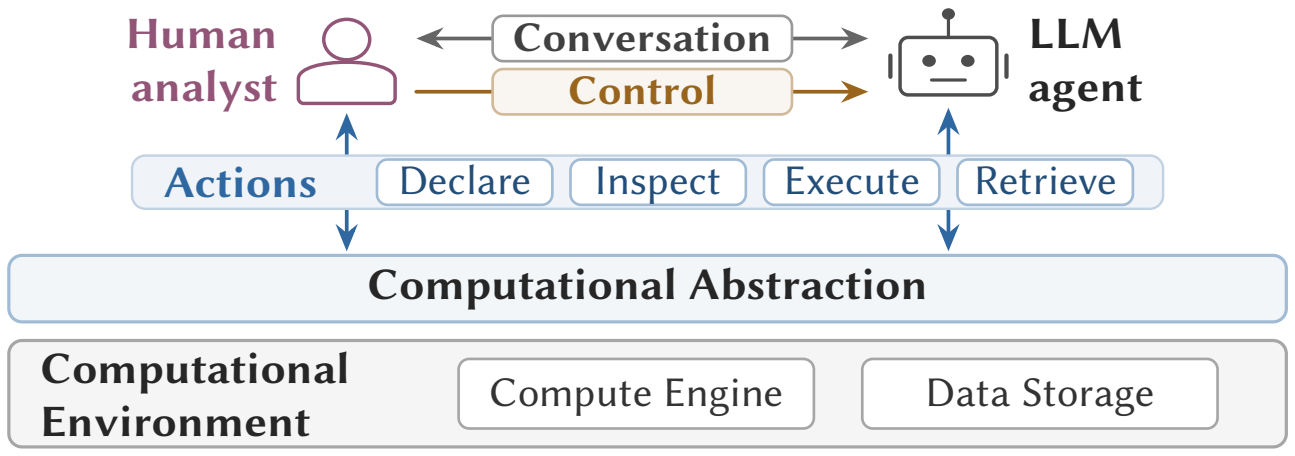

Similar to human analysts, LLM agents solve data science tasks by declaring analytical steps through a computational abstraction, requesting their execution, and examining the results.
For a given abstraction, we describe LLM agents through their environment, actions, and action paradigm~\cite{sumers2024cognitive,wang2025openhands}, as shown in \autoref{fig:problem-setting}.

\boldstart{Environment.}
The environment provides an engine for executing a declared analysis and storage for data and execution results.
For example, a Python interpreter executes scripts~\cite{pmlr-v235-wang24h}, with data files and intermediate results stored in memory or on disk for retrieval.

\boldstart{Actions.}
An agent interacts with the environment through predefined actions.
Each action specifies an operation and its arguments, such as a script to execute or a result to retrieve.
The available actions and their formats are described to the LLM.
The environment carries out each action and returns an observation, such as an execution result or an error message.

\boldstart{Action Paradigm.}
\looseness=-1
\textls[-5]{An action paradigm describes how an agent uses an LLM to reason about a task and generate actions.
ReAct~\cite{conf/iclr/YaoZYDSN023}, a common paradigm for data science agents, interleaves reasoning with actions~\cite{conf/acl/HongLLLWZLCZWZZ25,conf/emnlp/HuangLYZLWHHLZL24,journals/corr/abs-2509-21825,xu2026longdsbench,fan2026deepprep}.
In each step, the agent provides an LLM with a context including the task, available action descriptions, and a history of earlier reasoning, actions, and observations.
The LLM reasons over this context and generates actions to be executed by the environment.
The agent appends the generated reasoning, actions, and returned observations to the history for the next step.
The loop ends when the agent completes the task or reaches a step limit.}

\par
\subsection{Human--Agent Collaboration}
\label{sec:pre-collaboration}

Human--agent collaboration is a powerful paradigm because data science tasks often require iterative refinement of the analysis~\cite{freund2025flowco,dataformulator2}. We identify three collaboration primitives in \autoref{fig:problem-setting}.

\begin{itemize}[leftmargin=*, nosep]
\item \textbf{Action Primitives.}
Allow both the analyst and agent to create and modify data-processing steps, execute them, and inspect results.
For example, the analyst can change the rating condition to include 4-star ratings and retrieve the matching movie titles.
\item \textbf{Conversation Primitives.}
Enable two-way communication: the analyst provides tasks, feedback, and questions, while the agent reports its reasoning, actions, observations, and answers.
\item \textbf{Control Primitives.}
Allow the analyst to interrupt the agent, inspect history and earlier results, and resume from a previous version in a new branch while preserving subsequent analyses.

\end{itemize}

In this paper, we want to answer the following question:

\begin{tcolorbox}[colback=gray!8, colframe=gray!60, boxrule=0.4pt, arc=0pt, left=4pt, right=4pt, top=3pt, bottom=3pt]
What is the right abstraction to enable effective human--agent collaboration on data science tasks?
\end{tcolorbox}

\par
\section{Identifying the Right Abstraction}
\label{sec:abstraction}
In this section, we identify the requirements for an effective computational abstraction and examine how dataflows support human--agent collaboration in data science.

\par
\subsection{Requirements for the Right Abstraction}
\label{sec:motivation}
Data science tasks involve declaring and executing sequences of data-processing steps, which can be expressed through different computational abstractions.
For effective human--agent collaboration, we identify three requirements for the abstraction:
\begin{itemize}[leftmargin=*, nosep]
\item \textbf{Requirement 1: Precise Semantics.}
Each data-processing step should have well-defined semantics that unambiguously specify its operation, enabling agents to execute analyses consistently and analysts to verify alignment with their analytical objectives.
\item \textbf{Requirement 2: Intuitive Representation.}
The abstraction should represent analytical steps through high-level operations that analysts can easily understand and verify, without exposing low-level implementation details, while remaining concise for agents to generate and manipulate.
\item \textbf{Requirement 3: Step-Level Execution Transparency.}
Each analytical step should remain identifiable during execution, with intermediate results accessible to both analysts and agents.
The abstraction should expose step dependencies, allowing analysts to inspect individual operations, provide targeted feedback, and iteratively refine the analysis.

\end{itemize}

\begin{table}[!htbp]
\centering
\small
\setlength{\tabcolsep}{3pt}
\renewcommand{\arraystretch}{0.9}
\setlength{\aboverulesep}{0pt}
\setlength{\belowrulesep}{0pt}
\caption{Abstractions for human--agent collaboration.}
\label{tab:abstraction-comparison}
\begin{tabular*}{\linewidth}{@{\extracolsep{\fill}}lccc@{}}
\toprule\addlinespace[1pt]
\textbf{Abstraction}
& \shortstack[ c ]{\textbf{Precise}\\[-2pt]\textbf{Semantics}}
& \shortstack[ c ]{\textbf{Intuitive}\\[-2pt]\textbf{Representation}}
& \shortstack[ c ]{\textbf{Step-Level}\\[-2pt]\textbf{Transparency}} \\
\midrule
Natural Language       & \xmark     & \checkmark & \xmark \\
Programming Language   & \checkmark & \xmark     & \checkmark \\
SQL                    & \checkmark & \xmark     & \xmark \\
\midrule
\textbf{Dataflow}      & \checkmark & \checkmark & \checkmark \\
\bottomrule
\end{tabular*}
\end{table}
\autoref{tab:abstraction-comparison} compares common abstractions in terms of these requirements.
Natural language lacks precise semantics due to its inherent ambiguity and potential inconsistency across interpretations, while programming languages expose physical implementation details.
SQL provides a precise, declarative specification, but does not expose the intermediate result of each declared step.
Moreover, the optimizer may substantially rewrite the physical plan, so even a small declaration change can invalidate the execution structure and its intermediate results, complicating step-by-step iteration.

These limitations motivate an abstraction with precise semantics, intuitive representation, and step-level execution transparency.

\par
\subsection{The Dataflow Abstraction}
\label{sec:dataflow}

We propose \textit{dataflows} as an abstraction satisfying these requirements. We next specify this abstraction and explain its benefits.

\boldstart{Declarative Operators.}
We define a \emph{declarative operator} (or ``operator'' for short) as a logical specification of a data-processing operation, independent of its physical implementation.
Its inputs and outputs are structured tables, and its declaration specifies how to produce the outputs from the inputs.
A \emph{source operator} takes no input tables; its declaration instead specifies how to produce output tables by loading data from files or external sources.

At execution time, each declarative operator can map to one physical implementation, which may differ across dataflow engines while preserving its data-processing semantics.
We define a set of declarative operators that capture common data-processing semantics, as shown in \autoref{tab:operators}, along with example physical implementations in Pandas or PySpark.
We also define a \emph{custom} declarative operator to represent data-processing semantics not covered by the predefined operators in the above set.
A custom operator specifies the intended processing logic with a generated description, while the dataflow engine registers its physical implementation separately.
For example, \texttt{custom("extract entities from text")} describes an operator that extracts entities from its input text, without specifying how the entity extraction is implemented.

\begin{table}[htbp]
\centering
\footnotesize
\setlength{\tabcolsep}{2pt}
\renewcommand{\arraystretch}{0.9}
\setlength{\aboverulesep}{1pt}
\setlength{\belowrulesep}{1pt}
\newcommand{\opplaceholder}[1]{\texttt{\{}\emph{#1}\texttt{\}}}

\caption{Declarative operators and example implementations.}
\label{tab:operators}

\begin{tabularx}{\columnwidth}{
    @{}
    >{\raggedright\arraybackslash}p{0.335\columnwidth}
    >{\raggedright\arraybackslash}p{0.31\columnwidth}
    >{\raggedright\arraybackslash}X
    @{}
}
\toprule
\textbf{Declarative operator}
& \textbf{Pandas}
& \textbf{PySpark} \\
\midrule

\texttt{scan(}\opplaceholder{path}, \opplaceholder{options}\texttt{)}
& \texttt{pd.read\_csv(...)}
& \texttt{spark.read.csv(...)} \\

\texttt{filter(}\opplaceholder{condition}\texttt{)}
& \texttt{query(...)}
& \texttt{filter(...)} \\

\texttt{distinct(}\opplaceholder{columns}\texttt{)}
& \texttt{drop\_duplicates(...)}
& \texttt{dropDuplicates(...)} \\

\texttt{sort(}\opplaceholder{columns}, \opplaceholder{orders}\texttt{)}
& \texttt{sort\_values(...)}
& \texttt{orderBy(...)} \\

\texttt{limit(}\opplaceholder{offset}, \opplaceholder{count}\texttt{)}
& \texttt{iloc[...]}
& \texttt{offset(...).limit(...)} \\

\texttt{project(}\opplaceholder{mapping}\texttt{)}
& \texttt{[[...]].rename(...)}
& \texttt{select(...)} \\

\texttt{cast(}\opplaceholder{types}\texttt{)}
& \texttt{astype(...)}
& \texttt{select(...)} \\

\texttt{join(}\opplaceholder{keys}, \opplaceholder{mode}\texttt{)}
& \texttt{merge(...)}
& \texttt{join(...)} \\

\texttt{union()}
& \texttt{pd.concat(...)}
& \texttt{unionByName(...)} \\

\texttt{aggregate(}\opplaceholder{groups},
\opplaceholder{measures}\texttt{)}
& \texttt{groupby(...)}\newline
  \texttt{.agg(...)}
& \texttt{groupBy(...)}\newline
  \texttt{.agg(...)} \\

\midrule
\texttt{custom(}\opplaceholder{description}\texttt{)}
& \texttt{df.apply(...)}
& \texttt{udf(...)} \\
\bottomrule
\end{tabularx}
\end{table}

\boldstart{Data Edges.}
Declarative operators are connected through directed \emph{data edges}, which specify how one operator's output supplies another operator's input.
Formally, an edge $e_{ij} = (v_i, v_j, \sigma)$ denotes that the output table of operator $v_i$, with schema $\sigma$ specifying its column names and types, is supplied to operator $v_j$. The edge thus records a data dependency and the table's schema.

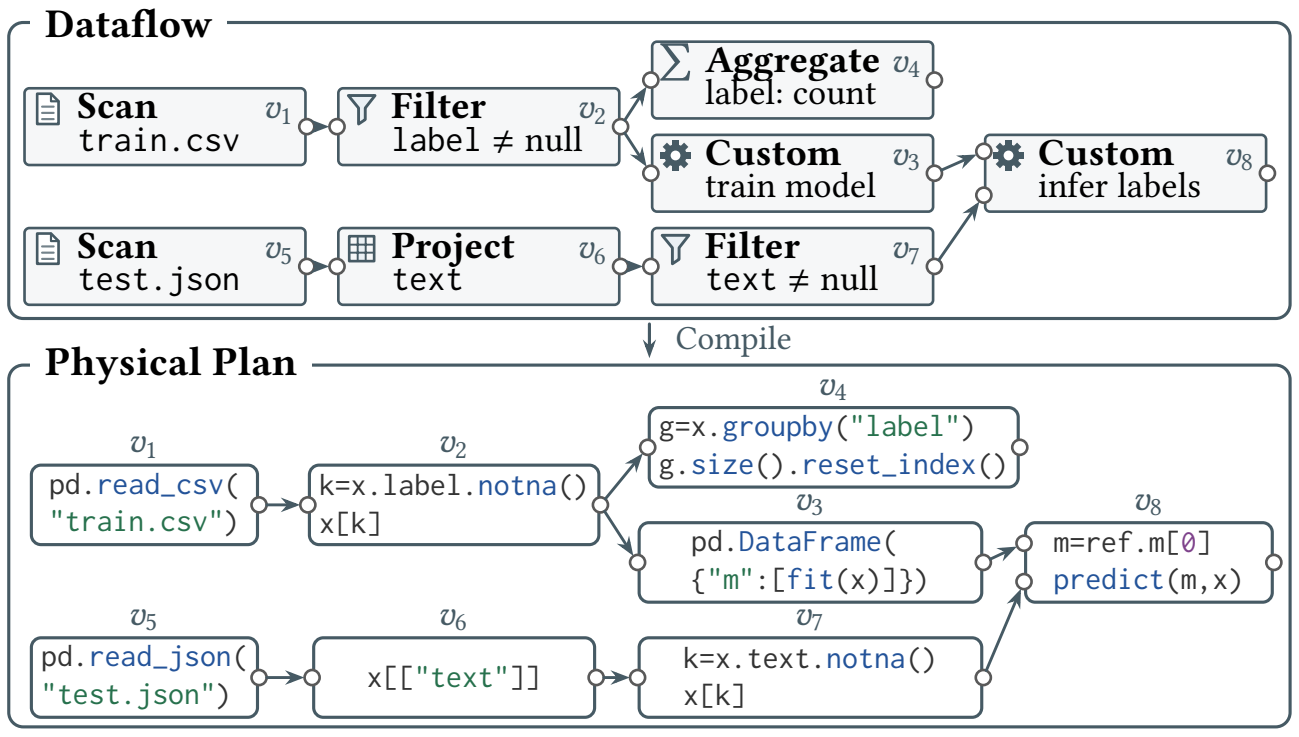
\begin{figure}[!htbp]
  \centering
  \resizebox{\linewidth}{!}{\begingroup%
\definecolor{mlfunction}{RGB}{35,83,153}%
\definecolor{mlstring}{RGB}{39,113,77}%
\definecolor{mlnumber}{RGB}{126,66,139}%

\newcommand{\mlcard}[6]{%
  \node[semop] (#1) at #2 {};
  \node[opid,anchor=base east]
    at ([xshift=-.08cm,yshift=-.23cm]#1.north east) {$v_{#3}$};
  \pic[scale=.65]
    at ([xshift=.18cm,yshift=-.16cm]#1.north west) {#4};
  \node[semtype]
    at ([xshift=.39cm,yshift=-.23cm]#1.north west) {#5};
  \node[semparam]
    at ([xshift=.39cm,yshift=-.465cm]#1.north west) {#6};}%

\newcommand{\mlcode}[1]{\detokenize{#1}}%
\newcommand{\mlfn}[1]{\textcolor{mlfunction}{\detokenize{#1}}}%
\newcommand{\mlstr}[1]{\textcolor{mlstring}{\detokenize{#1}}}%
\newcommand{\mlnum}[1]{\textcolor{mlnumber}{\detokenize{#1}}}%

\newcommand{\mlphysical}[4]{%
  \node[phyop,minimum width=#3] (p#1) at #2 {#4};
  \node[opid,anchor=south] at ([yshift=1pt]p#1.north) {$v_{#1}$};}%

\begin{tikzpicture}[
  x=1cm,y=1cm,text=black,line cap=round,line join=round,
  font=\normalfont\fontsize{8}{9}\selectfont,
  frame/.style={
    draw=cardline,line width=.65pt,fill=white,rounded corners=3pt
  },
  heading/.style={
    anchor=west,fill=white,inner xsep=3pt,inner ysep=1pt,
    font=\normalfont\bfseries\fontsize{8.5}{10}\selectfont
  },
  semop/.style={
    draw=cardline,line width=.55pt,rounded corners=1pt,
    fill=cardback,minimum width=2.05cm,minimum height=.56cm,
    inner sep=0pt
  },
  semtype/.style={
    anchor=base west,inner sep=0pt,
    font=\normalfont\bfseries\fontsize{8}{9}\selectfont
  },
  semparam/.style={
    anchor=base west,inner sep=0pt,
    font=\normalfont\fontsize{7.5}{8.5}\selectfont
  },
  opid/.style={
    inner sep=0pt,text=cardline,
    font=\normalfont\fontsize{7.5}{8.5}\selectfont
  },
  phyop/.style={
    draw=cardline,line width=.65pt,fill=white,rounded corners=2pt,
    minimum height=.58cm,inner xsep=2pt,inner ysep=1.5pt,
    align=left,text=black!80,
    font=\normalfont\ttfamily\fontsize{6.6}{7.8}\selectfont
  },
  flow/.style={
    -{Stealth[length=1.35mm,width=.95mm]},
    draw=cardline,line width=.65pt
  },
  smalltext/.style={
    inner sep=0pt,text=cardline,
    font=\normalfont\fontsize{7}{8}\selectfont
  },
  port/.style={
    draw=black!65,fill=white,line width=.55pt
  },
  pics/scan/.style={code={
    \draw[cardline,fill=white,line width=.55pt]
      (-.12,-.17)--(-.12,.17)--(.035,.17)--(.12,.085)
      --(.12,-.17)--cycle;
    \draw[cardline,line width=.45pt]
      (.035,.17)--(.035,.085)--(.12,.085);
    \foreach \y in {.025,-.045,-.115}{
      \draw[cardline!75,line width=.45pt]
        (-.065,\y)--(.065,\y);
    }
  }},
  pics/filter/.style={code={
    \draw[cardline,fill=opheader,line width=.6pt]
      (-.15,.15)--(.15,.15)--(.04,.01)--(.04,-.12)
      --(-.035,-.17)--(-.035,.01)--cycle;
  }},
  pics/project/.style={code={
    \fill[opheader] (-.045,-.14) rectangle (.045,.14);
    \draw[cardline,line width=.55pt]
      (-.14,-.14) rectangle (.14,.14);
    \draw[cardline,line width=.4pt]
      (-.045,-.14)--(-.045,.14)
      (.045,-.14)--(.045,.14)
      (-.14,.045)--(.14,.045)
      (-.14,-.045)--(.14,-.045);
  }},
  pics/model/.style={code={
    \foreach \angle in {0,45,...,315}{
      \fill[cardline,rotate=\angle]
        (-.034,.08) rectangle (.034,.17);
    }
    \fill[cardline] (0,0) circle[radius=.125];
    \fill[white] (0,0) circle[radius=.055];
  }},
  pics/aggregate/.style={code={
    \node[inner sep=0pt,text=cardline,
      font=\normalfont\fontsize{12}{12}\selectfont] {$\Sigma$};
  }}
]

\path[use as bounding box] (0,-.05) rectangle (9.35,5.28);

\draw[frame] (0,2.98) rectangle (9.35,5.14);
\node[heading] at (.15,5.14) {Dataflow};

\mlcard{s1}{(1.14,4.38)}{1}{scan}{Scan}{\texttt{train.csv}}
\mlcard{filter}{(3.43,4.38)}{2}{filter}{Filter}
  {$\texttt{label}\ne\mathrm{null}$}
\mlcard{counts}{(5.72,4.72)}{4}{aggregate}{Aggregate}
  {label: count}

\mlcard{train}{(5.72,4.04)}{3}{model}{Custom}{train model}

\mlcard{s2}{(1.14,3.36)}{5}{scan}{Scan}{\texttt{test.json}}
\mlcard{project}{(3.43,3.36)}{6}{project}{Project}
  {\texttt{text}}
\mlcard{clean}{(5.72,3.36)}{7}{filter}{Filter}
  {$\texttt{text}\ne\mathrm{null}$}

\mlcard{infer}{(8.15,4.04)}{8}{model}{Custom}{infer labels}

\bfports{s1}{0}
\bfports{s2}{0}
\bfports{filter}{1}
\bfports{train}{1}
\bfports{counts}{1}
\bfports{project}{1}
\bfports{clean}{1}
\bfports{infer}{2}

\bfconnect{s1}{filter}{1}
\bfconnect{filter}{train}{1}
\bfconnect{filter}{counts}{1}
\bfconnect{s2}{project}{1}
\bfconnect{project}{clean}{1}
\bfconnect{train}{infer}{1}
\bfconnect{clean}{infer}{2}

\draw[flow] (4.675,2.94)--(4.675,2.69);
\node[smalltext,anchor=west] at (4.86,2.81) {Compile};

\draw[frame] (0,0) rectangle (9.35,2.64);
\node[heading] at (.15,2.64) {Physical Plan};

\mlphysical{1}{(.99,1.62)}{1.65cm}
  {\mlcode{pd.}\mlfn{read_csv}\mlcode{(}\\
   \mlstr{"train.csv"}\mlcode{)}}

\mlphysical{2}{(3.25,1.62)}{2.05cm}
  {\mlcode{k=x.label.}\mlfn{notna}\mlcode{()}\\
   \mlcode{x[k]}}

\mlphysical{4}{(6.02,2.04)}{2.50cm}
  {\mlcode{g=x.}\mlfn{groupby}\mlcode{(}\mlstr{"label"}\mlcode{)}\\
   \mlcode{g.}\mlfn{size}\mlcode{().}\mlfn{reset_index}\mlcode{()}}

\mlphysical{3}{(5.85,1.20)}{2.50cm}
  {\mlcode{pd.}\mlfn{DataFrame}\mlcode{(}\\
   \{\mlstr{"m"}\mlcode{:[}\mlfn{fit}\mlcode{(x)]}\}\mlcode{)}}

\mlphysical{5}{(.99,.36)}{1.65cm}
  {\mlcode{pd.}\mlfn{read_json}\mlcode{(}\\
   \mlstr{"test.json"}\mlcode{)}}

\mlphysical{6}{(3.25,.36)}{2.05cm}
  {\mlcode{x[[}\mlstr{"text"}\mlcode{]]}}

\mlphysical{7}{(5.85,.36)}{2.50cm}
  {\mlcode{k=x.text.}\mlfn{notna}\mlcode{()}\\
   \mlcode{x[k]}}

\mlphysical{8}{(8.32,1.20)}{1.80cm}
  {\mlcode{m=ref.m[}\mlnum{0}\mlcode{]}\\
   \mlfn{predict}\mlcode{(m,x)}}

\foreach \op in {p1,p2,p3,p4,p5,p6,p7,p8}{
  \coordinate (\op-out) at (\op.east);
  \draw[port] (\op-out) circle[radius=.055];
}

\foreach \op in {p2,p3,p4,p6,p7}{
  \coordinate (\op-in1) at (\op.west);
  \draw[port] (\op-in1) circle[radius=.055];
}

\foreach \input/\offset in {1/.14,2/-.14}{
  \coordinate (p8-in\input) at ([yshift=\offset cm]p8.west);
  \draw[port] (p8-in\input) circle[radius=.055];
}

\bfconnect{p1}{p2}{1}
\bfconnect{p2}{p4}{1}
\bfconnect{p2}{p3}{1}
\bfconnect{p5}{p6}{1}
\bfconnect{p6}{p7}{1}
\bfconnect{p3}{p8}{1}
\bfconnect{p7}{p8}{2}

\end{tikzpicture}%
\endgroup%
}
\par
  \caption{A dataflow for ML model training and inference, with a corresponding physical plan.}
  \label{fig:dataflow-definition}
\end{figure}

\boldstart{Dataflow.}
A \emph{dataflow} is a directed acyclic graph (DAG), denoted as $D = (V, E)$, where $V$ is a set of declarative operators and $E$ is a set of data edges.
An operator with no outgoing data edges is a \emph{sink}.
\autoref{fig:dataflow-definition} shows a dataflow with eight declarative operators for model training and inference.
Operators $v_1$--$v_3$ load labeled text from \texttt{train.csv}, remove rows with missing labels, and train a text classifier.
The output of $v_2$ also feeds $v_4$, which counts training examples by label.
Operators $v_5$--$v_7$ load \texttt{test.json}, select its \texttt{text} column, and remove rows with missing text.
The inference operator $v_8$ consumes this cleaned inference data and a one-cell table containing the trained model produced by $v_3$, then outputs predicted labels.
Operators $v_4$ and $v_8$ are sinks of this dataflow.

\boldstart{Dataflow Execution.}
To execute a dataflow, the engine maps each declarative operator to one or more \emph{physical operators}, connected by \emph{physical edges} that carry data.
Together, they form a DAG called a \emph{physical plan}.
\autoref{fig:dataflow-definition} illustrates an example dataflow and its physical plan mapped by a dataflow engine.
The dataflow abstraction is independent of the physical implementation's programming language (e.g., Python, Java, or Scala) and execution model (e.g., operator-by-operator or pipelined execution).
The engine must preserve the dataflow's declared semantics and data dependencies regardless of implementation.
The engine also collects \emph{runtime profiles} for physical operators, including execution status, processing latency and rate, and data volumes and transfer rates.
These profiles are linked to the corresponding declarative operators, letting analysts and agents monitor individual analytical steps.

The engine retains output tables, including sink outputs, for later inspection.
We define a \emph{dataflow execution} as $X=(D,R)$, where $D$ is the submitted dataflow and $R$ contains execution results indexed by declarative operator.
For each $v_i \in V$, its result $r_i \in R$ includes output tables on success or error traces on failure, together with the available runtime profiles of its physical operators.

\par
\subsection{The Dataflow Abstraction in Data Science}
\label{sec:collaboration}

We illustrate how dataflows support humans and agents in constructing, refining, executing, and inspecting data science analyses.

\boldstart{Declaring and Refining an Analysis.}
Humans and agents express data-processing logic using the predefined declarative operators listed in \autoref{tab:operators} or \texttt{custom} operators with natural-language descriptions and separately registered physical implementations.
They compose complex analyses by connecting operators through data edges, allowing branches to share upstream results and combine outputs using operators such as \texttt{join} and \texttt{union}.
The resulting DAG makes data dependencies explicit, enabling humans and agents to understand individual steps by inspecting their operator declarations and tracing upstream dependencies.
For example, understanding the label counts produced by $v_4$ in \autoref{fig:dataflow-definition} requires examining $v_1 \rightarrow v_2 \rightarrow v_4$.
Similarly, downstream dependencies reveal which results may be affected by an edit.
Changing the filter condition in $v_2$ may affect label counts, model training, and predictions, while leaving the test-data preparation branch unaffected.

\boldstart{Specifying and Inspecting Execution.}
Humans and agents execute selected declarative operators along with their required upstream dependencies.
During execution, they can monitor operator-level runtime profiles, including execution status, processing rates, and data-transfer rates and volumes.
After execution, they can inspect result tables and retained runtime profiles to verify results and evaluate performance, or examine error traces to diagnose failures.

\par
\section{\sys: A Dataflow ReAct System}
\label{sec:components}

In this section, we present \sys, a system that uses dataflows as a shared abstraction for human--agent collaboration in data science.

\par
\subsection{System Overview}
\label{sec:architecture}

\begin{figure}[htbp]
  \centering
  \input{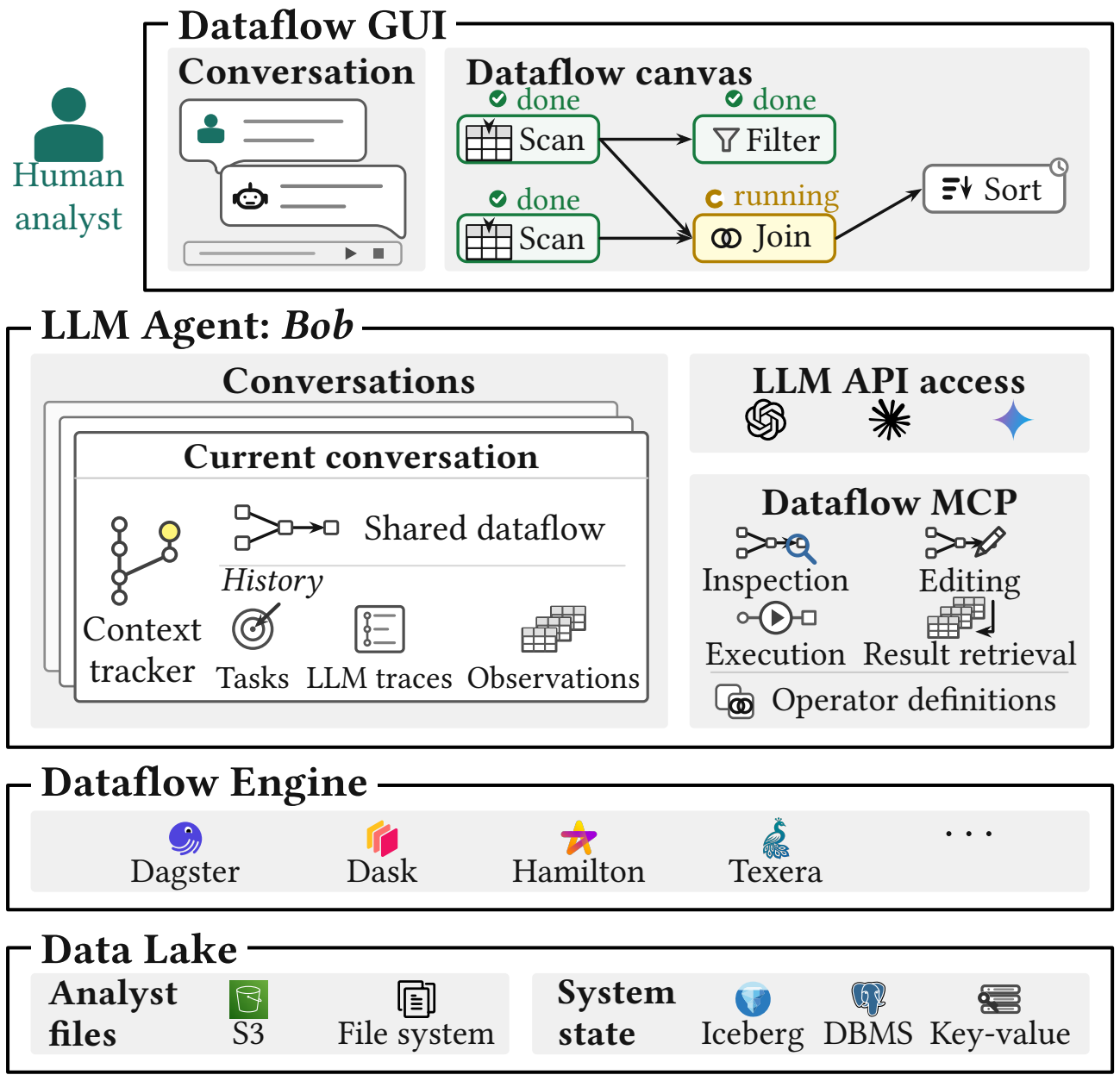}
\par
  \caption{System architecture of \sys.}
  \label{fig:architecture}
\end{figure}

\autoref{fig:architecture} shows \sys's four components: a data lake, dataflow engine, LLM agent, and dataflow graphical user interface (GUI).

\boldstart{Data Lake.} We use a data lake as the persistent storage layer of \sys.
It stores the data for analysis, such as files in object storage or file systems.
It also stores system states, including dataflow execution results and metadata.
Storage options include Apache Iceberg~\cite{website:apache-iceberg} tables, DBMSs~\cite{website:postgresql}, and key-value stores~\cite{website:rocksdb}.

\boldstart{Dataflow Engine.}
A dataflow is executed by an engine, such as Apache Texera\footnote{Apache Hamilton and Apache Texera are currently undergoing incubation at the Apache Software Foundation.}~\cite{website:apache-texera}, Apache Hamilton{\NoHyper\footnotemark[\value{footnote}]\endNoHyper}~\cite{website:apache-hamilton}, Dagster~\cite{website:dagster}, or Dask~\cite{conf/scipy/Rocklin15}.
During execution, the engine reads source data from the data lake and stores intermediate results.

\boldstart{LLM Agent.}
\agent is an LLM agent that collaborates with analysts through \emph{conversations}, each maintaining a shared dataflow and interaction history.
It connects to LLMs through APIs and uses ReAct to solve tasks with the dataflow abstraction.
A \emph{Dataflow MCP} component, based on the Model Context Protocol (MCP)~\cite{website:mcp}, executes LLM-generated actions, supporting dataflow inspection and editing, custom operator implementation management, dataflow execution, and result retrieval.
A \emph{context tracker} records conversation history and dataflow changes, enabling analysts to inspect reasoning and results, revisit earlier states, and branch from historical steps.

\boldstart{Dataflow GUI.}
The GUI supports human--agent collaboration on a shared dataflow and conversation.
Its \emph{dataflow canvas} lets analysts construct and execute dataflows, inspect operator runtime status and results, and select operators or edges for feedback to \agent.
The \emph{chat interface} lets them discuss tasks with \agent, review its reasoning, and provide feedback in natural language.

\par
\subsection{Solving a Task Using Dataflow ReAct}
\label{sec:dataflow-react}

\begin{figure}[!htbp]
\centering
\input{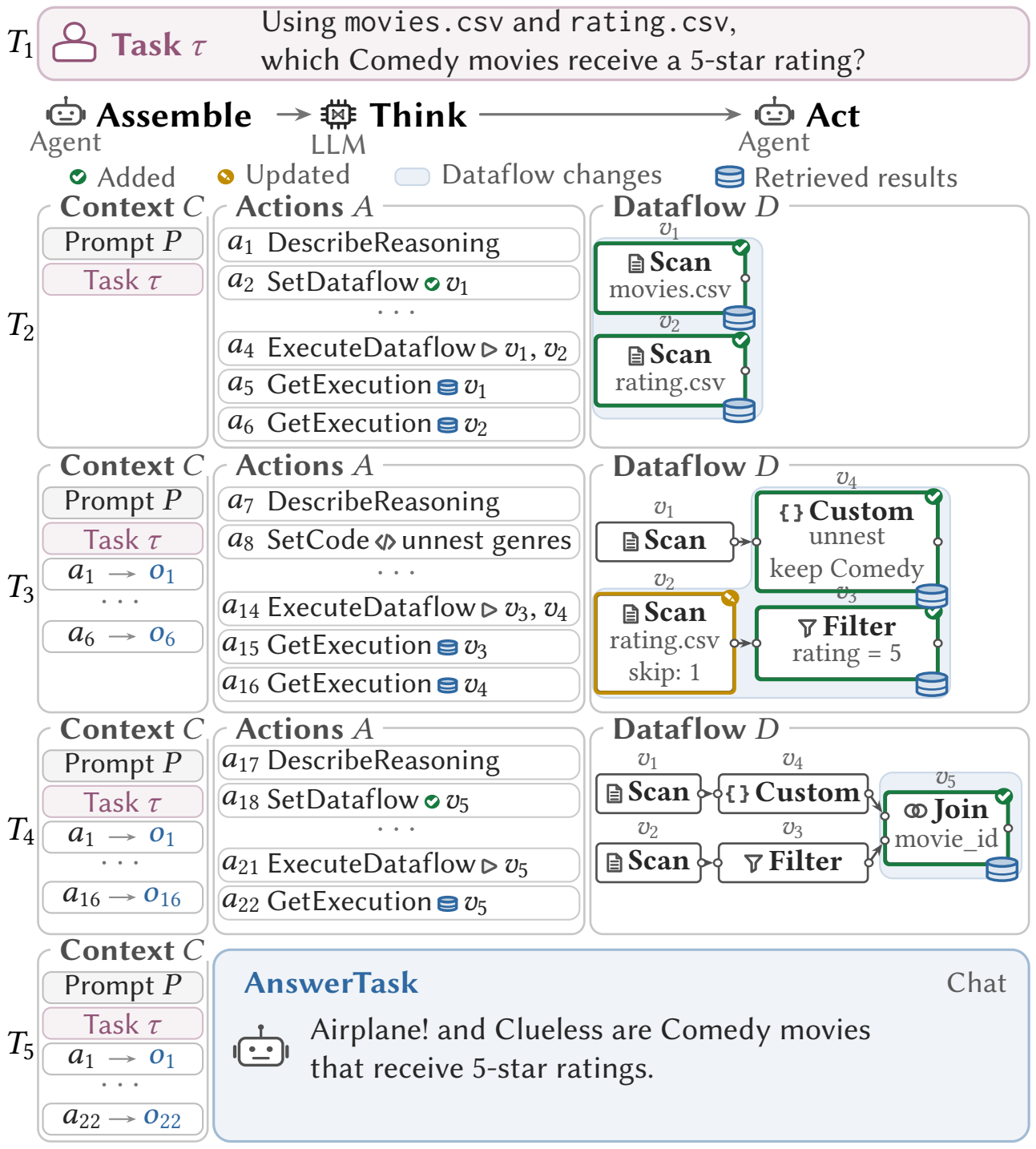}
\par
\caption{Dataflow ReAct life cycle for a data science task.}
\label{fig:full-steps}
\end{figure}

Solving a data science task requires multiple iterations in which an analyst and an agent reason about a dataflow and its execution results, revise the operators, and execute the updated dataflow.
To support this iterative process, \sys adopts ReAct~\cite{conf/iclr/YaoZYDSN023} and uses the dataflow abstraction for the agent's reasoning and actions.
We call this framework \emph{Dataflow ReAct}.
\autoref{fig:full-steps} illustrates how the agent solves the task in \autoref{sec:pre-task} using Dataflow ReAct.
Using this example, we describe a step's three phases (Assemble, Think, and Act), then explain how the agent iterates to compute an answer.

\Needspace{8\baselineskip}
\begin{definition}[Agent Actions]
We define an \emph{action} as a primitive operation exposed by \sys that an agent invokes with specific arguments to interact with the system.
\autoref{tab:react-actions} lists the supported action types, which allow the agent to inspect and modify the dataflow and custom operator implementations, execute dataflows, and retrieve execution results.
\end{definition}

\begin{table}[!htbp]
\centering
\normalsize
\setlength{\tabcolsep}{3pt}
\renewcommand{\arraystretch}{1.0}
\setlength{\aboverulesep}{1pt}
\setlength{\belowrulesep}{1pt}
\caption{Actions for solving tasks using dataflows.}
\label{tab:react-actions}

\resizebox{\linewidth}{!}{%
\begin{tabular}{@{}lll@{}}
\toprule
\textbf{Action} & \textbf{Specifies} & \textbf{Application} \\
\midrule
\emph{DescribeReasoning}
& Reasoning
& Display reasoning in chat. \\
\emph{AnswerTask}
& Answer
& Display answer in chat. \\
\emph{GetDataflow}
& Operators, edges
& Inspect operators, schemas, and edges. \\
\emph{SetDataflow}
& Dataflow changes
& Add, delete, or update operators and edges. \\
\emph{GetCode}
& $v \in V$
& Retrieve registered physical implementation. \\
\emph{SetCode}
& Code
& Register implementation for custom operators. \\
\emph{ExecuteDataflow}
& $D' \subseteq D$
& Execute the specified dataflow. \\
\emph{GetExecution}
& $v \in V$
& Retrieve full results, samples, statistics, or errors. \\
\bottomrule
\end{tabular}%
}
\end{table}

\boldstart{Initialization.}
\looseness=-1
When the agent receives the first data science task $\tau$ in a new conversation, it initializes an empty dataflow $D$ to represent the analysis throughout the Dataflow ReAct loop.
This dataflow is shared by the agent and the analyst, allowing both to inspect and revise the same analysis.
To retain the analyst's instructions and the actions carried out during the analysis, the agent maintains a shared conversation history $N$.
We model $N$ as a chronological sequence of entries and initialize it with the built-in system prompt $P$, followed by the task $\tau$: $N=[P,\tau]$.
The system prompt $P$ contains:

\noindent\parlabel{A goal}
that directs the LLM to solve the given data science task using the dataflow abstraction.

\noindent\parlabel{Available resources} that \agent can access, including the dataflow, the declarative operator specifications in \autoref{tab:operators}, the dataflow engine, the data lake, and the available actions in \autoref{tab:react-actions}.

\noindent\parlabel{End-to-end examples}
that show how the LLM can solve a task by reasoning over the current conversation and generating actions in the expected formats.

\boldstart{Phase 1 (Assemble): Preparing the Context.}
In this phase, \agent constructs context $C$ from conversation history $N$ so that the LLM can decide what actions to take next.
For each step, $A=[a_1,a_2,\ldots,a_k]$ and $O=[o_1,o_2,\ldots,o_k]$ denote the actions and their corresponding observations.
History $N$ records action--observation pairs $(a_i,o_i)$ chronologically:
\[
N = [P,\tau,(a_1,o_1),(a_2,o_2),\ldots].
\]

\looseness=-1
Throughout this section, we assume that each step's context contains the full conversation history, i.e., $C=N$.
Note that the dataflow $D$ is not included directly in $C$, since recorded actions suffice to reconstruct it.
In \autoref{fig:full-steps}, $C=[P,\tau]$ at $T_2$; at $T_3$, it also contains the actions and observations from the initial scans.

\boldstart{Phase 2 (Think): Reasoning over the Context.}
In this phase, \agent sends $C$ to the LLM, which generates action sequence $A$ to advance the analysis.
The LLM can include \emph{DescribeReasoning} to communicate its reasoning to the analyst.
In our running example, at $T_2$, it plans to inspect the input schemas and data through scans $v_1$ and $v_2$ before constructing the analysis.
It generates actions to create scan operators $v_1$ and $v_2$, execute both operators, and retrieve their results.
Through \emph{DescribeReasoning}, the LLM explains how the scans guide the search for Comedy movies with 5-star ratings.

\Needspace{4\baselineskip}
\boldstart{Phase 3 (Act): Carrying Out the Actions.}
\nopagebreak[4]
\agent executes $A$ sequentially and appends each action--observation pair $(a_i,o_i)$ to $N$ when $a_i$ completes.
Observations may include dataflow changes, execution status, or retrieved results.
At $T_2$, \agent displays the reasoning in chat, creates and executes the two scans through Dataflow MCP, and retrieves their schemas, sample tuples, and runtime profiles.
The schemas reveal a shared join key, \texttt{movie\_id}.
The samples reveal an extra title row to skip in $v_2$ and both single and pipe-separated genres in $v_1$, such as \texttt{Drama} and \texttt{Comedy|Romance}.

\begin{figure}[!b]
\centering
\includegraphics[width=\linewidth]{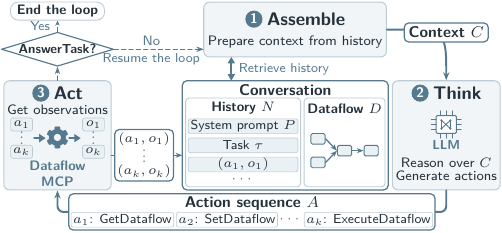}
\par
\caption{\protect Assemble, Think, and Act in Dataflow ReAct.}
\label{fig:dataflow-react-loop}
\end{figure}

\boldstart{The Loop: Iterating Assemble--Think--Act Steps.}
Together, the Assemble, Think, and Act phases form one step, as shown in \autoref{fig:dataflow-react-loop}.
Across steps, \agent maintains the evolving dataflow and uses history $N$ to assemble the next context.
The loop ends when the agent applies \emph{AnswerTask} or reaches the step limit.

In \autoref{fig:full-steps}, at $T_3$, the LLM uses the initial scan results to plan data cleaning and filtering.
The agent registers a custom implementation via \emph{SetCode} to split and unnest movie genres and retain Comedy movies.
It also updates scan $v_2$ to skip the extra title row in \texttt{rating.csv}, adds filter $v_3$ for 5-star ratings, and adds custom operator $v_4$ to process the output of movie scan $v_1$.
The agent then executes $v_3$ and $v_4$ and retrieves their results.
At $T_4$, the results reveal that $v_3$ identifies movies by \texttt{movie\_id} without titles, while $v_4$ provides Comedy movies with titles.
The agent joins their outputs on \texttt{movie\_id} using $v_5$ and retrieves the titles.
At $T_5$, the LLM generates an \emph{AnswerTask} action based on the retrieved titles, which \agent applies to return the answer and terminate the loop.

\par
\section{Human Interactions with the Agent}
\label{sec:context-tracking}

Human analysts may need to interact with the agent during a Dataflow ReAct loop, for example, to redirect the analysis, ask a question, or modify the dataflow by adding or editing an operator.
\sys supports these interactions.

\par
\subsection{Capturing Human Interactions in Context}
\begin{definition}[Human Interactions]
Human analysts can interact with \agent through two interfaces in \sys: a chat interface and a dataflow canvas.
On the chat interface, \sys allows analysts to send tasks to \agent, which starts a Dataflow ReAct loop to handle them.
On the dataflow canvas, \sys allows analysts to perform the actions related to the dataflow in \autoref{tab:react-actions} on the shared dataflow $D$.
To handle these actions from the analyst, the canvas supports inspecting and editing operators and edges, while the dataflow engine handles execution requests and the data lake handles execution result retrieval.
We call each task submission or canvas action a \emph{human interaction}.
As described in \autoref{sec:dataflow-react}, the initial task $\tau$ is stored directly in $N$.
After handling a subsequent interaction, the system appends an entry $(u_i,o_{u_i})$ to the history $N$, where $u_i$ records the submitted task or canvas action and $o_{u_i}$ records the resulting observation.
The extended history is in the form
\[
N = [P, \tau, (a_1,o_1), \boldsymbol{(u_1,o_{u_1})}, (a_2,o_2), \boldsymbol{(u_2,o_{u_2})}, \ldots].
\]
\par

This extension allows \agent to incorporate the analyst's interactions into subsequent Dataflow ReAct loops.
\end{definition}

\begin{definition}[Context Snapshot]
A \emph{context snapshot} $s_i=(N_i,D_i)$ logically captures the state of human--agent collaboration at a particular point in time.
Its \emph{conversation snapshot} $N_i$ contains the history up to that point, and its \emph{dataflow snapshot} $D_i$ captures the corresponding state of the shared dataflow.
\end{definition}

\boldstart{Context Tracker.}
To capture and manage the evolving context of human--agent collaboration, a \emph{context tracker} maintains snapshots as a tree $\mathcal{H}=(V_H,E_H)$ and tracks a \emph{current snapshot} from which the analyst and agent continue their analysis.
The root $s_0$ represents the initial state with conversation history $N_0=[P]$ and an empty dataflow, and is set as the current snapshot when a conversation begins.
Each completed human interaction or agent action advances the collaboration from the current snapshot $s_i$ to a new snapshot $s_j$, adding $(s_i,s_j)$ to $E_H$ and making $s_j$ the new current snapshot.
The edge $(s_i,s_j)$ records the conversation entry introduced by the human interaction or agent action together with its changes to the dataflow.
Rather than materializing the complete $N_j$ and $D_j$ for every snapshot, the context tracker records these incremental changes along the edges of $\mathcal{H}$.

\par
\subsection{Commands to Enable Human Interactions}
\begin{figure}[!htbp]
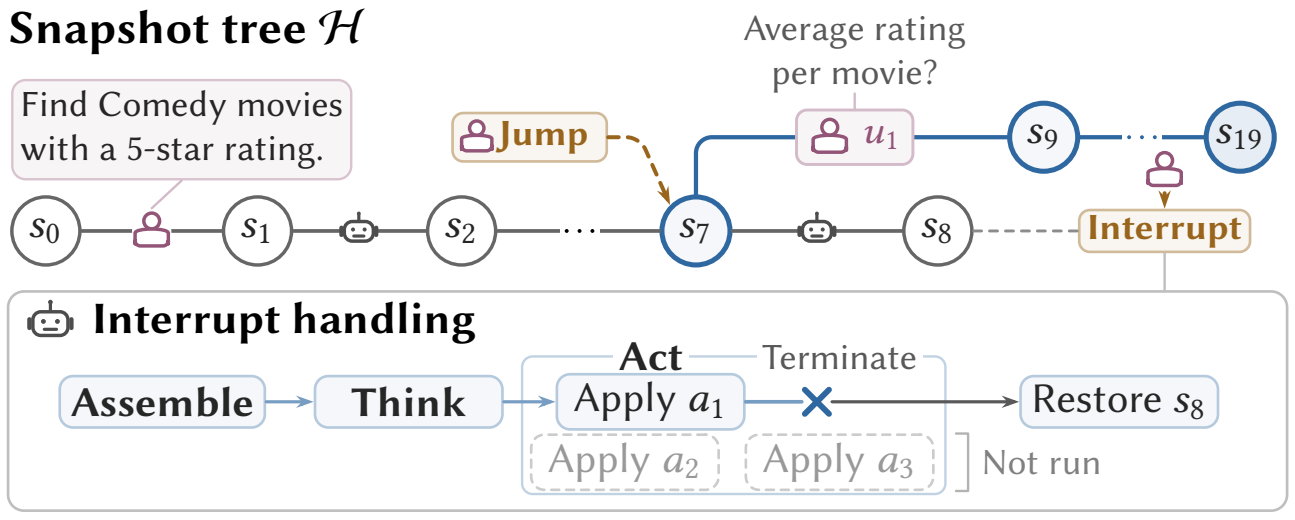

\centering
\begingroup%
\color{black}%
\endlinechar=-1\relax%
\input{figures/context-tracker-style.tex}%
\def\ctrw{8.46}%
\endlinechar=13\relax
\begin{tikzpicture}[ctrfigure]
\tikzset{
  ctrsmallcontrol/.style={ctrcontrol,inner xsep=1.5pt,inner ysep=.8pt,
    font=\normalfont\sffamily\bfseries\fontsize{7}{8.2}\selectfont},
  pics/ctrhorizontaldots/.style={code={
    \fill[white] (-.13,-.045) rectangle (.13,.045);
    \fill (-.085,0) circle[radius=.013];
    \fill (0,0) circle[radius=.013];
    \fill (.085,0) circle[radius=.013];
  }}
}
\path[use as bounding box] (0,0) rectangle (\ctrw,-3.32);

\node[ctrheading,anchor=base west] at (0,-.23) {Snapshot tree $\mathcal{H}$};
\node[ctrsnapshot] (s0) at (.25,-1.47) {$s_0$};
\node[ctrsnapshot] (s1) at (1.65,-1.47) {$s_1$};
\node[ctrsnapshot] (s2) at (3.00,-1.47) {$s_2$};
\node[ctrselected] (s7) at (4.55,-1.47) {$s_7$};
\node[ctrsnapshot] (s8) at (6.15,-1.47) {$s_8$};
\draw[ctrhistory] (s0.east)--(s1.west);
\draw[ctrhistory] (s1.east)--(s2.west);
\draw[ctrhistory] (s2.east)--(s7.west);
\draw[ctrhistory] (s7.east)--(s8.west);
\pic at (3.78,-1.47) {ctrhorizontaldots};
\foreach \x in {2.32,5.35} {
  \fill[white] (\x,-1.47) circle[radius=.12];
  \pic[scale=.48] at (\x,-1.47) {ctrrobot};
}
\fill[white] (.95,-1.47) circle[radius=.13];
\pic[scale=.55] at (.95,-1.47) {ctruser};
\node[ctrmessage,anchor=west] (taskmessage) at (.02,-.80)
  {Find Comedy movies\\with a 5-star rating.};
\draw[ctruser!40,line width=.45pt] (taskmessage.south)--(.95,-1.32);

\node[ctrsmallcontrol,minimum width=1.02cm,minimum height=.32cm] (jump)
  at (3.45,-.85) {\hspace{.18cm}Jump};
\pic[scale=.50] at (3.10,-.85) {ctruser};
\draw[ctrjump] (jump.east)--(4.18,-.85)--(s7.north west);
\node[ctrselected] (s9) at (6.85,-.85) {$s_9$};
\node[ctrselected,fill=ctrblue!12] (s19) at (8.15,-.85) {$s_{19}$};
\draw[ctrbranch,rounded corners=3pt] (s7.north)--(4.55,-.85)--(s9.west);
\node[draw=ctruser!40,fill=ctruser!6,line width=.55pt,rounded corners=2pt,
  minimum width=.77cm,minimum height=.36cm,inner sep=0pt] (u1) at (5.60,-.85) {};
\pic[scale=.55] at (5.435,-.85) {ctruser};
\node[ctrnote,text=ctruser] at (5.785,-.85) {$u_1$};
\node[ctrnote] (u1message) at (5.60,-.30) {Average rating\\per movie?};
\draw[ctruser!40,line width=.45pt] (u1.north)--(u1message.south);
\draw[ctrbranch] (s9.east)--(s19.west);
\pic[color=ctrblue] at (7.50,-.85) {ctrhorizontaldots};

\node[ctrsmallcontrol] (interrupt) at (7.65,-1.47) {Interrupt};
\draw[black!45,line width=.55pt,dash pattern=on 1.5pt off 1.5pt]
  (s8.east)--(interrupt.west);
\pic[scale=.55] at (7.65,-1.07) {ctruser};
\draw[ctrjump] (7.65,-1.22)--(interrupt.north);
\draw[black!25,line width=.45pt] (interrupt.south)--(7.65,-1.87);
\input{figures/context-tracker-interrupt.tex}
\end{tikzpicture}%
\endgroup%
\par
\caption{Human interactions in Dataflow ReAct.}
\label{fig:context-tracker}
\end{figure}

To enable human interactions throughout the analysis, \sys provides two commands: \emph{interrupt} and \emph{jump}.
As illustrated in \autoref{fig:context-tracker}, \emph{interrupt} allows an analyst to stop an ongoing Dataflow ReAct loop and interact with the current snapshot, while \emph{jump} allows the analyst to return to an earlier snapshot and interact from that point.

\boldstart{Command 1: Interrupting the Dataflow ReAct Loop.}
During a Dataflow ReAct loop, the analyst can monitor \agent's actions, evolving dataflow, and execution results through the GUI.
To provide feedback, modify the dataflow, or redirect the analysis, the analyst can issue an \emph{interrupt} command to the agent to stop the loop.
During \emph{Assemble}, \agent stops context preparation without modifying the conversation or dataflow.
During \emph{Think}, it discards the pending LLM response without applying its proposed actions.
During \emph{Act}, it cancels the ongoing action, discards uncommitted changes, and restores the latest completed snapshot.
In all cases, the context tracker ensures that the analyst and agent can resume from a consistent snapshot $s_i=(N_i,D_i)$.

For example, in \autoref{fig:context-tracker}, the analyst decides to redirect the analysis while \agent is identifying Comedy movies with 5-star ratings.
The interruption sequence illustrates how the analyst interrupts the Dataflow ReAct loop after $s_8$, while the agent is applying the first action $a_1$ of the illustrated step.
The agent cancels this action, discards its uncommitted changes, and terminates the loop.
The context tracker restores the conversation history $N_8$ and dataflow $D_8$, allowing the analyst to intervene from a consistent snapshot.
The agent then reports that it is ready.

\boldstart{Command 2: Jumping to a Snapshot.}
An analyst may revisit an earlier snapshot to provide feedback, ask questions, inspect results, or redirect the analysis.
\sys allows the analyst to jump to any snapshot in $\mathcal{H}$ when no Dataflow ReAct loop is running.

When the analyst selects a snapshot $s_i$, the context tracker restores its conversation history $N_i$ and dataflow $D_i$ using the incremental changes recorded in $\mathcal{H}$.
This can be achieved by reconstructing the selected snapshot from the root $s_0$ or reverting changes from the current snapshot.
Optionally, checkpointing can also be used to accelerate restoration.
The tracker then sets $s_i$ as the current snapshot and restores the corresponding conversation and dataflow in the GUI.
Subsequent interactions can create new branches without modifying existing analyses, allowing analysts to navigate between snapshots and explore alternative directions.

For example, after interrupting the loop at $s_8$ in \autoref{fig:context-tracker}, the analyst jumps to $s_7$ to inspect the two scans and their results.
The tracker restores $N_7$ and $D_7$, after which the analyst submits a new task ($u_1$) requesting each movie's average rating.
This creates $s_9$ as a child of $s_7$ and starts a new branch.
The agent continues from the restored context, computes average ratings using the existing scans, and returns the answer at $s_{19}$.
Both branches share the history up to $s_7$, while the original analysis at $s_8$ remains accessible.

\par
\section{Dataflow-Aware Context Optimizations}
\label{sec:optimization}
So far, we have assumed that context $C$ preserves the full conversation history $N$.
Context composition affects both decision accuracy and token cost: relevant information supports reasoning, while unnecessary information increases cost and may distract the LLM~\cite{yang2024sweagent,lindenbauer2025complexitytrapsimpleobservation}.
We relax this assumption and use two dataflow properties to optimize $C$: \emph{data properties} of operator outputs to improve result representation, and \emph{topology properties} of operator dependencies to select relevant historical results for analyst interactions.

\par
\subsection{Context Optimization with Data}
\label{sec:context-data}

During the Dataflow ReAct loop, the LLM inspects execution results to understand the data, check whether operators produce the intended output, and decide which actions to take next.
To inspect these results, the LLM generates \emph{GetExecution} actions, each specifying an operator.
\agent retrieves the requested results and records them as observations for the next step's context.
The challenge is to represent these results in context, preserving information needed for accurate decisions without excessive token costs.

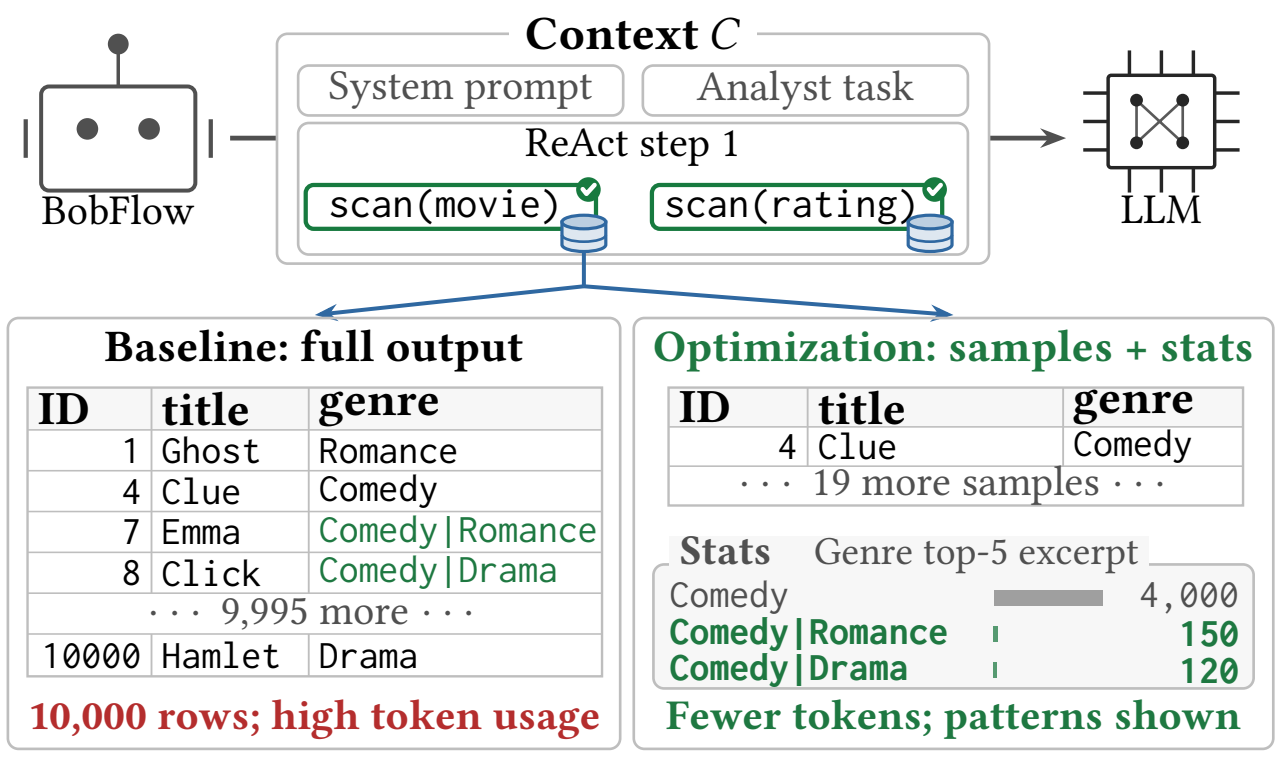
\begin{figure}[!b]
\centering
\begingroup%
\color{black}%

\definecolor{ctxink}{RGB}{80,80,80}%
\definecolor{ctxrule}{RGB}{190,190,190}%
\definecolor{ctxback}{RGB}{247,247,247}%
\definecolor{ctxblue}{RGB}{47,104,161}%
\definecolor{ctxgreen}{RGB}{30,120,65}%
\definecolor{ctxred}{RGB}{180,45,45}%

\begin{tikzpicture}[
  x=1cm,y=1cm,text=black,
  font=\normalfont\fontsize{8}{9}\selectfont,
  every node/.style={inner sep=0pt},
  ctxhead/.style={
    font=\normalfont\bfseries\fontsize{9}{10}\selectfont
  },
  ctxpanelhead/.style={
    font=\normalfont\bfseries\fontsize{8.5}{9.5}\selectfont
  },
  ctxnote/.style={
    font=\normalfont\bfseries\fontsize{8}{9}\selectfont
  },
  ctxdata/.style={
    font=\normalfont\ttfamily\fontsize{7.5}{8.5}\selectfont
  },
  ctxcard/.style={
    draw=ctxrule,line width=.5pt,rounded corners=2pt
  },
  ctxarrow/.style={
    -{Stealth[length=1.3mm]},draw=ctxink,line width=.6pt
  },
  pics/ctxrobot/.style={code={
    \draw[ctxink,line width=.65pt] (0,.13)--(0,.22);
    \fill[ctxink] (0,.24) circle[radius=.027];
    \draw[ctxink,fill=white,rounded corners=1.5pt,line width=.75pt]
      (-.20,-.14) rectangle (.20,.13);
    \fill[ctxink] (-.08,.02) circle[radius=.028];
    \fill[ctxink] (.08,.02) circle[radius=.028];
    \draw[ctxink,line width=.8pt]
      (-.24,-.055)--(-.24,.045)
      (.24,-.055)--(.24,.045);
  }},
  pics/ctxllm/.style={code={
    \draw[black!85,fill=white,line width=.65pt,rounded corners=.6pt]
      (-.15,-.13) rectangle (.15,.13);
    \foreach \x in {-.09,0,.09}
      \draw[black!85,line width=.65pt] (\x,.13)--(\x,.20) (\x,-.13)--(\x,-.20);
    \foreach \y in {-.08,0,.08}
      \draw[black!85,line width=.65pt] (-.15,\y)--(-.22,\y) (.15,\y)--(.22,\y);
    \draw[black!65,line width=.5pt] (-.07,-.06)--(.06,.06)--(.06,-.06)--(-.07,.06)--cycle;
    \foreach \x/\y in {-.07/-.06,.06/.06,.06/-.06,-.07/.06}
      \fill[black!85] (\x,\y) circle[radius=.018];
  }},
  pics/ctxresult/.style={code={
    \filldraw[ctxblue,fill=ctxblue!16,line width=.6pt]
      (-.145,.075)--(-.145,-.075)
      arc[start angle=180,end angle=360,
          x radius=.145,y radius=.045]
      --(.145,.075)--cycle;
    \draw[ctxblue,line width=.5pt]
      (-.145,-.01)
      arc[start angle=180,end angle=360,
          x radius=.145,y radius=.045];
    \filldraw[ctxblue,fill=ctxblue!32,line width=.6pt]
      (0,.075) ellipse[x radius=.145,y radius=.045];
  }}
]

\path[use as bounding box]
  (0,.96) rectangle (8.43,5.87);

\begin{scope}[yshift=-1.55cm]
\pic[scale=2.55] at (.76,6.56) {ctxrobot};
\node at (.76,6.08) {\sys};

\pic[scale=2.35] at (7.66,6.66) {ctxllm};
\node at (7.66,6.08) {LLM};

\draw[ctxarrow,line width=.9pt,
  -{Stealth[length=1.7mm]}]
  (1.50,6.55)--(7.03,6.55);

\draw[ctxcard,fill=white]
  (1.81,5.72) rectangle (6.52,7.24);

\node[ctxhead,fill=white,inner xsep=3pt]
  at (4.165,7.24) {Context $C$};

\foreach \left/\right/\label in {
  1.95/4.10/{System prompt},
  4.23/6.38/{Analyst task}
}{
  \draw[ctxcard,fill=white]
    (\left,6.70) rectangle (\right,7.03);
  \node[text=ctxink]
    at ({(\left+\right)/2},6.865) {\label};
}

\draw[ctxcard,fill=white]
  (1.95,5.78) rectangle (6.38,6.65);

\node at (4.165,6.49) {ReAct step 1};

\foreach \x/\name in {2.96/movie,5.25/rating}{
  \draw[opadded,line width=.8pt,
    rounded corners=1.5pt,fill=white]
    (\x-.96,5.95) rectangle (\x+.96,6.24);
  \node at (\x-.035,6.095)
    {\texttt{scan(\name)}};
  \begin{scope}[shift={(\x+.91,6.21)},scale=.65]
    \bfmarkadded
  \end{scope}
  \pic at (\x+.88,5.92) {ctxresult};
}
\end{scope}

\draw[ctxblue,line width=.75pt]
  (3.84,4.25)--(3.84,4.02);

\foreach \target in {2.06,6.285}{
  \draw[-{Stealth[length=1.5mm]},ctxblue,line width=.75pt]
    (3.84,4.02)--(\target,3.83);
}

\draw[ctxcard] (.03,.96) rectangle (4.08,3.81);
\draw[ctxcard] (4.17,.96) rectangle (8.40,3.81);

\node[ctxpanelhead] at (2.055,3.60) {Baseline: full output};

\node[ctxpanelhead,text=ctxgreen]
  at (6.285,3.60) {Optimization: samples + stats};

\fill[ctxback] (.16,3.07) rectangle (3.96,3.35);
\draw[ctxrule,line width=.45pt]
  (.16,1.44) rectangle (3.96,3.35);

\foreach \x in {.98,2.02}{
  \draw[ctxrule,line width=.35pt] (\x,1.99)--(\x,3.35);
  \draw[ctxrule,line width=.35pt] (\x,1.44)--(\x,1.72);
}
\foreach \y in {3.07,2.80,2.53,2.26,1.99,1.72}{
  \draw[ctxrule,line width=.3pt] (.16,\y)--(3.96,\y);
}
\node[ctxdata,anchor=west,font=\bfseries] at (.22,3.21) {ID};
\node[ctxdata,anchor=west,font=\bfseries] at (1.04,3.21) {title};
\node[ctxdata,anchor=west,font=\bfseries] at (2.08,3.21) {genre};

\foreach \id/\title/\genre/\y in {
  1/Ghost/Romance/2.935,
  4/Clue/Comedy/2.665,
  10000/Hamlet/Drama/1.580
}{
  \node[ctxdata,anchor=east] at (.92,\y) {\id};
  \node[ctxdata,anchor=west] at (1.04,\y) {\title};
  \node[ctxdata,anchor=west] at (2.08,\y) {\genre};
}
\foreach \id/\title/\genre/\y in {
  7/Emma/Romance/2.395,
  8/Click/Drama/2.125
}{
  \node[ctxdata,anchor=east] at (.92,\y) {\id};
  \node[ctxdata,anchor=west] at (1.04,\y) {\title};
  \node[ctxdata,anchor=west,text=ctxgreen]
    at (2.08,\y) {Comedy\textbar\genre};
}
\node[text=ctxink] at (2.06,1.855)
  {$\cdots$ 9,995 more $\cdots$};

\fill[ctxback] (4.40,3.09) rectangle (8.20,3.35);
\draw[ctxrule,line width=.45pt]
  (4.40,2.57) rectangle (8.20,3.35);

\foreach \x in {5.32,7.01}{
  \draw[ctxrule,line width=.35pt]
    (\x,2.83)--(\x,3.35);
}

\foreach \y in {3.09,2.83}{
  \draw[ctxrule,line width=.3pt]
    (4.40,\y)--(8.20,\y);
}

\node[ctxdata,anchor=west,font=\bfseries]
  at (4.46,3.22) {ID};
\node[ctxdata,anchor=west,font=\bfseries]
  at (5.38,3.22) {title};
\node[ctxdata,anchor=west,font=\bfseries]
  at (7.07,3.22) {genre};

\foreach \id/\title/\genre/\y in {
  4/Clue/Comedy/2.96
}{
  \node[ctxdata,anchor=east]
    at (5.26,\y) {\id};
  \node[ctxdata,anchor=west]
    at (5.38,\y) {\title};
  \node[ctxdata,anchor=west]
    at (7.07,\y) {\genre};
}

\node[text=ctxink]
  at (6.30,2.70)
  {$\cdots$ 19 more samples $\cdots$};

\draw[ctxcard,fill=ctxback]
  (4.30,1.36) rectangle (8.27,2.18);

\node[anchor=base west,fill=ctxback,inner xsep=2pt,text=ctxink]
  at (4.40,2.18)
  {\textbf{Stats}\quad\fontsize{7.5}{8.5}\selectfont Genre top-5 excerpt};

\foreach \label/\value/\width/\y in {
  Comedy/{4,000}/.72/1.96
}{
  \node[ctxdata,anchor=west,text=ctxink]
    at (4.40,\y) {\label};
  \fill[ctxink!55]
    (6.55,\y-.05) rectangle (6.55+\width,\y+.05);
  \node[ctxdata,anchor=east,text=ctxink]
    at (8.18,\y) {\value};
}

\foreach \label/\value/\width/\y in {
  {Comedy\textbar Romance}/{150}/.027/1.72,
  {Comedy\textbar Drama}/{120}/.022/1.48
}{
  \node[
    anchor=west,
    text=ctxgreen,
    font=\normalfont\ttfamily\bfseries
      \fontsize{7.5}{8.5}\selectfont
  ] at (4.40,\y) {\label};

  \fill[ctxgreen!75]
    (6.55,\y-.05) rectangle (6.55+\width,\y+.05);

  \node[
    anchor=east,
    text=ctxgreen,
    font=\normalfont\ttfamily\bfseries
      \fontsize{7.5}{8.5}\selectfont
  ] at (8.18,\y) {\value};
}

\node[ctxnote,text=ctxred]
  at (2.055,1.16) {10,000 rows; high token usage};

\node[ctxnote,text=ctxgreen]
  at (6.285,1.16) {Fewer tokens; patterns shown};

\end{tikzpicture}%
\endgroup%
\par
\caption{\protect Full output versus samples and result statistics.}
\label{fig:ctx-join-profile}
\end{figure}

We compare representations and their trade-offs with this task:
\begin{quote}
    ``Using \texttt{movies.csv} and \texttt{rating.csv}, which five Comedy movies have the highest average ratings?''
\end{quote}
To solve this task, the LLM first creates and executes two scan operators to inspect the movie and rating files.
\autoref{fig:ctx-join-profile} compares representations of the movie scan's output.

\boldstart{Baseline Approach.} Including an operator's full output in the context gives the LLM access to all records for identifying data patterns.
\looseness=-1
In the running example, the movie scan returns 10,000 records with single-genre values such as \texttt{Comedy} and multi-genre values such as \texttt{Comedy|Romance} and \texttt{Comedy|Drama}, helping the LLM recognize that filtering for Comedy movies must account for multiple genres.
However, the full output incurs high token costs and may exceed the LLM's context window.
Truncating it, for example, to 5,000 characters reduces its size but may hide important patterns, leading to incorrect assumptions about the data and inappropriate downstream operations.
If only single-genre values remain, the LLM may apply \texttt{filter(genre=\textquotesingle Comedy\textquotesingle)}, incorrectly excluding multi-genre Comedy movies.

\boldstart{Optimization: Samples and Result Statistics.}
In the dataflow abstraction, each operator produces a structured table with a schema specifying its columns and their types.
We refer to these properties as \emph{data properties}, which enable us to represent the output table with \emph{sample tuples} and \emph{result statistics}.
Sample tuples provide the LLM with examples of the output data.
Result statistics cover row and column counts, duplicate and all-null row counts, and per-column null, non-null, and distinct counts.
Type-specific statistics include numerical minima and maxima, the five most frequent string values and their counts, Boolean true and false counts, and earliest and latest datetime values.
Together, samples and statistics provide a compact, informative view of the output data, helping the LLM understand execution results accurately with fewer input tokens.

In the running example, \autoref{fig:ctx-join-profile} shows excerpts of an optimized representation with 20 sample tuples and result statistics.
The value distribution of the \texttt{genre} column reports 150 occurrences of \texttt{Comedy|Romance} and 120 of \texttt{Comedy|Drama}.
These values help the LLM recognize that it should split and unnest the genre column before filtering for Comedy movies.
This representation provides the information needed to specify the filter correctly at a lower token cost than including all 10,000 records.

\sys computes result statistics through the dataflow engine once an operator's output table is complete.
Using the table's schema, the engine computes type-specific column statistics and row-level statistics in a full-table scan, then stores them alongside the output in the data lake.
Samples are retrieved directly from the stored table without separate precomputation.

To support these representations, we extend the \emph{GetExecution} action in the Dataflow ReAct framework to allow the LLM to request either sample tuples with statistics or the full execution result.
Based on the request, \agent retrieves the corresponding data from the data lake and records it as the action's observation.

\par
\subsection{Context Optimization with Topology}
\label{sec:context-sequences}
\begin{figure}[!htbp]
\centering
\begingroup%
\color{black}%
\endlinechar=-1\relax
\definecolor{fcink}{RGB}{80,80,80}
\definecolor{fcuser}{RGB}{146,84,119}
\definecolor{fcresult}{RGB}{47,104,161}
\definecolor{fcgreen}{RGB}{30,120,65}

\newcommand{\fcoperator}[8]{%
  \node[fcop,minimum width=#7cm,minimum height=#8cm] (#1) at #2 {};
  \node[fcid,anchor=south] at ([yshift=.06cm]#1.north) {$v_{#3}$};
  \fcresultat{([xshift=-.05cm,yshift=-.04cm]#1.south east)}
  \pic[scale=.52] at ([xshift=.18cm,yshift=-.15cm]#1.north west) {fc#4};
  \node[fctype,anchor=base] at ([yshift=-.24cm]#1.north) {#5};
  \node[fcparam,anchor=base] at ([yshift=-.515cm]#1.north) {#6};}
\newcommand{\fcports}[2]{%
  \coordinate (#1-out) at (#1.east);
  \draw[fcport] (#1-out) circle[radius=.030];
  \ifnum#2=1
    \coordinate (#1-in1) at (#1.west);
    \draw[fcport] (#1-in1) circle[radius=.030];
  \fi
  \ifnum#2=2
    \coordinate (#1-in1) at ([yshift=.11cm]#1.west);
    \coordinate (#1-in2) at ([yshift=-.11cm]#1.west);
    \draw[fcport] (#1-in1) circle[radius=.030];
    \draw[fcport] (#1-in2) circle[radius=.030];
  \fi}
\newcommand{\fcedge}[3]{%
  \draw[fcflow,shorten <=.85pt,shorten >=.85pt] (#1-out)--(#2-in#3);}
\newcommand{\fcresultglyph}{%
  \filldraw[fcresult,fill=fcresult!16,line width=.65pt]
    (-.145,.075)--(-.145,-.075)
    arc[start angle=180,end angle=360,x radius=.145,y radius=.045]
    --(.145,.075)--cycle;
  \draw[fcresult,line width=.5pt]
    (-.145,-.01) arc[start angle=180,end angle=360,x radius=.145,y radius=.045];
  \filldraw[fcresult,fill=fcresult!32,line width=.65pt]
    (0,.075) ellipse[x radius=.145,y radius=.045];}
\newcommand{\fcresultat}[1]{%
  \begin{scope}[shift={#1},scale=.85]\fcresultglyph\end{scope}}

\endlinechar=13\relax
\begin{tikzpicture}[x=1cm,y=1cm,text=black!85,line cap=round,line join=round,
  font=\normalfont\sffamily\fontsize{7.2}{8.46}\selectfont,
  every node/.style={inner sep=0pt},
  fcop/.style={draw=fcink,fill=white,line width=.5pt,rounded corners=1pt},
  fctype/.style={font=\normalfont\bfseries\fontsize{7.2}{8.46}\selectfont},
  fcparam/.style={text=black!65,align=center,font=\normalfont\fontsize{6.75}{7.92}\selectfont},
  fcid/.style={text=black!65,font=\normalfont\fontsize{6.75}{7.92}\selectfont},
  fcport/.style={draw=fcink,fill=white,line width=.45pt},
  fcflow/.style={-{Stealth[length=.95mm,width=.65mm]},draw=fcink,line width=.55pt},
  fchead/.style={font=\normalfont\sffamily\bfseries\fontsize{8.5}{10}\selectfont},
  fcsmall/.style={font=\normalfont\sffamily\fontsize{7.5}{8.8}\selectfont},
  pics/fcuser/.style={code={
    \draw[fcuser,fill=white,line width=.65pt] (0,.12) circle[radius=.095];
    \draw[fcuser,fill=white,rounded corners=1pt,line width=.65pt]
      (-.20,-.18)--(-.20,-.09)
      .. controls (-.20,.04) and (.20,.04) .. (.20,-.09)
      --(.20,-.18)--cycle;
  }},
  pics/fcscan/.style={code={
    \draw[fcink,fill=white,line width=.55pt]
      (-.12,-.17)--(-.12,.17)--(.035,.17)--(.12,.085)--(.12,-.17)--cycle;
    \draw[fcink,line width=.45pt] (.035,.17)--(.035,.085)--(.12,.085);
    \foreach \y in {.025,-.045,-.115}
      \draw[fcink,line width=.45pt] (-.065,\y)--(.065,\y);
  }},
  pics/fcfilter/.style={code={
    \draw[fcink,fill=white,line width=.6pt]
      (-.15,.15)--(.15,.15)--(.04,.01)--(.04,-.12)
      --(-.035,-.17)--(-.035,.01)--cycle;
  }},
  pics/fcjoin/.style={code={
    \begin{scope}
      \clip (-.07,0) circle[radius=.12];
      \fill[fcink!20] (.07,0) circle[radius=.12];
    \end{scope}
    \draw[fcink,line width=.6pt]
      (-.07,0) circle[radius=.12] (.07,0) circle[radius=.12];
  }},
  pics/fccustom/.style={code={
    \draw[fcink,line width=.6pt]
      (-.07,.15)--(-.14,.15)--(-.14,.03)--(-.19,0)
      --(-.14,-.03)--(-.14,-.15)--(-.07,-.15)
      (.07,.15)--(.14,.15)--(.14,.03)--(.19,0)
      --(.14,-.03)--(.14,-.15)--(.07,-.15);
  }}]
\path[use as bounding box] (0,0) rectangle (8.45,2.49);

\path[fill=black!6,fill opacity=.4,draw=black!12,line width=.4pt,rounded corners=3pt]
  (.02,.02) rectangle (8.43,2.45);
\node[fchead,text=fcink,anchor=base east] at (8.20,2.11) {Baseline context};
\path[fill=fcgreen!12,draw=fcgreen!30,line width=.4pt,rounded corners=3pt]
  (.12,.90) rectangle (3.90,2.39);
\node[fchead,text=fcgreen,anchor=base west] at (.28,2.11) {Optimized context for $u$};

\fcoperator{movies}{(.95,1.45)}{1}{scan}{Scan}{movies.csv}{1.40}{.62}
\fcoperator{unnest}{(2.85,1.45)}{3}{custom}{Custom}{split+unnest}{1.80}{.62}
\fcoperator{filter}{(5.15,1.45)}{4}{filter}{Filter}{\texttt{genre=\textquotesingle Comedy\textquotesingle}}{2.20}{.62}
\fcoperator{join}{(7.425,1.02)}{5}{join}{Join}{movie\_id}{1.30}{.62}
\fcoperator{ratings}{(5.46,.54)}{2}{scan}{Scan}{rating.csv}{1.42}{.62}
\foreach \name/\inputs in {movies/0,unnest/1,filter/1,join/2,ratings/0}{
  \fcports{\name}{\inputs}}
\fcedge{movies}{unnest}{1}
\fcedge{unnest}{filter}{1}
\fcedge{filter}{join}{1}
\fcedge{ratings}{join}{2}
\draw[fcuser,line width=.95pt,rounded corners=1pt]
  (unnest.north west) rectangle (unnest.south east);

\path[draw=fcuser!65,fill=fcuser!8,line width=.65pt,rounded corners=2.5pt]
  (.15,.06)--(4.50,.06)--(4.50,.86)--(3.00,.86)
  --(2.85,1.10)--(2.70,.86)--(.15,.86)--cycle;
\pic[scale=.85] at (.45,.64) {fcuser};
\node[fcsmall,text=fcuser,anchor=base] at (.45,.24) {$u$};
\node[anchor=west,align=left,font=\normalfont\sffamily\fontsize{8}{9.2}\selectfont]
  at (.85,.52) {Why does the same movie\\appear in multiple rows?};

\end{tikzpicture}%
\endgroup%
\par
\caption{\protect Pruning irrelevant observations for $u$ at $v_3$.}
\label{fig:context-flow}
\end{figure}
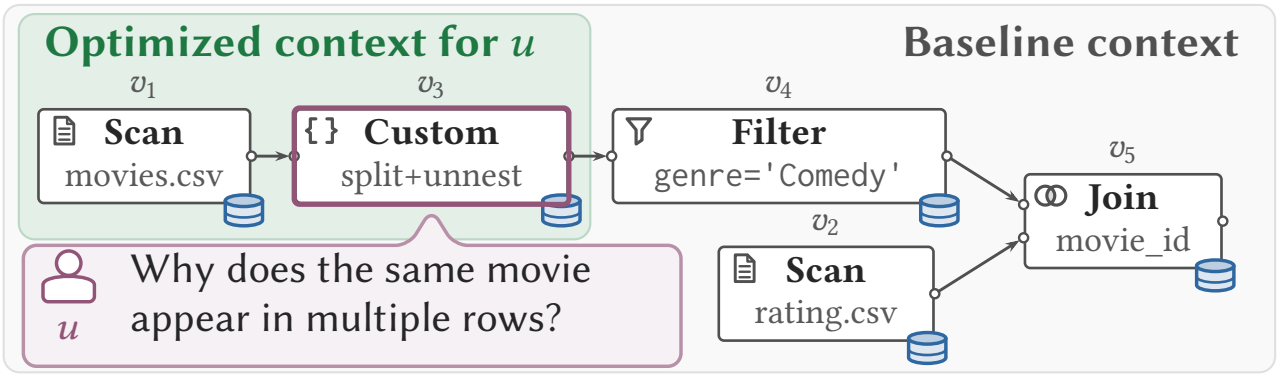

As described in \autoref{sec:architecture}, an analyst can inspect execution results and select operators on the dataflow canvas to ask \agent questions or provide feedback.
Such interactions may require only part of the history.
For example, while \agent identifies the five Comedy movies with the highest average ratings, the analyst selects $v_3$, which splits and unnests movie genres, and asks why the same movie appears in multiple output rows.
To answer, \agent can examine $v_3$'s genre rows and the output of its upstream movie scan $v_1$, as shown in \autoref{fig:context-flow}.
Since \emph{GetExecution} results are recorded as observations in the history, retaining relevant observations can give the LLM context needed to respond while reducing token cost and distractions.

\boldstart{Optimization: Pruning Irrelevant Observations.}
The directed edges in a dataflow encode dependencies between operators.
We refer to these structural relationships as \emph{topology properties}, which provide a way to identify which observations to retain.
The analyst's selection identifies the operators whose results the interaction concerns, and \agent uses these dependencies to identify all their upstream operators.
During the Assemble phase immediately after the interaction, \agent retains in $C$ the observations associated with relevant actions and interactions, replacing irrelevant observations with minimal markers of their occurrence.

Formally, let $D=(V,E)$ be the dataflow at the time of an analyst interaction $u$, and let $V_u\subseteq V$ be the nonempty set of operators selected by the analyst for this interaction.
An operator $v_i\in V$ is upstream of a selected operator $v_j\in V_u$ if there is a directed path from $v_i$ to $v_j$ in $D$.
We define $D_u$ as the sub-DAG containing the selected operators, all their upstream operators, and the edges in $D$ between these operators.
To construct $C$, \agent first derives $D_u$ from the analyst's selected operators $V_u$.
Starting with an empty context $C$, \agent traverses the conversation history $N$ in chronological order.
For each action--observation pair $(a_i,o_i)$, if $a_i$ is a \emph{GetExecution} action, \agent checks whether its specified operator $v$ belongs to $D_u$.
If so, \agent retains $(a_i,o_i)$ in $C$; otherwise, it retains $a_i$ and replaces $o_i$ with a marker.
All other history entries are appended to $C$ unchanged.
In the example, $D_u$ contains the selected operator $v_3$, its upstream movie scan $v_1$, and their connecting edge, as shown in \autoref{fig:context-flow}.
The context $C$ retains their \emph{GetExecution} observations and replaces those for the rating scan, Comedy filter, and join with minimal markers.
Together with the task, reasoning, and actions, these observations allow the LLM to explain that splitting and unnesting the genre column produces a row for each of a movie's genres.

\boldstart{Trade-off: Prompt Caching.}
Many LLM providers offer prompt caching, which reuses an unchanged initial token sequence and charges a lower rate for cached input tokens~\cite{openai2026promptcaching}.
Pruning modifies earlier observations and can shorten this reusable prefix, so more retained input tokens incur the uncached rate.
Consequently, the additional cost of uncached tokens may outweigh the savings from reducing the context size.
When earlier conversation history is no longer cached or prompt caching is disabled, pruning reduces uncached input tokens and can lower inference cost.

\providecommand{\modelhead}[2]{\textbf{\begin{tabular}[c]{@{}c@{}}#1\\#2\end{tabular}}}
\begingroup

\par
\section{Experiments}
\label{sec:experiments}

\looseness=-1
\textls[-5]{We evaluated the effectiveness of the dataflow abstraction as a shared medium for human--agent collaboration in data science.
We conducted a user study to evaluate how \sys and a script-based agent support analysts in understanding, verifying, and guiding agent-generated analyses during task completion and debugging.
For agents, we evaluated \sys's task accuracy and inference cost on three benchmarks, comparing it with the script-based agent, ablating both context optimizations, and adding representative case analyses and latency and memory measurements in human interactions.}
\par
\subsection{Experimental Setup}
\label{sec:experiments-setting}

\boldstart{Benchmarks.}
\looseness=-1\emergencystretch=1em
We used three benchmarks for table preparation, end-to-end analysis, and dependent interactions.
\emph{PrepBench}~\cite{xu2026prepbench} contains 306 tasks, each requiring 3--18 transformation steps, over 829 CSV files; we used its clarified requests to evaluate table preparation.
\emph{KramaBench}~\cite{journals/corr/abs-2506-06541} contains 104 tasks over 1,764 files from 24 heterogeneous sources, evaluating end-to-end data discovery, preparation, integration, and analysis.
\emph{LongDS-Bench}~\cite{xu2026longdsbench} evaluates multi-turn analysis; we used v1.1 Lite~\cite{longds2026release}, with 24 sessions and 777 interactions across six domains, to assess state reuse, historical revisiting, and alternative analyses.
Each benchmark turn corresponds to an interaction in which the analyst submits a request to the agent.

\boldstart{\sys Implementation.}
We built \sys based on Apache Texera~\cite{wang2024texera,website:apache-texera}, an open-source dataflow system, extending its components to implement the architecture in \autoref{fig:architecture}.
At the GUI layer, we added an interaction panel to each operator for analyst feedback and rendered the snapshot tree for revisiting snapshots.
We implemented Dataflow ReAct and added the context tracker with the data-property and topology-property optimizations.
We configured LiteLLM v1.83.10~\cite{website:litellm} as the LLM API proxy.
We built a Dataflow MCP component to let \agent apply dataflow actions to the dataflow engine Amber~\cite{journals/pvldb/KumarWNL20} and interact with the data lake.
At the data-lake layer, we stored analyst files in MinIO~\cite{website:minio}, operator outputs as Apache Iceberg~\cite{website:apache-iceberg} tables, and system metadata in PostgreSQL~\cite{website:postgresql}.

\boldstart{Baseline and Experimental Environment.}
We used smolagents CodeAgent v1.23.0~\cite{smolagents2025}, a Python-based ReAct agent, as a baseline (\emph{Script ReAct}).
In each step, the LLM generated an \emph{ExecuteCode} action containing Python code executed in a shared environment, with printed output recorded as an observation.
The baseline provides a built-in chat interface for submitting tasks and checking the agent's reasoning, generated code, and execution output.
Script ReAct follows the same ReAct paradigm as Dataflow ReAct but uses scripts to express and execute analyses, enabling comparison of the two abstractions.
Experiments ran on a server with an AMD Ryzen 9 7900X (12 cores, 24 threads), 31~GiB RAM, Ubuntu 20.04.5 LTS, and Python 3.12.12.
We accessed the LLMs through APIs.

\Needspace{5\baselineskip}
\subsection{Dataflow's Benefits for Human Analysts}
\label{sec:experiments-user-study}
\setlength{\intextsep}{4pt plus 1pt minus 1pt}
We conducted a user study to evaluate how the dataflow abstraction helps analysts collaborate with agents, understand and verify their analyses, and guide the analytical process.
We recruited 16 participants with varying levels of data science experience: six reported advanced experience, three intermediate, five beginner, and two no prior experience.
Fourteen reported familiarity with at least one programming language.
We asked participants to complete two human--agent collaboration scenarios using \sys and Script ReAct.
We used Claude Haiku 5.5~\cite{anthropic2026haiku55} in both systems in this study.

\boldstart{Scenario 1: Starting a Task from Scratch.}
This scenario examined whether dataflows can support human analysts in collaborating with agents end to end on a data science task.
Participants were given an Olympic history dataset~\cite{griffin2018olympics} and asked to collaborate with the agent on a task to identify the continents with the most gold medals.
Participants 1--9 used Script ReAct, while participants 10--16 used \sys.
By analyzing traces of participants using the agent, we found that the dataflow abstraction helped participants collaborate more effectively with the agent on solving the given task.
As shown in \autoref{fig:experiments-user-study}, all participants using \sys reached the correct answer with their first request, whereas a third of the participants using Script ReAct needed further requests, and the agent in \sys used fewer iterations and less model inference time per task.

\begin{figure}[!htbp]
\centering
\includegraphics[width=\linewidth]{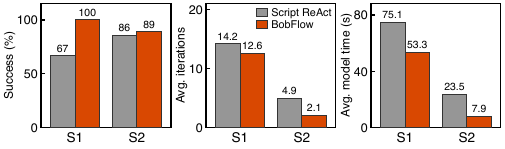}
\par
\caption{User study results for Scenarios 1 (S1) and 2 (S2).}
\label{fig:experiments-user-study}
\end{figure}

\boldstart{Scenario 2: Debugging an Existing Task.}
Next, we examined how dataflows support interaction with the agent on an existing analysis.
In this scenario, for each system, we prepared a pipeline generated by an agent for the Inside Airbnb dataset for San Francisco~\cite{insideairbnb2026sf} and introduced an equivalent price-parsing mistake, causing unexpectedly low average rental prices.
Each participant received the pipeline corresponding to their assigned system: participants 1--9 used \sys with a given dataflow, while participants 10--16 used Script ReAct with a given script.
Participants were asked to identify and correct the mistake by interacting with the corresponding agent, without prior knowledge of the underlying cause.
Our system traces again showed that the dataflow abstraction made interaction with the agent more effective.
As shown in \autoref{fig:experiments-user-study}, participants identified the bug at comparable rates in both systems, while the agent in \sys handled each interaction with less than half the iterations and about a third of the model inference time.

\boldstart{User Evaluations.}
To better understand participants' experiences with each system, we asked them to complete a questionnaire using a five-point Likert scale and analyzed their responses.
The results shown in \autoref{tab:experiments-user-study} echoed the benefits of dataflows for human--agent collaboration.
In particular, participants found it much easier to follow data lineage across analytical steps with \sys, rating this item 1.9 points higher than with Script ReAct.
They also found it easier to locate problematic steps, with a 0.8-point improvement, helping them identify where interaction with the agent was needed.

\begin{table}[!htbp]
\centering
\fontsize{6.8}{8}\selectfont
\setlength{\tabcolsep}{0.5pt}
\setlength{\aboverulesep}{1pt}
\setlength{\belowrulesep}{1pt}
\renewcommand{\arraystretch}{0.9}
\captionsetup{skip=1.5pt}
\caption{Participant ratings of intuitiveness, transparency, and control (1 = strongly disagree; 5 = strongly agree). Means are shown; bold indicates the more favorable score; $\Delta$ = \sys{} $-$ Script ReAct.}
\label{tab:experiments-user-study}
\begin{tabularx}{\columnwidth}{@{}c@{\hspace{5pt}}>{\raggedright\arraybackslash}X>{\centering\arraybackslash}p{0.157\columnwidth}>{\centering\arraybackslash}p{0.135\columnwidth}>{\centering\arraybackslash}p{0.055\columnwidth}@{}}
\toprule
\textbf{Dimension} & \textbf{Questionnaire item} & \textbf{Script ReAct} & \textbf{\sys} & \textbf{$\Delta$} \\
\midrule
Intuitiveness & Easy to collaborate with the agent & 3.9 & \textbf{4.0} & $\mathbf{+0.1}$ \\
\midrule
\multirow{7}{*}{Transparency} & Easy to grasp processed/produced tables & 3.5 & \textbf{4.3} & $\mathbf{+0.8}$ \\
& Confident about what each step does & 3.3 & \textbf{3.8} & $\mathbf{+0.5}$ \\
& Easy to view each step's output tables & 2.6 & \textbf{4.2} & $\mathbf{+1.6}$ \\
& Easy to follow data lineage & 2.3 & \textbf{4.2} & $\mathbf{+1.9}$ \\
& Clear final-answer vs.\ exploratory steps & 2.9 & \textbf{4.0} & $\mathbf{+1.1}$ \\
& Easy to verify each step's output tables & 2.7 & \textbf{3.7} & $\mathbf{+1.0}$ \\
& Hard to verify the agent's analysis & 3.8 & \textbf{3.6} & $\mathbf{-0.2}$ \\
\midrule
\multirow{3}{*}{Control} & Easy to locate problematic steps & 2.6 & \textbf{3.4} & $\mathbf{+0.8}$ \\
& Confident only requested edits occur & 3.3 & \textbf{3.8} & $\mathbf{+0.5}$ \\
& Easy to verify revised answers & 3.4 & \textbf{3.8} & $\mathbf{+0.4}$ \\
\bottomrule
\end{tabularx}
\end{table}

\Needspace{5\baselineskip}
\subsection{Dataflow's Benefits for Agents}
\label{sec:experiments-single-task}

To evaluate the dataflow abstraction's benefits for agents, we compared Script ReAct with \sys on KramaBench and PrepBench using the eight proprietary and open-weight LLMs~\cite{openai2026gpt56,openai2026gpt61,google2026gemini38flash,google2026gemini4argon,deepseek2026v41flash,anthropic2026opus55} listed in \autoref{tab:experiments-overall}.
\sys enabled the data-property and topology-property optimizations described in \autoref{sec:optimization}.
We computed accuracy scores using each benchmark's evaluator and expressed them on a 0--100 scale.
For PrepBench, correctness required all output tables to match their references under the official comparator, which checked row counts, key-aligned rows, and column-specific matching rules~\cite{xu2026prepbench}.
For KramaBench, we ran Script ReAct and \sys three times and reported the mean accuracy from the KramaBench evaluator~\cite{journals/corr/abs-2506-06541}.
We calculated inference cost from per-call token usage at OpenAI, Anthropic, and OpenRouter prices, reporting cost and token totals per benchmark run.

\begin{table}[htbp]
\centering
\scriptsize
\setlength{\tabcolsep}{2pt}
\renewcommand{\arraystretch}{0.95}
\setlength{\aboverulesep}{1pt}
\setlength{\belowrulesep}{1pt}
\caption{\protect Task accuracy scores and inference cost (\$); bold indicates the better value (higher accuracy, lower cost).}
\label{tab:experiments-overall}

\resizebox{\columnwidth}{!}{%
\begin{tabular}{@{}ll*{4}{cc}@{}}
\toprule
\multirow{2}{*}{\textbf{Benchmark}}
& \multirow{2}{*}{\textbf{Agent}}
& \multicolumn{2}{c}{\modelhead{GPT-5.6}{Luna}}
& \multicolumn{2}{c}{\modelhead{GPT-5.6}{Terra}}
& \multicolumn{2}{c}{\modelhead{GPT-5.6}{Sol}}
& \multicolumn{2}{c}{\modelhead{GPT-6.1}{Sol}} \\
\cmidrule(lr){3-4}
\cmidrule(lr){5-6}
\cmidrule(lr){7-8}
\cmidrule(lr){9-10}
& & Acc. & Cost & Acc. & Cost
& Acc. & Cost & Acc. & Cost \\
\midrule

\multirow{2}{*}{KramaBench}
& Script ReAct
& 62.3 & 0.40
& 64.3 & 6.99
& 68.9 & 10.22
& \textbf{78.9} & 3.22 \\

& \textbf{\sys}
& \textbf{76.4} & \textbf{0.30}
& \textbf{72.3} & \textbf{2.76}
& \textbf{76.4} & \textbf{5.09}
& 78.4 & \textbf{1.67} \\

\midrule

\multirow{2}{*}{PrepBench}
& Script ReAct
& 69.9 & 1.21
& 71.9 & 24.56
& 79.4 & 45.97
& 90.5 & 17.33 \\

& \textbf{\sys}
& \textbf{83.3} & \textbf{0.99}
& \textbf{86.3} & \textbf{8.67}
& \textbf{88.6} & \textbf{16.50}
& \textbf{90.8} & \textbf{6.33} \\

\midrule\midrule

\multirow{2}{*}{\textbf{Benchmark}}
& \multirow{2}{*}{\textbf{Agent}}
& \multicolumn{2}{c}{\modelhead{Gemini 3.8}{Flash}}
& \multicolumn{2}{c}{\modelhead{Gemini 4}{Argon}}
& \multicolumn{2}{c}{\modelhead{DeepSeek}{V4.1-Flash}}
& \multicolumn{2}{c}{\modelhead{Claude}{Opus 5.5}} \\
\cmidrule(lr){3-4}
\cmidrule(lr){5-6}
\cmidrule(lr){7-8}
\cmidrule(lr){9-10}
& & Acc. & Cost & Acc. & Cost
& Acc. & Cost & Acc. & Cost \\
\midrule

\multirow{2}{*}{KramaBench}
& Script ReAct
& 67.0 & 10.31
& \textbf{77.7} & 11.40
& 68.4 & \textbf{0.62}
& \textbf{77.9} & 4.76 \\

& \textbf{\sys}
& \textbf{81.9} & \textbf{6.38}
& 76.2 & \textbf{6.15}
& \textbf{73.1} & 0.66
& 77.1 & \textbf{4.35} \\

\midrule

\multirow{2}{*}{PrepBench}
& Script ReAct
& 70.9 & 37.05
& 88.6 & 33.41
& 85.6 & 1.40
& 90.2 & 17.76 \\

& \textbf{\sys}
& \textbf{89.5} & \textbf{13.22}
& \textbf{91.5} & \textbf{15.42}
& \textbf{88.9} & \textbf{1.31}
& \textbf{90.5} & \textbf{16.89} \\

\bottomrule
\end{tabular}%
}
\end{table}

\looseness=-1
As shown in \autoref{tab:experiments-overall}, \sys improved accuracy by 0.3--18.6 points in 13 of the 16 model--benchmark settings and reduced inference cost by 4.9--64.7\% in 15 settings compared with Script ReAct.
With Gemini 3.8 Flash on PrepBench, \sys improved accuracy by 18.6 points while reducing inference cost by 64.3\%.
\sys also decreased accuracy by 0.5--1.5 points in some settings, and these decreases fall within the random noise of LLM outputs.

\begin{figure}[!b]
\centering
\includegraphics[width=\linewidth]{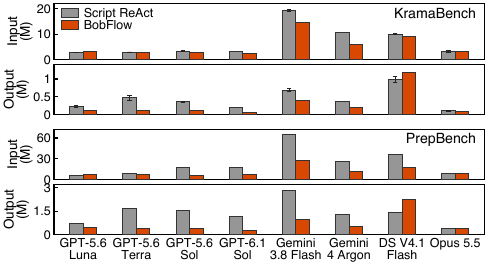}
\par
\caption{Input and output tokens per benchmark run.}
\label{fig:experiments-tokens}
\end{figure}

We broke down the input and output token usage of Script ReAct and \sys to understand the cost difference.
As shown in \autoref{fig:experiments-tokens}, \sys generated 40.2--77.3\% fewer output tokens than Script ReAct for GPT and Gemini across both benchmarks.
\sys also used 1.1--62.3\% fewer input tokens than Script ReAct in most settings.
On KramaBench with DeepSeek V4.1-Flash, \sys used 9.3\% fewer input tokens but generated 19.8\% more output tokens than Script ReAct, contributing to its 6.5\% higher inference cost.

\par
\subsection{Effect of Data-Property Optimization}
\label{sec:experiments-ablation}

We evaluated data-property optimization (``opt1'') by running PrepBench on \sys with and without opt1.
As shown in \autoref{tab:experiments-ablation}, \sys reduced inference cost by 14.2--31.1\% across all eight models, while accuracy changed by $-0.7$ to $+2.2$ points.

\begin{table}[!htbp]
\centering
\small
\setlength{\tabcolsep}{3pt}
\renewcommand{\arraystretch}{0.95}
\caption{Data-property optimization on PrepBench: \sys without opt1 $\to$ \sys.
Accuracy changes are absolute; cost changes are relative.}
\label{tab:experiments-ablation}
\begin{tabular*}{\linewidth}{@{\extracolsep{\fill}}lcccc@{}}
\toprule
\textbf{Model}
& \textbf{Accuracy}
& \textbf{$\Delta$ Acc.}
& \textbf{Cost (\$)}
& \textbf{$\Delta$ Cost} \\
\midrule
GPT-5.6 Luna
& 82.4 $\to$ 83.3
& $+0.9$
& 1.21 $\to$ 0.99
& $\mathbf{-18.2}$\textbf{\%} \\

GPT-5.6 Terra
& 85.9 $\to$ 86.3
& $+0.4$
& 10.66 $\to$ 8.67
& $\mathbf{-18.7}$\textbf{\%} \\

GPT-5.6 Sol
& 88.2 $\to$ 88.6
& $+0.4$
& 21.35 $\to$ 16.50
& $\mathbf{-22.7}$\textbf{\%} \\

GPT-6.1 Sol
& 88.6 $\to$ 90.8
& $+2.2$
& 8.39 $\to$ 6.33
& $\mathbf{-24.6}$\textbf{\%} \\

Gemini 3.8 Flash
& 90.2 $\to$ 89.5
& $-0.7$
& 16.19 $\to$ 13.22
& $\mathbf{-18.3}$\textbf{\%} \\

Gemini 4 Argon
& 90.2 $\to$ 91.5
& $+1.3$
& 17.97 $\to$ 15.42
& $\mathbf{-14.2}$\textbf{\%} \\

DeepSeek V4.1-Flash
& 86.9 $\to$ 88.9
& $+2.0$
& 1.90 $\to$ 1.31
& $\mathbf{-31.1}$\textbf{\%} \\

Claude Opus 5.5
& 88.9 $\to$ 90.5
& $+1.6$
& 24.48 $\to$ 16.89
& $\mathbf{-31.0}$\textbf{\%} \\
\bottomrule
\end{tabular*}
\end{table}

We also compared engine execution times with and without opt1 during end-to-end task solving to measure the overhead of computing result statistics.
For tasks processing more than 1~GB of input and producing large results, opt1 increased execution time by about 10\% with essentially unchanged worker memory usage, motivating further work on efficient statistics computation.

\par
\subsection{Effect of Topology-Property Optimization}
\label{sec:experiments-task-sequences}

\begin{figure*}[!t]
\centering
\resizebox{\textwidth}{!}{\input{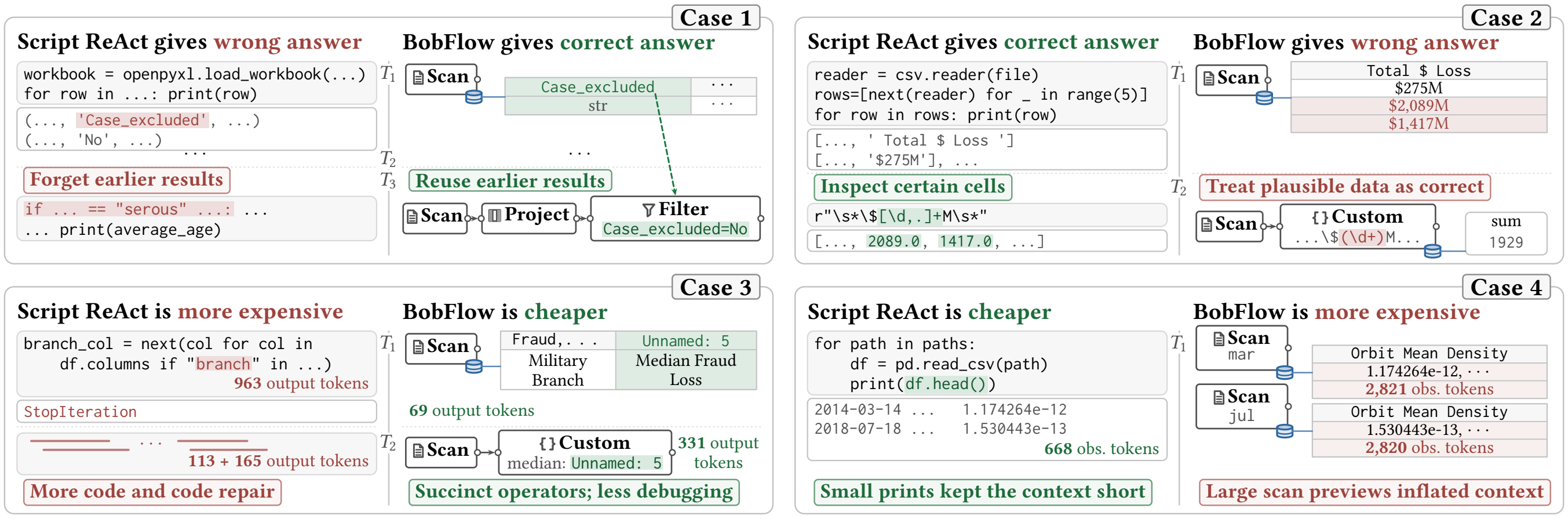}}
\par
\caption{Simplified traces of four KramaBench cases comparing \sys and Script ReAct.}
\label{fig:case-overview}
\end{figure*}

\begin{figure}[!t]
\centering
\includegraphics[width=\linewidth]{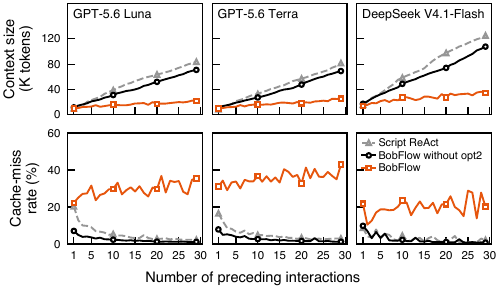}
\par
\caption{Median context size (top) and cache-miss rate (bottom) versus the number of preceding interactions.}
\label{fig:experiments-context-caching}
\end{figure}

We evaluated topology-property optimization (``opt2'') during human--agent interactions for task solving.
In the absence of a dedicated benchmark for this setting, we used LongDS-Bench, which evaluates multi-turn analysis, to simulate human interactions, treating each session as a task and its requests as analyst interactions.
We evaluated \sys with and without opt2 using three models: GPT-5.6 Luna, GPT-5.6 Terra, and DeepSeek V4.1-Flash.

During human--agent collaboration, analysts may continue an analysis while earlier prompts remain cached or resume it after the cache expires.
Therefore, we considered two conditions for reusing cached prompts across interactions.
In the \emph{warm-start} condition, earlier prompts could be reused, and costs reflected the recorded cache usage.
In the \emph{cold-start} condition, earlier prompts were assumed to be unavailable; we estimated costs by charging the first model call of each interaction at uncached input rates while retaining recorded cache usage for subsequent calls.

\begin{table}[!htbp]
\centering
\small
\setlength{\tabcolsep}{2pt}
\renewcommand{\arraystretch}{0.9}
\setlength{\aboverulesep}{1pt}
\setlength{\belowrulesep}{1.5pt}
\caption{LongDS-Bench v1.1 Lite results; mean tokens in thousands per interaction.}
\label{tab:experiments-sequences}
\resizebox{\linewidth}{!}{%
\begin{tabular}{@{}llccccc@{}}
\toprule
\multirow{2}{*}{\textbf{Model}} & \multirow{2}{*}{\textbf{Configuration}} & \multirow{2}{*}{\textbf{Accuracy}} & \textbf{Input} & \textbf{Output} & \multicolumn{2}{c}{\textbf{Cost (\$)}} \\
\cmidrule(l){6-7}
 & & & \textbf{(K)} & \textbf{(K)} & \textbf{Cold} & \textbf{Warm} \\
\midrule
\multirow{2}{*}{\shortstack[l]{GPT-5.6\\Luna}} & \sys without opt2 & \textbf{42.9} & 126 & \textbf{1.4} & 9.73 & \textbf{3.54} \\
 & \sys & \textbf{42.9} & \textbf{54} & 1.9 & \textbf{5.43} & 4.85 \\
\midrule
\multirow{2}{*}{\shortstack[l]{GPT-5.6\\Terra}} & \sys without opt2 & 54.6 & 99 & \textbf{1.2} & 88.33 & \textbf{29.52} \\
 & \sys & \textbf{58.4} & \textbf{45} & 1.5 & \textbf{49.37} & 43.47 \\
\midrule
\multirow{2}{*}{\shortstack[l]{DeepSeek\\V4.1-Flash}} & \sys without opt2 & 58.0 & 239 & \textbf{4.0} & 9.87 & \textbf{2.57} \\
 & \sys & \textbf{58.9} & \textbf{163} & 10.2 & \textbf{8.00} & 7.28 \\
\bottomrule
\end{tabular}%
}
\end{table}

\autoref{tab:experiments-sequences} compares accuracy, mean token usage, and total inference cost across all 777 interactions.
Compared with \sys without opt2, \sys maintained or improved accuracy by up to 3.8 points and reduced input-token usage by 31.7--56.9\%.
According to the traces, after pruning, the agent sometimes generated additional actions to inspect earlier results and relevant parts of the dataflow.
These additional actions contributed to the higher output-token usage with opt2.
Opt2 also reduced estimated cold-start cost by 18.9--44.2\% but increased warm-start cost by 37.0--182.9\%.

\boldstart{Context Size Across Interactions.}
To examine how opt2 reduced input-token usage, we measured context size as the input length of the first model call for each interaction.
The upper panel of \autoref{fig:experiments-context-caching} reports the median across 22 sessions with at least 30 interactions; the horizontal axis counts preceding interactions.
Opt2 limited context growth as interactions accumulated, reducing context size by 63.5--69.1\% across the three models at the start of interaction 30.

\par
\boldstart{Trade-off with Prompt Caching.}
We observed the trade-off discussed in \autoref{sec:context-sequences}: opt2 reduced estimated inference cost when earlier prompts were not cached, but raised it when they could be reused from the cache.
As shown in the lower panel of \autoref{fig:experiments-context-caching}, the cache-miss rate, the fraction of input tokens not served from the cache, rose at interaction 30 from 0.8--1.3\% without opt2 to 20.2--42.9\% with opt2.
The higher cost of uncached inputs offset savings from smaller contexts.
DeepSeek also generated 2.5 times as many output tokens with opt2, further increasing its cost.
These results suggest enabling opt2 to reduce inference cost when the cache expires or cannot be reused after a model switch, and leaving it disabled while the same model can reuse cached history.
\par

\par
\subsection{Case Studies of the Two Abstractions}
\label{sec:experiments-cases}

We studied LLM traces to see how agents' use of the dataflow abstraction affected accuracy and cost.
The four KramaBench cases in \autoref{fig:case-overview} help explain the results in \autoref{sec:experiments-single-task}.

\boldstart{Case 1: \sys Answered the Task Correctly.}
This task asked for the average age of patients whose serous tumor samples were analyzed in a study.
Both agents read the \texttt{Case\_excluded} column, but Script ReAct ignored it in subsequent computations and returned an incorrect average.
\sys kept it in an intermediate table and used it to exclude flagged samples, answering correctly.
\nopagebreak[4]
\begin{experimentobservation}{1}
\sys helped the agent accurately reuse earlier results from the context, improving answer accuracy.
\end{experimentobservation}

\boldstart{Case 2: Script ReAct Answered the Task Correctly.}
This task asked for the total amount defrauded across payment methods.
\sys used a custom operator whose parser skipped amounts containing commas.
The operator returned only the final sum, and \sys accepted this plausible but incorrect result without inspecting the parsed values.
Script ReAct handled these amounts correctly and printed the parsed values, obtaining the correct total.
\nopagebreak[4]
\begin{experimentobservation}{2}
\sys let the agent trust plausible results despite insufficient runtime inspection, reducing accuracy.
\end{experimentobservation}

\boldstart{Case 3: \sys Solved the Task at Lower Cost.}
This task asked which U.S.~military branch had the largest median fraud loss.
Script ReAct generated code using assumed column names; after the code failed, it inspected the table and corrected the code.
\sys first inspected the table and used its observed structure to generate the analysis.
Both agents answered correctly, but \sys avoided the failed attempt and reduced inference cost.
\nopagebreak[4]
\begin{experimentobservation}{3}
\sys inspected data first, reducing debugging cost.
\end{experimentobservation}

\boldstart{Case 4: Script ReAct Solved the Task at Lower Cost.}
This task asked for the ratio of peak atmospheric mass density during two periods.
Before computing the ratio, \sys gathered large table previews, whereas Script ReAct inspected only a few rows.
Both agents answered the question correctly, but the larger observations increased \sys's input-token usage and inference cost.
\begin{experimentobservation}{4}
\sys's excessive previews raised cost, not accuracy.
\end{experimentobservation}

\par
\subsection{Human Interaction Support}
\label{sec:experiments-snapshots}
\setlength{\intextsep}{4pt plus 1pt minus 1pt}

We evaluated how efficiently \sys maintains interaction histories and restores earlier analysis states as interactions accumulate.

\boldstart{Memory Utilization.}
We replayed 207 recorded KramaBench conversations in one session to measure memory growth as their histories accumulated.
After garbage collection, we measured JavaScript heap and external-buffer growth for \sys and Python memory for Script ReAct.
Both measurements excluded initial memory, execution-engine memory, and Python execution variables.
As shown in \autoref{fig:experiments-memory}, \sys's history memory grew by 110~MiB after 1,239 interactions, roughly 95\% less than Script ReAct's 2.3~GiB.
Script ReAct retained copies of model inputs at each step, while \sys recorded incremental changes.

\begin{figure}[!htbp]
\centering
\includegraphics[width=\linewidth]{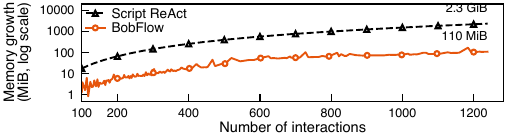}
\par
\caption{Memory growth from interaction 100 onward.}
\label{fig:experiments-memory}
\end{figure}

\boldstart{Jump Latency and Tracking Overhead.}
To measure jump latency as the snapshot tree grew, we replayed the same 207 conversations and jumped to 50 uniformly sampled snapshots at intervals of about 250 snapshots.
We measured each jump command's end-to-end latency, including restoration of the target snapshot.
As shown in \autoref{fig:experiments-tracker} (left), median latency was 1.2~ms across 1,150 jumps and remained below 1.9~ms in each 500-snapshot interval.

\begin{figure}[!htbp]
\centering
\includegraphics[width=\linewidth]{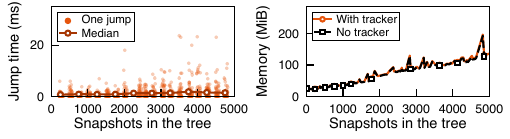}
\par
\caption{Jump latency (left) and memory with and without tracking (right) in replays of over 5,000 snapshots.}
\label{fig:experiments-tracker}
\end{figure}

Using the same snapshots, we measured tracking overhead by comparing heap memory with and without the tracker.
At 5,095 snapshots, tracking added 4.1~MiB of overhead (\autoref{fig:experiments-tracker}, right).

\endgroup

\par
\section{Related Work}

\boldstart{ReAct Framework.} ReAct~\cite{conf/iclr/YaoZYDSN023} enables LLM agents to solve diverse tasks by interleaving reasoning and actions, with observations guiding later steps.
For example, SWE-agent~\cite{yang2024sweagent} resolves software issues by navigating repositories, editing code, and running tests; WebVoyager~\cite{he2024webvoyager} completes web tasks through reasoning over webpage observations and browser actions; and ChemCrow~\cite{bran2024chemcrow} plans and executes syntheses with specialized chemistry tools.
We design \sys as a Dataflow ReAct system for solving data science tasks.

{\emergencystretch=2em
\boldstart{Agentic Systems for Data Processing.}
As LLMs improve in reasoning and tool use, more agentic systems are being built for data science tasks. Python scripts or SQL queries commonly serve as their computational abstraction.
For example, CodeAct~\cite{pmlr-v235-wang24h}, DA-Agent~\cite{conf/emnlp/HuangLYZLWHHLZL24}, and DA-Studio~\cite{dastudio2026} use Python, while Chat2Data~\cite{chat2data2024} and SiriusBI~\cite{journals/pvldb/JiangXSSZLZLLHWZYCC25} generate SQL queries.
\sys instead formalizes dataflows as a computational abstraction for data science agents.

Another line of work optimizes how agents solve specific data-processing tasks.
For data preparation, AutoPrep~\cite{fan2025autoprep} and DataFlow~\cite{liang2025dataflow} coordinate specialized agents to construct pipelines, while DeepPrep~\cite{fan2026deepprep} uses execution-guided tree search to explore and revise them.
With their emphasis on optimization, these systems focus on automated task solving.
In \sys, we emphasize analyst--agent interaction throughout a task and explore optimization in this setting.
\par}

\boldstart{Human--Agent Collaboration in Data Science.}
Human--agent collaboration has become a focus of many proprietary systems and research prototypes, with interfaces for analysts often centered on dialogue or code.
Dialogue-oriented interfaces in Claude Science~\cite{claude-science}, OpenAI's in-house data agent~\cite{openai-data-agent}, SiriusBI~\cite{journals/pvldb/JiangXSSZLZLLHWZYCC25}, and DataChat~\cite{conf/sigmod/JohnBCRLDCKP23} let analysts express analytical requests in natural language.
Code-oriented interfaces expose analytical steps for inspection and revision through generated code, notebooks, or diagrams derived from code.
For example, Stepwise~\cite{conf/uist/Kazemitabaar0DG24} and ChatPipe~\cite{chatpipe2024} expose analytical steps and generated code.
Through notebooks, DocWrangler~\cite{shankar2025docwrangler} lets analysts annotate outputs and refine document-processing pipelines, while Carnot~\cite{carnot2026} lets them inspect, modify, and execute operator DAGs.
WaitGPT~\cite{xie2024waitgpt} presents interactive operation diagrams from code.
Compared with these systems, \sys centers interaction on a shared visual dataflow that analysts and agents can inspect and modify, with a chat interface for conversation and context management.
{\emergencystretch=2em
\looseness=-1
Systems such as Flowco~\cite{freund2025flowco} and DataFlow-Harness~\cite{he2026dataflowharness} use executable dataflow graphs to help analysts construct and revise analyses with agents.
Compared with these works, we also study how agents can use dataflow properties to construct and optimize their context during collaboration.
Our demonstration paper~\cite{bai2026bobflowdemo} showcased \sys's interface and scenarios, whereas this paper presents the dataflow abstraction, Dataflow ReAct, the context optimizations, and their evaluation.
\par}

\par
\section{Conclusion}
\label{sec:conclusion}
In this paper, we identify requirements for human--agent collaboration in data science and formalize dataflows as a shared abstraction.
Dataflow ReAct in \sys lets agents iteratively construct, modify, and execute dataflows to solve data science tasks.
We support human interactions by tracking conversation and dataflow snapshots for interruption and exploration.
We optimize agent context with data properties and prune irrelevant observations using topology properties.
Our user study shows that dataflows make collaboration and interaction with agents more effective and intuitive.
Our experiments demonstrate that dataflows can substantially improve task accuracy while reducing inference cost in most evaluated settings.
These findings support dataflows as a shared abstraction for human--agent collaboration.

\FloatBarrier
\begin{acks}
We thank Dr. Xi Lu for her help.
We thank the participants in our user study.
This work was supported by NSF awards 2200274 and 2609582, NIH NIDDK award 2U24DK097771-11, and an SAP grant.
\end{acks}

\clearpage
\bibliographystyle{ACM-Reference-Format}
\bibliography{global-references,local-references}


\begin{thebibliography}{64}


\ifx \showCODEN    \undefined \def \showCODEN     #1{\unskip}     \fi
\ifx \showDOI      \undefined \def \showDOI       #1{#1}\fi
\ifx \showISBNx    \undefined \def \showISBNx     #1{\unskip}     \fi
\ifx \showISBNxiii \undefined \def \showISBNxiii  #1{\unskip}     \fi
\ifx \showISSN     \undefined \def \showISSN      #1{\unskip}     \fi
\ifx \showLCCN     \undefined \def \showLCCN      #1{\unskip}     \fi
\ifx \shownote     \undefined \def \shownote      #1{#1}          \fi
\ifx \showarticletitle \undefined \def \showarticletitle #1{#1}   \fi
\ifx \showURL      \undefined \def \showURL       {\relax}        \fi
\providecommand\bibfield[2]{#2}
\providecommand\bibinfo[2]{#2}
\providecommand\natexlab[1]{#1}
\providecommand\showeprint[2][]{arXiv:#2}

\bibitem[\protect\citeauthoryear{{Anthropic}}{{Anthropic}}{2024}]%
        {website:mcp}
\bibfield{author}{\bibinfo{person}{{Anthropic}}.} \bibinfo{year}{2024}\natexlab{}.
\newblock \bibinfo{title}{Introducing the {Model Context Protocol}}.
\newblock
\newblock
\urldef\tempurl%
\url{https://www.anthropic.com/news/model-context-protocol}
\showURL{%
\tempurl}
\newblock
\shownote{Accessed October 18, 2026.}


\bibitem[\protect\citeauthoryear{{Anthropic}}{{Anthropic}}{2026a}]%
        {claude-science}
\bibfield{author}{\bibinfo{person}{{Anthropic}}.} \bibinfo{year}{2026}\natexlab{a}.
\newblock \bibinfo{title}{{Claude Science, an AI workbench for scientists, is now available}}.
\newblock
\newblock
\urldef\tempurl%
\url{https://www.anthropic.com/news/claude-science-ai-workbench}
\showURL{%
\tempurl}
\newblock
\shownote{Accessed November 1, 2026.}


\bibitem[\protect\citeauthoryear{{Anthropic}}{{Anthropic}}{2026b}]%
        {anthropic2026haiku55}
\bibfield{author}{\bibinfo{person}{{Anthropic}}.} \bibinfo{year}{2026}\natexlab{b}.
\newblock \bibinfo{title}{Introducing {Claude Haiku 5.5}}.
\newblock
\newblock
\urldef\tempurl%
\url{https://www.anthropic.com/claude-haiku-5-5}
\showURL{%
\tempurl}
\newblock
\shownote{Accessed November 1, 2026.}


\bibitem[\protect\citeauthoryear{{Anthropic}}{{Anthropic}}{2026c}]%
        {anthropic2026opus55}
\bibfield{author}{\bibinfo{person}{{Anthropic}}.} \bibinfo{year}{2026}\natexlab{c}.
\newblock \bibinfo{title}{Introducing {Claude Opus 5.5}}.
\newblock
\newblock
\urldef\tempurl%
\url{https://www.anthropic.com/claude-opus-5-5}
\showURL{%
\tempurl}
\newblock
\shownote{Accessed November 1, 2026.}


\bibitem[\protect\citeauthoryear{{Apache Software Foundation}}{{Apache Software Foundation}}{[n.d.]a}]%
        {website:apache-hamilton}
\bibfield{author}{\bibinfo{person}{{Apache Software Foundation}}.} \bibinfo{year}{[n.d.]}\natexlab{a}.
\newblock \bibinfo{title}{{Apache Hamilton}}.
\newblock
\newblock
\urldef\tempurl%
\url{https://hamilton.apache.org/}
\showURL{%
\tempurl}
\newblock
\shownote{Accessed November 1, 2026.}


\bibitem[\protect\citeauthoryear{{Apache Software Foundation}}{{Apache Software Foundation}}{[n.d.]b}]%
        {website:apache-iceberg}
\bibfield{author}{\bibinfo{person}{{Apache Software Foundation}}.} \bibinfo{year}{[n.d.]}\natexlab{b}.
\newblock \bibinfo{title}{{Apache Iceberg}}.
\newblock
\newblock
\urldef\tempurl%
\url{https://iceberg.apache.org/}
\showURL{%
\tempurl}
\newblock
\shownote{Accessed September 28, 2026.}


\bibitem[\protect\citeauthoryear{{Apache Software Foundation}}{{Apache Software Foundation}}{[n.d.]c}]%
        {website:apache-texera}
\bibfield{author}{\bibinfo{person}{{Apache Software Foundation}}.} \bibinfo{year}{[n.d.]}\natexlab{c}.
\newblock \bibinfo{title}{{Apache Texera}}.
\newblock
\newblock
\urldef\tempurl%
\url{https://texera.apache.org/}
\showURL{%
\tempurl}
\newblock
\shownote{Accessed October 31, 2026.}


\bibitem[\protect\citeauthoryear{Bai, Huang, and Li}{Bai et~al\mbox{.}}{2026}]%
        {bai2026bobflowdemo}
\bibfield{author}{\bibinfo{person}{Jiadong Bai}, \bibinfo{person}{Yicong Huang}, {and} \bibinfo{person}{Chen Li}.} \bibinfo{year}{2026}\natexlab{}.
\newblock \showarticletitle{Demonstration of {BobFlow}: Human-Agent Collaboration Using Dataflows on Data Science Tasks}.
\newblock \bibinfo{journal}{\emph{Proc. {VLDB} Endow.}} \bibinfo{volume}{19}, \bibinfo{number}{12} (\bibinfo{year}{2026}), \bibinfo{pages}{4718--4721}.
\newblock
\urldef\tempurl%
\url{https://doi.org/10.14778/3827998.3828105}
\showDOI{\tempurl}


\bibitem[\protect\citeauthoryear{Bran, Cox, Schilter, Baldassari, White, and Schwaller}{Bran et~al\mbox{.}}{2024}]%
        {bran2024chemcrow}
\bibfield{author}{\bibinfo{person}{Andres~M. Bran}, \bibinfo{person}{Sam Cox}, \bibinfo{person}{Oliver Schilter}, \bibinfo{person}{Carlo Baldassari}, \bibinfo{person}{Andrew~D. White}, {and} \bibinfo{person}{Philippe Schwaller}.} \bibinfo{year}{2024}\natexlab{}.
\newblock \showarticletitle{Augmenting large language models with chemistry tools}.
\newblock \bibinfo{journal}{\emph{Nature Machine Intelligence}} \bibinfo{volume}{6}, \bibinfo{number}{5} (\bibinfo{year}{2024}), \bibinfo{pages}{525--535}.
\newblock
\urldef\tempurl%
\url{https://doi.org/10.1038/s42256-024-00832-8}
\showDOI{\tempurl}


\bibitem[\protect\citeauthoryear{Chen, Liu, Jin, Sun, Feng, Fan, Du, and Tang}{Chen et~al\mbox{.}}{2024}]%
        {chatpipe2024}
\bibfield{author}{\bibinfo{person}{Sibei Chen}, \bibinfo{person}{Hanbing Liu}, \bibinfo{person}{Waiting Jin}, \bibinfo{person}{Xiangyu Sun}, \bibinfo{person}{Xiaoyao Feng}, \bibinfo{person}{Ju Fan}, \bibinfo{person}{Xiaoyong Du}, {and} \bibinfo{person}{Nan Tang}.} \bibinfo{year}{2024}\natexlab{}.
\newblock \showarticletitle{{ChatPipe}: Orchestrating Data Preparation Pipelines by Optimizing Human-{ChatGPT} Interactions}. In \bibinfo{booktitle}{\emph{Companion of the 2024 International Conference on Management of Data ({SIGMOD/PODS} 2024)}}. \bibinfo{publisher}{{ACM}}, \bibinfo{pages}{484--487}.
\newblock
\urldef\tempurl%
\url{https://doi.org/10.1145/3626246.3654727}
\showDOI{\tempurl}


\bibitem[\protect\citeauthoryear{{Dagster Labs}}{{Dagster Labs}}{[n.d.]}]%
        {website:dagster}
\bibfield{author}{\bibinfo{person}{{Dagster Labs}}.} \bibinfo{year}{[n.d.]}\natexlab{}.
\newblock \bibinfo{title}{{Dagster} Documentation}.
\newblock
\newblock
\urldef\tempurl%
\url{https://docs.dagster.io/}
\showURL{%
\tempurl}
\newblock
\shownote{Accessed September 29, 2026.}


\bibitem[\protect\citeauthoryear{{DeepSeek}}{{DeepSeek}}{2026}]%
        {deepseek2026v41flash}
\bibfield{author}{\bibinfo{person}{{DeepSeek}}.} \bibinfo{year}{2026}\natexlab{}.
\newblock \bibinfo{title}{{DeepSeek-V4.1-Flash}: Pushing the Limits of {KV} Cache Compression}.
\newblock
\newblock
\urldef\tempurl%
\url{https://huggingface.co/deepseek-ai/DeepSeek-V4.1-Flash}
\showURL{%
\tempurl}
\newblock
\shownote{Accessed October 16, 2026.}


\bibitem[\protect\citeauthoryear{Fan, Fan, Tang, Cao, Li, and Du}{Fan et~al\mbox{.}}{2025}]%
        {fan2025autoprep}
\bibfield{author}{\bibinfo{person}{Meihao Fan}, \bibinfo{person}{Ju Fan}, \bibinfo{person}{Nan Tang}, \bibinfo{person}{Lei Cao}, \bibinfo{person}{Guoliang Li}, {and} \bibinfo{person}{Xiaoyong Du}.} \bibinfo{year}{2025}\natexlab{}.
\newblock \showarticletitle{{AutoPrep}: Natural Language Question-Aware Data Preparation with a Multi-Agent Framework}.
\newblock \bibinfo{journal}{\emph{Proc. {VLDB} Endow.}} \bibinfo{volume}{18}, \bibinfo{number}{10} (\bibinfo{year}{2025}), \bibinfo{pages}{3504--3517}.
\newblock
\urldef\tempurl%
\url{https://doi.org/10.14778/3748191.3748211}
\showDOI{\tempurl}


\bibitem[\protect\citeauthoryear{Fan, Fan, Zhang, Zhang, Du, Song, Li, Jiang, Zhang, and Chen}{Fan et~al\mbox{.}}{2026}]%
        {fan2026deepprep}
\bibfield{author}{\bibinfo{person}{Meihao Fan}, \bibinfo{person}{Ju Fan}, \bibinfo{person}{Yuxin Zhang}, \bibinfo{person}{Shaolei Zhang}, \bibinfo{person}{Xiaoyong Du}, \bibinfo{person}{Jie Song}, \bibinfo{person}{Peng Li}, \bibinfo{person}{Fuxin Jiang}, \bibinfo{person}{Tieying Zhang}, {and} \bibinfo{person}{Jianjun Chen}.} \bibinfo{year}{2026}\natexlab{}.
\newblock \showarticletitle{{DeepPrep}: An {LLM}-Powered Agentic System for Autonomous Data Preparation}.
\newblock \bibinfo{journal}{\emph{Proc. {VLDB} Endow.}} \bibinfo{volume}{19}, \bibinfo{number}{11} (\bibinfo{year}{2026}), \bibinfo{pages}{3371--3384}.
\newblock
\urldef\tempurl%
\url{https://doi.org/10.14778/3836663.3836695}
\showDOI{\tempurl}


\bibitem[\protect\citeauthoryear{Freund, Simon, Berger, and Jun}{Freund et~al\mbox{.}}{2025}]%
        {freund2025flowco}
\bibfield{author}{\bibinfo{person}{Stephen~N. Freund}, \bibinfo{person}{Brooke Simon}, \bibinfo{person}{Emery~D. Berger}, {and} \bibinfo{person}{Eunice Jun}.} \bibinfo{year}{2025}\natexlab{}.
\newblock \showarticletitle{{Flowco}: Mixed-Initiative Authoring of Reliable End-to-End Data Analyses via Dataflow Graphs and {LLMs}}. In \bibinfo{booktitle}{\emph{Proceedings of the 38th Annual ACM Symposium on User Interface Software and Technology}}. \bibinfo{publisher}{ACM}, \bibinfo{pages}{1--20}.
\newblock
\urldef\tempurl%
\url{https://doi.org/10.1145/3746059.3747636}
\showDOI{\tempurl}


\bibitem[\protect\citeauthoryear{{Google}}{{Google}}{2026a}]%
        {google2026gemini4argon}
\bibfield{author}{\bibinfo{person}{{Google}}.} \bibinfo{year}{2026}\natexlab{a}.
\newblock \bibinfo{title}{{Gemini 4 Argon}: Built to Solve Your Hardest Problems}.
\newblock
\newblock
\urldef\tempurl%
\url{https://blog.google/innovation-and-ai/models-and-research/gemini-models/introducing-gemini-4-argon/}
\showURL{%
\tempurl}
\newblock
\shownote{Accessed November 1, 2026.}


\bibitem[\protect\citeauthoryear{{Google}}{{Google}}{2026b}]%
        {google2026gemini38flash}
\bibfield{author}{\bibinfo{person}{{Google}}.} \bibinfo{year}{2026}\natexlab{b}.
\newblock \bibinfo{title}{Introducing {Gemini 3.8 Flash} and 3.8 {Flash Cyber}}.
\newblock
\newblock
\urldef\tempurl%
\url{https://blog.google/innovation-and-ai/models-and-research/gemini-models/3-8-flash-and-3-8-flash-cyber/}
\showURL{%
\tempurl}
\newblock
\shownote{Accessed November 1, 2026.}


\bibitem[\protect\citeauthoryear{Gottweis, Weng, Daryin, Tu, et~al\mbox{.}}{Gottweis et~al\mbox{.}}{2026}]%
        {gottweis2025towards}
\bibfield{author}{\bibinfo{person}{Juraj Gottweis}, \bibinfo{person}{Wei-Hung Weng}, \bibinfo{person}{Alexander Daryin}, \bibinfo{person}{Tao Tu}, {et~al\mbox{.}}} \bibinfo{year}{2026}\natexlab{}.
\newblock \showarticletitle{Accelerating scientific discovery with {Co-Scientist}}.
\newblock \bibinfo{journal}{\emph{Nature}} \bibinfo{volume}{655}, \bibinfo{number}{8122} (\bibinfo{year}{2026}), \bibinfo{pages}{487--496}.
\newblock
\urldef\tempurl%
\url{https://doi.org/10.1038/s41586-026-10644-y}
\showDOI{\tempurl}


\bibitem[\protect\citeauthoryear{Griffin}{Griffin}{2018}]%
        {griffin2018olympics}
\bibfield{author}{\bibinfo{person}{Randi Griffin}.} \bibinfo{year}{2018}\natexlab{}.
\newblock \bibinfo{title}{120 Years of {Olympic} History: Athletes and Results}.
\newblock \bibinfo{howpublished}{Kaggle}.
\newblock
\urldef\tempurl%
\url{https://www.kaggle.com/datasets/heesoo37/120-years-of-olympic-history-athletes-and-results}
\showURL{%
\tempurl}
\newblock
\shownote{Accessed October 28, 2026.}


\bibitem[\protect\citeauthoryear{He, Yao, Ma, Yu, Dai, Zhang, Lan, and Yu}{He et~al\mbox{.}}{2024}]%
        {he2024webvoyager}
\bibfield{author}{\bibinfo{person}{Hongliang He}, \bibinfo{person}{Wenlin Yao}, \bibinfo{person}{Kaixin Ma}, \bibinfo{person}{Wenhao Yu}, \bibinfo{person}{Yong Dai}, \bibinfo{person}{Hongming Zhang}, \bibinfo{person}{Zhenzhong Lan}, {and} \bibinfo{person}{Dong Yu}.} \bibinfo{year}{2024}\natexlab{}.
\newblock \showarticletitle{{WebVoyager}: Building an End-to-End Web Agent with Large Multimodal Models}. In \bibinfo{booktitle}{\emph{Proceedings of the 62nd Annual Meeting of the Association for Computational Linguistics (Volume 1: Long Papers)}}. \bibinfo{publisher}{Association for Computational Linguistics}, \bibinfo{pages}{6864--6890}.
\newblock
\urldef\tempurl%
\url{https://doi.org/10.18653/v1/2024.acl-long.371}
\showDOI{\tempurl}


\bibitem[\protect\citeauthoryear{He, Wong, Liang, Meng, Shen, Ma, and Zhang}{He et~al\mbox{.}}{2026}]%
        {he2026dataflowharness}
\bibfield{author}{\bibinfo{person}{Runming He}, \bibinfo{person}{Zhen~Hao Wong}, \bibinfo{person}{Hao Liang}, \bibinfo{person}{Zimo Meng}, \bibinfo{person}{Chengyu Shen}, \bibinfo{person}{Xiaochen Ma}, {and} \bibinfo{person}{Wentao Zhang}.} \bibinfo{year}{2026}\natexlab{}.
\newblock \bibinfo{title}{{DataFlow-Harness}: A Grounded Code-Agent Platform for Constructing Editable {LLM} Data Pipelines}.
\newblock
\newblock
\urldef\tempurl%
\url{https://doi.org/10.48550/arXiv.2607.16617}
\showDOI{\tempurl}
\showeprint[arxiv]{2607.16617}~[cs.SE]


\bibitem[\protect\citeauthoryear{Hong, Lin, Liu, Liu, Wu, Zhang, Li, Chen, Zhang, Wang, Zhang, Zhang, Yang, Zhuge, Guo, Zhou, Tao, Tang, Lu, Zheng, Liang, Fei, Cheng, Ni, Gou, Xu, Luo, and Wu}{Hong et~al\mbox{.}}{2025}]%
        {conf/acl/HongLLLWZLCZWZZ25}
\bibfield{author}{\bibinfo{person}{Sirui Hong}, \bibinfo{person}{Yizhang Lin}, \bibinfo{person}{Bang Liu}, \bibinfo{person}{Bangbang Liu}, \bibinfo{person}{Binhao Wu}, \bibinfo{person}{Ceyao Zhang}, \bibinfo{person}{Danyang Li}, \bibinfo{person}{Jiaqi Chen}, \bibinfo{person}{Jiayi Zhang}, \bibinfo{person}{Jinlin Wang}, \bibinfo{person}{Li Zhang}, \bibinfo{person}{Lingyao Zhang}, \bibinfo{person}{Min Yang}, \bibinfo{person}{Mingchen Zhuge}, \bibinfo{person}{Taicheng Guo}, \bibinfo{person}{Tuo Zhou}, \bibinfo{person}{Wei Tao}, \bibinfo{person}{Robert Tang}, \bibinfo{person}{Xiangtao Lu}, \bibinfo{person}{Xiawu Zheng}, \bibinfo{person}{Xinbing Liang}, \bibinfo{person}{Yaying Fei}, \bibinfo{person}{Yuheng Cheng}, \bibinfo{person}{Yongxin Ni}, \bibinfo{person}{Zhibin Gou}, \bibinfo{person}{Zongze Xu}, \bibinfo{person}{Yuyu Luo}, {and} \bibinfo{person}{Chenglin Wu}.} \bibinfo{year}{2025}\natexlab{}.
\newblock \showarticletitle{Data Interpreter: An {LLM} Agent for Data Science}. In \bibinfo{booktitle}{\emph{Findings of the Association for Computational Linguistics, {ACL} 2025, Vienna, Austria, July 27 - August 1, 2025}} \emph{(\bibinfo{series}{Findings of {ACL}})}, \bibfield{editor}{\bibinfo{person}{Wanxiang Che}, \bibinfo{person}{Joyce Nabende}, \bibinfo{person}{Ekaterina Shutova}, {and} \bibinfo{person}{Mohammad~Taher Pilehvar}} (Eds.). \bibinfo{publisher}{Association for Computational Linguistics}, \bibinfo{pages}{19796--19821}.
\newblock
\urldef\tempurl%
\url{https://doi.org/10.18653/v1/2025.findings-acl.1016}
\showDOI{\tempurl}


\bibitem[\protect\citeauthoryear{Huang, Luo, Yu, Zhang, Lei, Wei, He, Huang, Liu, Zhao, and Liu}{Huang et~al\mbox{.}}{2024}]%
        {conf/emnlp/HuangLYZLWHHLZL24}
\bibfield{author}{\bibinfo{person}{Yiming Huang}, \bibinfo{person}{Jianwen Luo}, \bibinfo{person}{Yan Yu}, \bibinfo{person}{Yitong Zhang}, \bibinfo{person}{Fangyu Lei}, \bibinfo{person}{Yifan Wei}, \bibinfo{person}{Shizhu He}, \bibinfo{person}{Lifu Huang}, \bibinfo{person}{Xiao Liu}, \bibinfo{person}{Jun Zhao}, {and} \bibinfo{person}{Kang Liu}.} \bibinfo{year}{2024}\natexlab{}.
\newblock \showarticletitle{DA-Code: Agent Data Science Code Generation Benchmark for Large Language Models}. In \bibinfo{booktitle}{\emph{Proceedings of the 2024 Conference on Empirical Methods in Natural Language Processing, {EMNLP} 2024, Miami, FL, USA, November 12-16, 2024}}, \bibfield{editor}{\bibinfo{person}{Yaser Al{-}Onaizan}, \bibinfo{person}{Mohit Bansal}, {and} \bibinfo{person}{Yun{-}Nung Chen}} (Eds.). \bibinfo{publisher}{Association for Computational Linguistics}, \bibinfo{pages}{13487--13521}.
\newblock
\urldef\tempurl%
\url{https://doi.org/10.18653/V1/2024.EMNLP-MAIN.748}
\showDOI{\tempurl}


\bibitem[\protect\citeauthoryear{{Inside Airbnb}}{{Inside Airbnb}}{2026}]%
        {insideairbnb2026sf}
\bibfield{author}{\bibinfo{person}{{Inside Airbnb}}.} \bibinfo{year}{2026}\natexlab{}.
\newblock \bibinfo{title}{{San Francisco} Listings, Reviews, and Neighbourhoods}.
\newblock
\newblock
\urldef\tempurl%
\url{https://insideairbnb.com/get-the-data/}
\showURL{%
\tempurl}
\newblock
\shownote{September 14, 2026 snapshot. Accessed October 28, 2026.}


\bibitem[\protect\citeauthoryear{Jiang, Xie, Shen, Shen, Zhang, Lei, Zheng, Li, Li, Huang, Wu, Zhang, Cui, and Chen}{Jiang et~al\mbox{.}}{2025}]%
        {journals/pvldb/JiangXSSZLZLLHWZYCC25}
\bibfield{author}{\bibinfo{person}{Jie Jiang}, \bibinfo{person}{Haining Xie}, \bibinfo{person}{Siqi Shen}, \bibinfo{person}{Yu Shen}, \bibinfo{person}{Zihan Zhang}, \bibinfo{person}{Meng Lei}, \bibinfo{person}{Yifeng Zheng}, \bibinfo{person}{Yang Li}, \bibinfo{person}{Chunyou Li}, \bibinfo{person}{Danqing Huang}, \bibinfo{person}{Yinjun Wu}, \bibinfo{person}{Wentao Zhang}, \bibinfo{person}{Bin Cui}, {and} \bibinfo{person}{Peng Chen}.} \bibinfo{year}{2025}\natexlab{}.
\newblock \showarticletitle{SiriusBI: {A} Comprehensive {LLM}-Powered Solution for Data Analytics in Business Intelligence}.
\newblock \bibinfo{journal}{\emph{Proc. {VLDB} Endow.}} \bibinfo{volume}{18}, \bibinfo{number}{12} (\bibinfo{year}{2025}), \bibinfo{pages}{4860--4873}.
\newblock
\urldef\tempurl%
\url{https://doi.org/10.14778/3750601.3750610}
\showDOI{\tempurl}


\bibitem[\protect\citeauthoryear{John, Bacon, Chen, Ramesh, Li, Das, Claus, Kendall, and Patel}{John et~al\mbox{.}}{2023}]%
        {conf/sigmod/JohnBCRLDCKP23}
\bibfield{author}{\bibinfo{person}{Rogers Jeffrey~Leo John}, \bibinfo{person}{Dylan Bacon}, \bibinfo{person}{Junda Chen}, \bibinfo{person}{Ushmal Ramesh}, \bibinfo{person}{Jiatong Li}, \bibinfo{person}{Deepan Das}, \bibinfo{person}{Robert~V. Claus}, \bibinfo{person}{Amos Kendall}, {and} \bibinfo{person}{Jignesh~M. Patel}.} \bibinfo{year}{2023}\natexlab{}.
\newblock \showarticletitle{DataChat: An Intuitive and Collaborative Data Analytics Platform}. In \bibinfo{booktitle}{\emph{Companion of the 2023 International Conference on Management of Data, {SIGMOD/PODS} 2023, Seattle, WA, USA, June 18-23, 2023}}, \bibfield{editor}{\bibinfo{person}{Sudipto Das}, \bibinfo{person}{Ippokratis Pandis}, \bibinfo{person}{K.~Sel{\c{c}}uk Candan}, {and} \bibinfo{person}{Sihem Amer{-}Yahia}} (Eds.). \bibinfo{publisher}{{ACM}}, \bibinfo{pages}{203--215}.
\newblock
\urldef\tempurl%
\url{https://doi.org/10.1145/3555041.3589678}
\showDOI{\tempurl}


\bibitem[\protect\citeauthoryear{Kazemitabaar, Williams, Drosos, Grossman, Henley, Negreanu, and Sarkar}{Kazemitabaar et~al\mbox{.}}{2024}]%
        {conf/uist/Kazemitabaar0DG24}
\bibfield{author}{\bibinfo{person}{Majeed Kazemitabaar}, \bibinfo{person}{Jack Williams}, \bibinfo{person}{Ian Drosos}, \bibinfo{person}{Tovi Grossman}, \bibinfo{person}{Austin~Zachary Henley}, \bibinfo{person}{Carina Negreanu}, {and} \bibinfo{person}{Advait Sarkar}.} \bibinfo{year}{2024}\natexlab{}.
\newblock \showarticletitle{Improving Steering and Verification in AI-Assisted Data Analysis with Interactive Task Decomposition}. In \bibinfo{booktitle}{\emph{Proceedings of the 37th Annual {ACM} Symposium on User Interface Software and Technology, {UIST} 2024, Pittsburgh, PA, USA, October 13-16, 2024}}, \bibfield{editor}{\bibinfo{person}{Lining Yao}, \bibinfo{person}{Mayank Goel}, \bibinfo{person}{Alexandra Ion}, {and} \bibinfo{person}{Pedro Lopes}} (Eds.). \bibinfo{publisher}{{ACM}}, \bibinfo{pages}{92:1--92:19}.
\newblock
\urldef\tempurl%
\url{https://doi.org/10.1145/3654777.3676345}
\showDOI{\tempurl}


\bibitem[\protect\citeauthoryear{Kumar, Wang, Ni, and Li}{Kumar et~al\mbox{.}}{2020}]%
        {journals/pvldb/KumarWNL20}
\bibfield{author}{\bibinfo{person}{Avinash Kumar}, \bibinfo{person}{Zuozhi Wang}, \bibinfo{person}{Shengquan Ni}, {and} \bibinfo{person}{Chen Li}.} \bibinfo{year}{2020}\natexlab{}.
\newblock \showarticletitle{Amber: {A} Debuggable Dataflow System Based on the Actor Model}.
\newblock \bibinfo{journal}{\emph{Proc. {VLDB} Endow.}} \bibinfo{volume}{13}, \bibinfo{number}{5} (\bibinfo{year}{2020}), \bibinfo{pages}{740--753}.
\newblock
\urldef\tempurl%
\url{https://doi.org/10.14778/3377369.3377381}
\showDOI{\tempurl}


\bibitem[\protect\citeauthoryear{Lai, Vitagliano, Zhang, Chabra, Sudhir, Zeng, Zabreyko, Li, Kossmann, Ding, Chen, Markakis, Russo, Wang, Wu, Cafarella, Cao, Madden, and Kraska}{Lai et~al\mbox{.}}{2026}]%
        {journals/corr/abs-2506-06541}
\bibfield{author}{\bibinfo{person}{Eugenie Lai}, \bibinfo{person}{Gerardo Vitagliano}, \bibinfo{person}{Ziyu Zhang}, \bibinfo{person}{Om Chabra}, \bibinfo{person}{Sivaprasad Sudhir}, \bibinfo{person}{Anna Zeng}, \bibinfo{person}{Anton~A. Zabreyko}, \bibinfo{person}{Chenning Li}, \bibinfo{person}{Ferdi Kossmann}, \bibinfo{person}{Jialin Ding}, \bibinfo{person}{Jun Chen}, \bibinfo{person}{Markos Markakis}, \bibinfo{person}{Matthew Russo}, \bibinfo{person}{Weiyang Wang}, \bibinfo{person}{Ziniu Wu}, \bibinfo{person}{Michael~J. Cafarella}, \bibinfo{person}{Lei Cao}, \bibinfo{person}{Samuel Madden}, {and} \bibinfo{person}{Tim Kraska}.} \bibinfo{year}{2026}\natexlab{}.
\newblock \showarticletitle{KramaBench: {A} Benchmark for {AI} Systems on Data-to-Insight Pipelines over Data Lakes}. In \bibinfo{booktitle}{\emph{The Fourteenth International Conference on Learning Representations, {ICLR} 2026}}.
\newblock
\urldef\tempurl%
\url{https://openreview.net/forum?id=fZfUdeCC5X}
\showURL{%
\tempurl}


\bibitem[\protect\citeauthoryear{Liang, Ma, Liu, Wong, Zhao, Meng, He, Shen, Cai, Han, Qiang, Feng, Bai, Pan, Guo, Jiang, Deng, You, Lai, Guo, Tsai, Feng, Hu, Yu, Niu, Zeng, An, Ma, Huang, Zheng, He, Tang, Cui, E, and Zhang}{Liang et~al\mbox{.}}{2025}]%
        {liang2025dataflow}
\bibfield{author}{\bibinfo{person}{Hao Liang}, \bibinfo{person}{Xiaochen Ma}, \bibinfo{person}{Zhou Liu}, \bibinfo{person}{Zhen~Hao Wong}, \bibinfo{person}{Zhengyang Zhao}, \bibinfo{person}{Zimo Meng}, \bibinfo{person}{Runming He}, \bibinfo{person}{Chengyu Shen}, \bibinfo{person}{Qifeng Cai}, \bibinfo{person}{Zhaoyang Han}, \bibinfo{person}{Meiyi Qiang}, \bibinfo{person}{Yalin Feng}, \bibinfo{person}{Tianyi Bai}, \bibinfo{person}{Zewei Pan}, \bibinfo{person}{Ziyi Guo}, \bibinfo{person}{Yizhen Jiang}, \bibinfo{person}{Jingwen Deng}, \bibinfo{person}{Qijie You}, \bibinfo{person}{Peichao Lai}, \bibinfo{person}{Tianyu Guo}, \bibinfo{person}{Chi~Hsu Tsai}, \bibinfo{person}{Hengyi Feng}, \bibinfo{person}{Rui Hu}, \bibinfo{person}{Wenkai Yu}, \bibinfo{person}{Junbo Niu}, \bibinfo{person}{Bohan Zeng}, \bibinfo{person}{Ruichuan An}, \bibinfo{person}{Lu Ma}, \bibinfo{person}{Jihao Huang}, \bibinfo{person}{Yaowei Zheng}, \bibinfo{person}{Conghui He}, \bibinfo{person}{Linpeng Tang}, \bibinfo{person}{Bin Cui}, \bibinfo{person}{Weinan E}, {and} \bibinfo{person}{Wentao Zhang}.} \bibinfo{year}{2025}\natexlab{}.
\newblock \bibinfo{title}{{DataFlow}: An {LLM}-Driven Framework for Unified Data Preparation and Workflow Automation in the Era of Data-Centric {AI}}.
\newblock
\newblock
\urldef\tempurl%
\url{https://doi.org/10.48550/arXiv.2512.16676}
\showDOI{\tempurl}
\showeprint[arxiv]{2512.16676}~[cs.LG]


\bibitem[\protect\citeauthoryear{Lindenbauer, Slinko, Felder, Bogomolov, and Zharov}{Lindenbauer et~al\mbox{.}}{2025}]%
        {lindenbauer2025complexitytrapsimpleobservation}
\bibfield{author}{\bibinfo{person}{Tobias Lindenbauer}, \bibinfo{person}{Igor Slinko}, \bibinfo{person}{Ludwig Felder}, \bibinfo{person}{Egor Bogomolov}, {and} \bibinfo{person}{Yaroslav Zharov}.} \bibinfo{year}{2025}\natexlab{}.
\newblock \bibinfo{title}{The Complexity Trap: Simple Observation Masking Is as Efficient as {LLM} Summarization for Agent Context Management}.
\newblock
\newblock
\urldef\tempurl%
\url{https://doi.org/10.48550/arXiv.2508.21433}
\showDOI{\tempurl}
\showeprint[arxiv]{2508.21433}~[cs.SE]


\bibitem[\protect\citeauthoryear{{LiteLLM}}{{LiteLLM}}{[n.d.]}]%
        {website:litellm}
\bibfield{author}{\bibinfo{person}{{LiteLLM}}.} \bibinfo{year}{[n.d.]}\natexlab{}.
\newblock \bibinfo{title}{{LiteLLM} AI Gateway}.
\newblock
\newblock
\urldef\tempurl%
\url{https://docs.litellm.ai/docs/simple_proxy}
\showURL{%
\tempurl}
\newblock
\shownote{Accessed October 31, 2026.}


\bibitem[\protect\citeauthoryear{Liu, Zhang, and Fan}{Liu et~al\mbox{.}}{2026}]%
        {dastudio2026}
\bibfield{author}{\bibinfo{person}{Yizhe Liu}, \bibinfo{person}{Shaolei Zhang}, {and} \bibinfo{person}{Ju Fan}.} \bibinfo{year}{2026}\natexlab{}.
\newblock \showarticletitle{{DA-Studio}: An Agentic System for End-to-End Data Analysis}.
\newblock \bibinfo{journal}{\emph{Proc. {VLDB} Endow.}} \bibinfo{volume}{19}, \bibinfo{number}{12} (\bibinfo{year}{2026}), \bibinfo{pages}{4766--4769}.
\newblock
\urldef\tempurl%
\url{https://doi.org/10.14778/3827998.3828117}
\showDOI{\tempurl}


\bibitem[\protect\citeauthoryear{Luo, Li, Fan, and Tang}{Luo et~al\mbox{.}}{2026}]%
        {luo2026dataagents}
\bibfield{author}{\bibinfo{person}{Yuyu Luo}, \bibinfo{person}{Guoliang Li}, \bibinfo{person}{Ju Fan}, {and} \bibinfo{person}{Nan Tang}.} \bibinfo{year}{2026}\natexlab{}.
\newblock \showarticletitle{Data Agents: Levels, State of the Art, and Open Problems}. In \bibinfo{booktitle}{\emph{Companion of the 2026 International Conference on Management of Data ({SIGMOD/PODS} 2026)}}. \bibinfo{publisher}{{ACM}}, \bibinfo{pages}{571--579}.
\newblock
\urldef\tempurl%
\url{https://doi.org/10.1145/3788853.3801878}
\showDOI{\tempurl}


\bibitem[\protect\citeauthoryear{{MinIO}}{{MinIO}}{[n.d.]}]%
        {website:minio}
\bibfield{author}{\bibinfo{person}{{MinIO}}.} \bibinfo{year}{[n.d.]}\natexlab{}.
\newblock \bibinfo{title}{{MinIO}}.
\newblock
\newblock
\urldef\tempurl%
\url{https://www.min.io/}
\showURL{%
\tempurl}
\newblock
\shownote{Accessed November 1, 2026.}


\bibitem[\protect\citeauthoryear{Nam, Yoon, Chen, Sinha, Shin, and Pfister}{Nam et~al\mbox{.}}{2025}]%
        {journals/corr/abs-2509-21825}
\bibfield{author}{\bibinfo{person}{Jaehyun Nam}, \bibinfo{person}{Jinsung Yoon}, \bibinfo{person}{Jiefeng Chen}, \bibinfo{person}{Raj Sinha}, \bibinfo{person}{Jinwoo Shin}, {and} \bibinfo{person}{Tomas Pfister}.} \bibinfo{year}{2025}\natexlab{}.
\newblock \bibinfo{title}{{DS-STAR:} Data Science Agent for Solving Diverse Tasks across Heterogeneous Formats and Open-Ended Queries}.
\newblock
\newblock
\urldef\tempurl%
\url{https://doi.org/10.48550/arXiv.2509.21825}
\showDOI{\tempurl}
\showeprint[arxiv]{2509.21825}~[cs.AI]


\bibitem[\protect\citeauthoryear{{OpenAI}}{{OpenAI}}{2026a}]%
        {openai2026gpt56}
\bibfield{author}{\bibinfo{person}{{OpenAI}}.} \bibinfo{year}{2026}\natexlab{a}.
\newblock \bibinfo{title}{{GPT-5.6}: Frontier Intelligence That Scales with Your Ambition}.
\newblock
\newblock
\urldef\tempurl%
\url{https://openai.com/index/gpt-5-6/}
\showURL{%
\tempurl}
\newblock
\shownote{Accessed October 16, 2026.}


\bibitem[\protect\citeauthoryear{{OpenAI}}{{OpenAI}}{2026b}]%
        {openai-data-agent}
\bibfield{author}{\bibinfo{person}{{OpenAI}}.} \bibinfo{year}{2026}\natexlab{b}.
\newblock \bibinfo{title}{{Inside OpenAI's in-house data agent}}.
\newblock
\newblock
\urldef\tempurl%
\url{https://openai.com/index/inside-our-in-house-data-agent/}
\showURL{%
\tempurl}
\newblock
\shownote{Accessed November 1, 2026.}


\bibitem[\protect\citeauthoryear{{OpenAI}}{{OpenAI}}{2026c}]%
        {openai2026gpt61}
\bibfield{author}{\bibinfo{person}{{OpenAI}}.} \bibinfo{year}{2026}\natexlab{c}.
\newblock \bibinfo{title}{Introducing {GPT-6.1 Sol}}.
\newblock
\newblock
\urldef\tempurl%
\url{https://openai.com/index/introducing-gpt-6-1-sol/}
\showURL{%
\tempurl}
\newblock
\shownote{Accessed November 1, 2026.}


\bibitem[\protect\citeauthoryear{{OpenAI}}{{OpenAI}}{2026d}]%
        {openai2026promptcaching}
\bibfield{author}{\bibinfo{person}{{OpenAI}}.} \bibinfo{year}{2026}\natexlab{d}.
\newblock \bibinfo{title}{Prompt Caching}.
\newblock
\newblock
\urldef\tempurl%
\url{https://developers.openai.com/api/docs/guides/prompt-caching}
\showURL{%
\tempurl}
\newblock
\shownote{Accessed September 13, 2026.}


\bibitem[\protect\citeauthoryear{Rocklin}{Rocklin}{2015}]%
        {conf/scipy/Rocklin15}
\bibfield{author}{\bibinfo{person}{Matthew Rocklin}.} \bibinfo{year}{2015}\natexlab{}.
\newblock \showarticletitle{Dask: Parallel Computation with Blocked algorithms and Task Scheduling}. In \bibinfo{booktitle}{\emph{Proceedings of the 14th Python in Science Conference 2015 (SciPy 2015), Austin, Texas, July 6 - 12, 2015}}, \bibfield{editor}{\bibinfo{person}{Kathryn Huff} {and} \bibinfo{person}{James Bergstra}} (Eds.). \bibinfo{publisher}{scipy.org}, \bibinfo{pages}{126--132}.
\newblock
\urldef\tempurl%
\url{https://doi.org/10.25080/MAJORA-7B98E3ED-013}
\showDOI{\tempurl}


\bibitem[\protect\citeauthoryear{{RocksDB}}{{RocksDB}}{[n.d.]}]%
        {website:rocksdb}
\bibfield{author}{\bibinfo{person}{{RocksDB}}.} \bibinfo{year}{[n.d.]}\natexlab{}.
\newblock \bibinfo{title}{{RocksDB}: A Persistent Key-Value Store}.
\newblock
\newblock
\urldef\tempurl%
\url{https://rocksdb.org/}
\showURL{%
\tempurl}
\newblock
\shownote{Accessed September 28, 2026.}


\bibitem[\protect\citeauthoryear{Roucher, del Moral, Wolf, von Werra, and Kaunism{\"a}ki}{Roucher et~al\mbox{.}}{2025}]%
        {smolagents2025}
\bibfield{author}{\bibinfo{person}{Aymeric Roucher}, \bibinfo{person}{Albert~Villanova del Moral}, \bibinfo{person}{Thomas Wolf}, \bibinfo{person}{Leandro von Werra}, {and} \bibinfo{person}{Erik Kaunism{\"a}ki}.} \bibinfo{year}{2025}\natexlab{}.
\newblock \bibinfo{title}{\texttt{smolagents}: a smol library to build great agentic systems}.
\newblock
\newblock
\urldef\tempurl%
\url{https://github.com/huggingface/smolagents}
\showURL{%
\tempurl}


\bibitem[\protect\citeauthoryear{Russo, Agarwal, Li, Gu, Cafarella, Khattab, Kraska, and Madden}{Russo et~al\mbox{.}}{2026}]%
        {carnot2026}
\bibfield{author}{\bibinfo{person}{Matthew Russo}, \bibinfo{person}{Yash Agarwal}, \bibinfo{person}{Tianyu Li}, \bibinfo{person}{Zhuohan Gu}, \bibinfo{person}{Michael Cafarella}, \bibinfo{person}{Omar Khattab}, \bibinfo{person}{Tim Kraska}, {and} \bibinfo{person}{Samuel Madden}.} \bibinfo{year}{2026}\natexlab{}.
\newblock \showarticletitle{{Carnot}: Interpretable, Interactive, and Optimized Execution of Deep Research Queries}.
\newblock \bibinfo{journal}{\emph{Proc. {VLDB} Endow.}} \bibinfo{volume}{19}, \bibinfo{number}{12} (\bibinfo{year}{2026}), \bibinfo{pages}{4642--4645}.
\newblock
\urldef\tempurl%
\url{https://doi.org/10.14778/3827998.3828086}
\showDOI{\tempurl}


\bibitem[\protect\citeauthoryear{Schmidgall, Su, Wang, Sun, Wu, Yu, Liu, Moor, Liu, and Barsoum}{Schmidgall et~al\mbox{.}}{2025}]%
        {schmidgall2025agentlaboratory}
\bibfield{author}{\bibinfo{person}{Samuel Schmidgall}, \bibinfo{person}{Yusheng Su}, \bibinfo{person}{Ze Wang}, \bibinfo{person}{Ximeng Sun}, \bibinfo{person}{Jialian Wu}, \bibinfo{person}{Xiaodong Yu}, \bibinfo{person}{Jiang Liu}, \bibinfo{person}{Michael Moor}, \bibinfo{person}{Zicheng Liu}, {and} \bibinfo{person}{Emad Barsoum}.} \bibinfo{year}{2025}\natexlab{}.
\newblock \showarticletitle{Agent Laboratory: Using LLM Agents as Research Assistants}. In \bibinfo{booktitle}{\emph{Findings of the Association for Computational Linguistics: EMNLP 2025}}. \bibinfo{publisher}{Association for Computational Linguistics}, \bibinfo{pages}{5977--6043}.
\newblock
\urldef\tempurl%
\url{https://doi.org/10.18653/v1/2025.findings-emnlp.320}
\showDOI{\tempurl}


\bibitem[\protect\citeauthoryear{Shankar, Chopra, Hasan, Lee, Hartmann, Hellerstein, Parameswaran, and Wu}{Shankar et~al\mbox{.}}{2025}]%
        {shankar2025docwrangler}
\bibfield{author}{\bibinfo{person}{Shreya Shankar}, \bibinfo{person}{Bhavya Chopra}, \bibinfo{person}{Mawil Hasan}, \bibinfo{person}{Stephen Lee}, \bibinfo{person}{Bj{\"o}rn Hartmann}, \bibinfo{person}{Joseph~M. Hellerstein}, \bibinfo{person}{Aditya~G. Parameswaran}, {and} \bibinfo{person}{Eugene Wu}.} \bibinfo{year}{2025}\natexlab{}.
\newblock \showarticletitle{Steering Semantic Data Processing With {DocWrangler}}. In \bibinfo{booktitle}{\emph{Proceedings of the 38th Annual ACM Symposium on User Interface Software and Technology}}. \bibinfo{publisher}{Association for Computing Machinery}, \bibinfo{pages}{84:1--84:18}.
\newblock
\urldef\tempurl%
\url{https://doi.org/10.1145/3746059.3747625}
\showDOI{\tempurl}


\bibitem[\protect\citeauthoryear{Sumers, Yao, Narasimhan, and Griffiths}{Sumers et~al\mbox{.}}{2024}]%
        {sumers2024cognitive}
\bibfield{author}{\bibinfo{person}{Theodore Sumers}, \bibinfo{person}{Shunyu Yao}, \bibinfo{person}{Karthik~R. Narasimhan}, {and} \bibinfo{person}{Thomas~L. Griffiths}.} \bibinfo{year}{2024}\natexlab{}.
\newblock \showarticletitle{Cognitive Architectures for Language Agents}.
\newblock \bibinfo{journal}{\emph{Transactions on Machine Learning Research}} (\bibinfo{year}{2024}).
\newblock
\showISSN{2835-8856}
\urldef\tempurl%
\url{https://openreview.net/forum?id=1i6ZCvflQJ}
\showURL{%
\tempurl}


\bibitem[\protect\citeauthoryear{Taylor, Huang, Hao, Lin, Chen, Wang, and Li}{Taylor et~al\mbox{.}}{2024}]%
        {conf/icde/TaylorHHLCWL24}
\bibfield{author}{\bibinfo{person}{Alexander~K. Taylor}, \bibinfo{person}{Yicong Huang}, \bibinfo{person}{Junheng Hao}, \bibinfo{person}{Xinyuan Lin}, \bibinfo{person}{Xiusi Chen}, \bibinfo{person}{Wei Wang}, {and} \bibinfo{person}{Chen Li}.} \bibinfo{year}{2024}\natexlab{}.
\newblock \showarticletitle{Data Science Tasks Implemented with Scripts versus GUI-Based Workflows: The Good, the Bad, and the Ugly}. In \bibinfo{booktitle}{\emph{40th International Conference on Data Engineering, {ICDE} 2024 - Workshops, Utrecht, Netherlands, May 13-16, 2024}}. \bibinfo{publisher}{{IEEE}}, \bibinfo{pages}{267--277}.
\newblock
\urldef\tempurl%
\url{https://doi.org/10.1109/ICDEW61823.2024.00040}
\showDOI{\tempurl}


\bibitem[\protect\citeauthoryear{{The PostgreSQL Global Development Group}}{{The PostgreSQL Global Development Group}}{[n.d.]}]%
        {website:postgresql}
\bibfield{author}{\bibinfo{person}{{The PostgreSQL Global Development Group}}.} \bibinfo{year}{[n.d.]}\natexlab{}.
\newblock \bibinfo{title}{About {PostgreSQL}}.
\newblock
\newblock
\urldef\tempurl%
\url{https://www.postgresql.org/about/}
\showURL{%
\tempurl}
\newblock
\shownote{Accessed September 28, 2026.}


\bibitem[\protect\citeauthoryear{Wang, Lee, Drucker, Marshall, and Gao}{Wang et~al\mbox{.}}{2025a}]%
        {dataformulator2}
\bibfield{author}{\bibinfo{person}{Chenglong Wang}, \bibinfo{person}{Bongshin Lee}, \bibinfo{person}{Steven~M. Drucker}, \bibinfo{person}{Dan Marshall}, {and} \bibinfo{person}{Jianfeng Gao}.} \bibinfo{year}{2025}\natexlab{a}.
\newblock \showarticletitle{{Data Formulator 2}: Iterative Creation of Data Visualizations, with {AI} Transforming Data Along the Way}. In \bibinfo{booktitle}{\emph{Proceedings of the 2025 CHI Conference on Human Factors in Computing Systems}}. \bibinfo{publisher}{ACM}, Article \bibinfo{articleno}{677}, \bibinfo{numpages}{17}~pages.
\newblock
\urldef\tempurl%
\url{https://doi.org/10.1145/3706598.3713296}
\showDOI{\tempurl}


\bibitem[\protect\citeauthoryear{Wang, Chen, Yuan, Zhang, Li, Peng, and Ji}{Wang et~al\mbox{.}}{2024a}]%
        {pmlr-v235-wang24h}
\bibfield{author}{\bibinfo{person}{Xingyao Wang}, \bibinfo{person}{Yangyi Chen}, \bibinfo{person}{Lifan Yuan}, \bibinfo{person}{Yizhe Zhang}, \bibinfo{person}{Yunzhu Li}, \bibinfo{person}{Hao Peng}, {and} \bibinfo{person}{Heng Ji}.} \bibinfo{year}{2024}\natexlab{a}.
\newblock \showarticletitle{Executable Code Actions Elicit Better {LLM} Agents}. In \bibinfo{booktitle}{\emph{Proceedings of the 41st International Conference on Machine Learning}} \emph{(\bibinfo{series}{Proceedings of Machine Learning Research})}, Vol.~\bibinfo{volume}{235}. \bibinfo{publisher}{PMLR}, \bibinfo{pages}{50208--50232}.
\newblock
\urldef\tempurl%
\url{https://proceedings.mlr.press/v235/wang24h.html}
\showURL{%
\tempurl}


\bibitem[\protect\citeauthoryear{Wang, Li, Song, Xu, Tang, Zhuge, Pan, Song, Li, Singh, Tran, Li, Ma, Zheng, Qian, Shao, Muennighoff, Zhang, Hui, Lin, Brennan, Peng, Ji, and Neubig}{Wang et~al\mbox{.}}{2025b}]%
        {wang2025openhands}
\bibfield{author}{\bibinfo{person}{Xingyao Wang}, \bibinfo{person}{Boxuan Li}, \bibinfo{person}{Yufan Song}, \bibinfo{person}{Frank~F. Xu}, \bibinfo{person}{Xiangru Tang}, \bibinfo{person}{Mingchen Zhuge}, \bibinfo{person}{Jiayi Pan}, \bibinfo{person}{Yueqi Song}, \bibinfo{person}{Bowen Li}, \bibinfo{person}{Jaskirat Singh}, \bibinfo{person}{Hoang~H. Tran}, \bibinfo{person}{Fuqiang Li}, \bibinfo{person}{Ren Ma}, \bibinfo{person}{Mingzhang Zheng}, \bibinfo{person}{Bill Qian}, \bibinfo{person}{Yanjun Shao}, \bibinfo{person}{Niklas Muennighoff}, \bibinfo{person}{Yizhe Zhang}, \bibinfo{person}{Binyuan Hui}, \bibinfo{person}{Junyang Lin}, \bibinfo{person}{Robert Brennan}, \bibinfo{person}{Hao Peng}, \bibinfo{person}{Heng Ji}, {and} \bibinfo{person}{Graham Neubig}.} \bibinfo{year}{2025}\natexlab{b}.
\newblock \showarticletitle{{OpenHands}: An Open Platform for {AI} Software Developers as Generalist Agents}. In \bibinfo{booktitle}{\emph{The Thirteenth International Conference on Learning Representations, {ICLR} 2025}}.
\newblock
\urldef\tempurl%
\url{https://arxiv.org/abs/2407.16741}
\showURL{%
\tempurl}


\bibitem[\protect\citeauthoryear{Wang, Huang, Ni, Kumar, Alsudais, Liu, Lin, Ding, and Li}{Wang et~al\mbox{.}}{2024b}]%
        {wang2024texera}
\bibfield{author}{\bibinfo{person}{Zuozhi Wang}, \bibinfo{person}{Yicong Huang}, \bibinfo{person}{Shengquan Ni}, \bibinfo{person}{Avinash Kumar}, \bibinfo{person}{Sadeem Alsudais}, \bibinfo{person}{Xiaozhen Liu}, \bibinfo{person}{Xinyuan Lin}, \bibinfo{person}{Yunyan Ding}, {and} \bibinfo{person}{Chen Li}.} \bibinfo{year}{2024}\natexlab{b}.
\newblock \showarticletitle{{Texera}: A System for Collaborative and Interactive Data Analytics Using Workflows}.
\newblock \bibinfo{journal}{\emph{Proc. {VLDB} Endow.}} \bibinfo{volume}{17}, \bibinfo{number}{11} (\bibinfo{year}{2024}), \bibinfo{pages}{3580--3588}.
\newblock
\urldef\tempurl%
\url{https://doi.org/10.14778/3681954.3682022}
\showDOI{\tempurl}


\bibitem[\protect\citeauthoryear{Wu, Bansal, Zhang, Wu, Li, Zhu, Jiang, Zhang, Zhang, Liu, Awadallah, White, Burger, and Wang}{Wu et~al\mbox{.}}{2024}]%
        {autogen2024}
\bibfield{author}{\bibinfo{person}{Qingyun Wu}, \bibinfo{person}{Gagan Bansal}, \bibinfo{person}{Jieyu Zhang}, \bibinfo{person}{Yiran Wu}, \bibinfo{person}{Beibin Li}, \bibinfo{person}{Erkang Zhu}, \bibinfo{person}{Li Jiang}, \bibinfo{person}{Xiaoyun Zhang}, \bibinfo{person}{Shaokun Zhang}, \bibinfo{person}{Jiale Liu}, \bibinfo{person}{Ahmed~Hassan Awadallah}, \bibinfo{person}{Ryen~W. White}, \bibinfo{person}{Doug Burger}, {and} \bibinfo{person}{Chi Wang}.} \bibinfo{year}{2024}\natexlab{}.
\newblock \showarticletitle{{AutoGen}: Enabling Next-Gen {LLM} Applications via Multi-Agent Conversations}. In \bibinfo{booktitle}{\emph{Proceedings of the First Conference on Language Modeling}}.
\newblock
\urldef\tempurl%
\url{https://openreview.net/forum?id=BAakY1hNKS}
\showURL{%
\tempurl}


\bibitem[\protect\citeauthoryear{Wu, Hellerstein, and Satyanarayan}{Wu et~al\mbox{.}}{2020}]%
        {b2notebooks2020}
\bibfield{author}{\bibinfo{person}{Yifan Wu}, \bibinfo{person}{Joseph~M. Hellerstein}, {and} \bibinfo{person}{Arvind Satyanarayan}.} \bibinfo{year}{2020}\natexlab{}.
\newblock \showarticletitle{{B2}: Bridging Code and Interactive Visualization in Computational Notebooks}. In \bibinfo{booktitle}{\emph{Proceedings of the 33rd Annual ACM Symposium on User Interface Software and Technology}}. \bibinfo{publisher}{ACM}, \bibinfo{pages}{152--165}.
\newblock
\urldef\tempurl%
\url{https://doi.org/10.1145/3379337.3415851}
\showDOI{\tempurl}


\bibitem[\protect\citeauthoryear{Xie, Zheng, Xia, Qu, and Zhu-Tian}{Xie et~al\mbox{.}}{2024}]%
        {xie2024waitgpt}
\bibfield{author}{\bibinfo{person}{Liwenhan Xie}, \bibinfo{person}{Chengbo Zheng}, \bibinfo{person}{Haijun Xia}, \bibinfo{person}{Huamin Qu}, {and} \bibinfo{person}{Chen Zhu-Tian}.} \bibinfo{year}{2024}\natexlab{}.
\newblock \showarticletitle{{WaitGPT}: Monitoring and Steering Conversational {LLM} Agent in Data Analysis with On-the-Fly Code Visualization}. In \bibinfo{booktitle}{\emph{Proceedings of the 37th Annual ACM Symposium on User Interface Software and Technology}}. \bibinfo{publisher}{ACM}, \bibinfo{pages}{1--14}.
\newblock
\urldef\tempurl%
\url{https://doi.org/10.1145/3654777.3676374}
\showDOI{\tempurl}


\bibitem[\protect\citeauthoryear{Xu, Wang, Wang, and Li}{Xu et~al\mbox{.}}{2026b}]%
        {xu2026prepbench}
\bibfield{author}{\bibinfo{person}{Jingzhe Xu}, \bibinfo{person}{Rui Wang}, \bibinfo{person}{Jiannan Wang}, {and} \bibinfo{person}{Guoliang Li}.} \bibinfo{year}{2026}\natexlab{b}.
\newblock \showarticletitle{PrepBench: How Far Are We from Natural-Language-Driven Data Preparation?}
\newblock \bibinfo{journal}{\emph{Proc. {VLDB} Endow.}} \bibinfo{volume}{19}, \bibinfo{number}{10} (\bibinfo{year}{2026}), \bibinfo{pages}{2866--2879}.
\newblock
\urldef\tempurl%
\url{https://doi.org/10.14778/3828612.3828638}
\showDOI{\tempurl}


\bibitem[\protect\citeauthoryear{Xu, Lu, Qiao, Ding, Xu, Liang, and Zhang}{Xu et~al\mbox{.}}{2026a}]%
        {xu2026longdsbench}
\bibfield{author}{\bibinfo{person}{Kewei Xu}, \bibinfo{person}{Xiaoben Lu}, \bibinfo{person}{Shuofei Qiao}, \bibinfo{person}{Zihan Ding}, \bibinfo{person}{Haoming Xu}, \bibinfo{person}{Lei Liang}, {and} \bibinfo{person}{Ningyu Zhang}.} \bibinfo{year}{2026}\natexlab{a}.
\newblock \showarticletitle{{LongDS-Bench}: On the Failure of Long-Horizon Agentic Data Analysis}. In \bibinfo{booktitle}{\emph{Proceedings of the 2026 Conference on Empirical Methods in Natural Language Processing}}. \bibinfo{publisher}{Association for Computational Linguistics}, \bibinfo{pages}{13351--13383}.
\newblock
\urldef\tempurl%
\url{https://doi.org/10.18653/2026.emnlp-main.584}
\showDOI{\tempurl}


\bibitem[\protect\citeauthoryear{Yang, Jimenez, Wettig, Lieret, Yao, Narasimhan, and Press}{Yang et~al\mbox{.}}{2024}]%
        {yang2024sweagent}
\bibfield{author}{\bibinfo{person}{John Yang}, \bibinfo{person}{Carlos~E. Jimenez}, \bibinfo{person}{Alexander Wettig}, \bibinfo{person}{Kilian Lieret}, \bibinfo{person}{Shunyu Yao}, \bibinfo{person}{Karthik Narasimhan}, {and} \bibinfo{person}{Ofir Press}.} \bibinfo{year}{2024}\natexlab{}.
\newblock \showarticletitle{{SWE-agent}: Agent-Computer Interfaces Enable Automated Software Engineering}. In \bibinfo{booktitle}{\emph{Advances in Neural Information Processing Systems}}, Vol.~\bibinfo{volume}{37}.
\newblock
\urldef\tempurl%
\url{https://doi.org/10.52202/079017-1601}
\showDOI{\tempurl}


\bibitem[\protect\citeauthoryear{Yao, Zhao, Yu, Du, Shafran, Narasimhan, and Cao}{Yao et~al\mbox{.}}{2023}]%
        {conf/iclr/YaoZYDSN023}
\bibfield{author}{\bibinfo{person}{Shunyu Yao}, \bibinfo{person}{Jeffrey Zhao}, \bibinfo{person}{Dian Yu}, \bibinfo{person}{Nan Du}, \bibinfo{person}{Izhak Shafran}, \bibinfo{person}{Karthik~R. Narasimhan}, {and} \bibinfo{person}{Yuan Cao}.} \bibinfo{year}{2023}\natexlab{}.
\newblock \showarticletitle{{ReAct:} Synergizing Reasoning and Acting in Language Models}. In \bibinfo{booktitle}{\emph{The Eleventh International Conference on Learning Representations, {ICLR} 2023, Kigali, Rwanda, May 1-5, 2023}}. \bibinfo{publisher}{OpenReview.net}.
\newblock
\urldef\tempurl%
\url{https://openreview.net/forum?id=WE_vluYUL-X}
\showURL{%
\tempurl}


\bibitem[\protect\citeauthoryear{You, Zhang, Xu, Lou, Yan, Wang, Zhang, and Huang}{You et~al\mbox{.}}{2025}]%
        {you2025datawiseagent}
\bibfield{author}{\bibinfo{person}{Ziming You}, \bibinfo{person}{Yumiao Zhang}, \bibinfo{person}{Dexuan Xu}, \bibinfo{person}{Yiwei Lou}, \bibinfo{person}{Yandong Yan}, \bibinfo{person}{Wei Wang}, \bibinfo{person}{Huamin Zhang}, {and} \bibinfo{person}{Yu Huang}.} \bibinfo{year}{2025}\natexlab{}.
\newblock \showarticletitle{DatawiseAgent: A Notebook-Centric LLM Agent Framework for Adaptive and Robust Data Science Automation}. In \bibinfo{booktitle}{\emph{Proceedings of the 2025 Conference on Empirical Methods in Natural Language Processing}}. \bibinfo{publisher}{Association for Computational Linguistics}, \bibinfo{pages}{1099--1123}.
\newblock
\urldef\tempurl%
\url{https://doi.org/10.18653/v1/2025.emnlp-main.58}
\showDOI{\tempurl}


\bibitem[\protect\citeauthoryear{Zeng, Jin, Wang, Zheng, and Li}{Zeng et~al\mbox{.}}{2025}]%
        {zeng2025airepr}
\bibfield{author}{\bibinfo{person}{Qiuhai Zeng}, \bibinfo{person}{Claire Jin}, \bibinfo{person}{Xinyue Wang}, \bibinfo{person}{Yuhan Zheng}, {and} \bibinfo{person}{Qunhua Li}.} \bibinfo{year}{2025}\natexlab{}.
\newblock \showarticletitle{AIRepr: An Analyst-Inspector Framework for Evaluating Reproducibility of LLMs in Data Science}. In \bibinfo{booktitle}{\emph{Findings of the Association for Computational Linguistics: EMNLP 2025}}. \bibinfo{publisher}{Association for Computational Linguistics}, \bibinfo{pages}{10170--10201}.
\newblock
\urldef\tempurl%
\url{https://doi.org/10.18653/v1/2025.findings-emnlp.539}
\showDOI{\tempurl}


\bibitem[\protect\citeauthoryear{Zhao, Zhou, and Li}{Zhao et~al\mbox{.}}{2024}]%
        {chat2data2024}
\bibfield{author}{\bibinfo{person}{Xinyang Zhao}, \bibinfo{person}{Xuanhe Zhou}, {and} \bibinfo{person}{Guoliang Li}.} \bibinfo{year}{2024}\natexlab{}.
\newblock \showarticletitle{{Chat2Data}: An Interactive Data Analysis System with {RAG}, Vector Databases and {LLMs}}.
\newblock \bibinfo{journal}{\emph{Proc. {VLDB} Endow.}} \bibinfo{volume}{17}, \bibinfo{number}{12} (\bibinfo{year}{2024}), \bibinfo{pages}{4481--4484}.
\newblock
\urldef\tempurl%
\url{https://doi.org/10.14778/3685800.3685905}
\showDOI{\tempurl}


\bibitem[\protect\citeauthoryear{{ZJUNLP}}{{ZJUNLP}}{2026}]%
        {longds2026release}
\bibfield{author}{\bibinfo{person}{{ZJUNLP}}.} \bibinfo{year}{2026}\natexlab{}.
\newblock \bibinfo{title}{{LongDS-Bench}}.
\newblock
\newblock
\urldef\tempurl%
\url{https://github.com/zjunlp/DataMind/blob/main/longds/README.md}
\showURL{%
\tempurl}
\newblock
\shownote{Version 1.1 Lite. Accessed October 16, 2026.}


\end{thebibliography}

\end{document}